\documentclass[preprint,floatfix] {revtex4} 
\newcommand{\rvec}{\mathrm {\mathbf {r}}} 
\newcommand{\rpvec}{\mathrm {\mathbf {r'}}} 
\usepackage{graphicx}
\usepackage{subfigure}
\begin{document}

\title{A density functional method for general excited states in atoms}
\author{Amlan K.\ Roy}
\email{akroy@iiserkol.ac.in, akroy@chem.ucla.edu} 
\affiliation{Division of Chemical Sciences, Indian Institute of Science Education and \\
Research (IISER), Block FC, Sector III, Salt Lake, Kolkata-700106, India.}

\begin{abstract}
This chapter presents the development of a density functional theory (DFT)-based method for accurate, 
reliable treatment of various resonances in atoms. Many of these are known to be notorious for their 
strong correlation, proximity to more than one thresholds, degeneracy with more than one minima. Therefore 
these pose unusual challenges to both theoreticians and experimentalists. It is well-known that DFT has 
been one of the most powerful and successful tools for electronic structure calculation of atoms, 
molecules, solids in the past two decades. While it has witnessed diverse spectacular applications for 
ground states, the same for excited states, unfortunately, has come rather lately and remains somehow    
less conspicuous relatively. Our method uses a work-function-based exchange potential in conjunction with 
the popular gradient-corrected Lee-Yang-Parr correlation functional. The resulting Kohn-Sham equation, in the 
non-relativistic framework, is numerically solved accurately and efficiently by means of a generalized 
pseudospectral method through a non-uniform, optimal spatial discretization. This has been applied to a 
variety of excited states, such as low and high states; single, double, triple as well as multiple 
excitations; valence and core excitations; autoionizing states; satellites; hollow and doubly-hollow 
states; very high-lying Rydberg resonances; etc., of atoms and ions, with remarkable success. A thorough 
and systematic comparison with literature data reveals that, for all these systems, the exchange-only results 
are practically of Hartree-Fock quality; while with inclusion of correlation, this offers excellent agreement 
with available experimental data as well as those obtained from other sophisticated theoretical methods. 
Properties such as individual state energies, excitation energies, radial densities as well as various 
expectation values are studied. This helps us in predicting many states for the first time. In summary, 
we have presented an accurate, simple, general DFT method for description of arbitrary excited states of 
atoms and ions. 
\end{abstract}
\maketitle

\section{Introduction}
A central objective in modern quantum chemistry, much of materials science, condensed matter physics and
various other branches in science today, is, stated simply, to understand structure, dynamics, energetics
through the application of rigorous principles of quantum mechanics. While these results usually complement 
the information obtained from physical, chemical and biological experiments, in many cases, this could also 
be used to predict and unveil many hitherto unobserved phenomena. Some of the most commonly used properties
are: calculating the potential energy surface (energy changes as a function of structural parameters), 
interaction energies (absolute as well as relative), electronic charge distribution, dipole and higher multipole
moments, spectroscopic quantities such as vibrational frequencies, NMR chemical shifts, ESR {\it g} tensors, 
hyperfine coupling constants, chemical reactivity, mechanistic routes of a reaction, cross sections for collision
with other particles, behavior of atoms/molecules under an external field, such as a strong laser field, etc.
The starting point for most such quantum mechanical studies is the non-relativistic Schr\"odinger equation: 
\begin{equation} 
H\Psi(\rvec, t) = E \Psi(\rvec, t)
\end{equation}
H is the Hamiltonian operator including various energy components whereas the solution of this equation yields 
total energy E, as well as the many-particle wave function, which contains all relevant informations about the 
system under investigation. The ultimate goal of achieving \emph{exact} solution of this equation is essentially 
beyond our reach except for a small number of highly simplified model cases. Our systems of interest contain 
many atoms and electrons, where the solution easily becomes unmanageable. The main difficulty arises due to the
presence of electron-electron interaction terms, and approximation methods must be invoked. This lead to the
development of multitude of \emph{ab initio} methods in today's electronic structure theory, one of the very
first and successful being the Hartree-Fock (HF) method.
 
\emph{Every} conceivable property of a many-electron interacting system can be obtained as a \emph{functional}
of the \emph{ground-state electron density}, $\rho(\rvec)$, thus replacing the complicated, intractable 
many-particle wave function. Stated otherwise, in principle, this scalar function of position, 
determines all the informations embedded in the many-body wave function of ground and all excited states. 
Existence of such functionals lies at the heart of density functional theory (DFT) and as such, the electron 
density is defined as,
\begin{equation}
\rho(\rvec_1) = N \int \cdots \int |\Psi(\mathrm{\mathbf{x_1}},\mathrm{\mathbf{x_2}},\cdots,
\mathrm{\mathbf{x_N}}) |^2 \ \mathrm{ds_1} \mathrm{d\mathbf{x_2}} \cdots 
\mathrm{d\mathbf{x_N}}
\end{equation}
The multiple integration involves integral over spin coordinates of all electrons (spin integration implies
summation over the possible spin states) and all but one of the spatial variables. Clearly $\rho(\rvec)$ is a 
real, non-negative, visualizable function of only three spatial coordinates (in contrast to the 4N dimensional
many-electron wave function for an N electron system) with direct physical significance (can be measured 
experimentally unlike the complex-valued, wave function), vanishes at infinity and integrates to the total 
number of electrons:
\begin{eqnarray}
\rho (\rvec \rightarrow \infty) & = & 0 \nonumber \\
\int \rho (\rvec) \mathrm{d} \rvec & = & N  
\end{eqnarray}
Because of its transparency in dealing with the problematic inter-electronic repulsion in a rigorous quantitative 
way plus favorable computational cost, DFT has been the most popular and beloved of quantum mechanical 
methods for atoms, molecules and solids for more than three decades or so.

The first attempt to use electron density as basic variable in the context of atoms/molecules, rather than 
wave function is almost as old as quantum mechanics itself. In the original quantum statistical model of Thomas
and Fermi \cite{thomas27, fermi27}, kinetic energy of electrons is approximated as an explicit functional 
of density, whereas nuclear-electron attraction and electron-electron repulsion contributions are treated in a 
classical manner. The following simple expression of kinetic energy is derived by assuming electrons to be
in the background of an idealized, non-interacting homogeneous electron gas, i.e., a fictitious model of system 
of constant electron density: 
\begin{equation}
T_{TF}[\rho(\rvec)]= \frac{3}{10}(3\pi^2)^{2/3} \int \rho(\rvec)^{5/3} \ \mathrm{d}\rvec
\end{equation}
Combining this with the classical terms, one can obtain the so-called celebrated Thomas-Fermi energy functional 
for electrons in an external potential $v_{ext}(\rvec)$ as follows,
\begin{equation}
E_{TF}[\rho(\rvec)] = T_{TF}[\rho(\rvec)] + \int v_{ext}(\rvec) \rho (\rvec) \ \mathrm{d} \rvec +
\frac{1}{2} \int \int \frac{\rho(\rvec) \rho(\rpvec)}{|\rvec-\rpvec|} \ \mathrm{d}\rvec \mathrm{d}\rpvec
\end{equation}
Here the last term represents classical electrostatic repulsion. Now, minimization of this functional 
$E[\rho(\rvec)]$ for all possible $\rho(\rvec)$ subject to the constraint on total number of electrons,
\begin{equation}
\int \rho(\rvec) \ \mathrm{d}\rvec = N
\end{equation}
leads to the ground-state density and energy. Since $T_{TF}[\rho]$ is only a very coarse approximation to 
true kinetic energy, as well as exchange and correlation effects are completely ignored, results obtained 
using this approach are rather crude. It misses the essential physics and chemistry, such as shell structure of
atoms and molecular binding. However, this illustrates the important fact that energy of an interacting system 
can be written \emph{completely} in terms of single-particle density \emph{alone}. This was further extended 
by Dirac \cite{dirac30} to incorporate exchange effects, leading to the so-called local density approximation 
(LDA). This gives significant improvements over the original TF method and very much is in use, still today, 
\begin{equation}
E_{TFD}[\rho(\rvec)] = T_{TF}[\rho(\rvec)] + \int v_{ext}(\rvec) \rho (\rvec) \ \mathrm{d} \rvec 
-\frac{3}{4} \left( \frac{3}{\pi} \right)^{1/3} \int \rho(\rvec)^{4/3} \mathrm{d}\rvec +
\frac{1}{2} \int \int \frac{\rho(\rvec) \rho(\rpvec)}{|\rvec-\rpvec|} \ \mathrm{d}\rvec \mathrm{d}\rpvec.
\end{equation}

Soon, the provocative simplicity of above density approach, compared to traditional wave-function-based 
methods influenced a considerably large number of calculations. However, due 
to the lack of a rigorous foundation (e.g., no variational principle was established as yet) combined with 
the fact that fairly large errors were encountered in solid-state and molecular calculations, the theory 
somehow lost its appeal and charm; realistic electronic structure calculation within such a deceptively 
simple route seemed a far cry, leading to very little practical impact on chemistry. However, the situation
was about to change after the landmark paper by Hohenberg and Kohn \cite{hohenberg64}, where this was put
on a firm theoretical footing. This earns it the status of an \emph{exact} theory of many-body system and 
eventually, laid the groundwork of all of modern DFT. The first HK theorem simply states that the external 
potential $v_{ext}(\rvec)$ 
in a many-electron interacting system is \emph{uniquely} determined, to within a constant, by ground-state 
density $\rho(\rvec)$. Now, since $H$ is fully determined (except to a constant), it easily follows that many-body
wave functions for \emph{all} states (ground and excited) are also determined. Thus all properties of the
system are completely determined by $\rho(\rvec)$ only. The proof is based on \emph{reductio ad absurdum} and
skipped here for brevity. 

Note that this is only an existence theorem and as such, completely unhelpful in providing any 
indication of how to predict the density of a system. The answer lies in the second theorem which states that, for 
any external potential $v_{ext} (\rvec) $, one can define a functional, $E[\rho]$ in terms of $\rho (\rvec)$, 
as follows,
\begin{equation}
E[\rho]= E_{ne}[\rho]+(T[\rho]+E_{ee}[\rho])=\int v_{ext}(\rvec) \rho (\rvec) \ \mathrm{d} \rvec +F_{HK} [\rho]
\end{equation} 
Here, the energy has been separated into two parts: one that depending on the actual system, i.e., the potential energy 
because of nuclear-electron attraction, and a universal term (in a sense that the form is independent of 
$N, R_A, Z_A$, or in other words, \emph{same for all electrons}), consisting of the kinetic and electron repulsion 
energy components. For a given $v_{ext} (\rvec)$, global minimum of this functional provides ground-state energy 
while, the density minimizing this functional corresponds to exact ground-state density as,
\begin{equation}
E_0= \min_{\rho \rightarrow N} \left( F[\rho] + \int \rho(\rvec) v_{ext}(\rvec) \ \mathrm{d}{\rvec} \right );
\ \ \ \ \  F[\rho]= \min_{\Psi \rightarrow \rho} \langle \Psi|T+E_{ee} | \Psi \rangle .
\end{equation}

It is worth noting that, this apparently simple-looking \emph{universal} or \emph{Hohenberg-Kohn} functional, 
$F_{HK}[\rho]$ is the holy grail of DFT. It remains absolutely silent about the explicit forms of functionals for
both $T[\rho]$ and $E_{ee}[\rho]$. Design of their accurate forms remains one of the major challenges and lies at 
the forefront of modern development works in DFT. Further partitioning of energy is possible through the following 
realization,
\begin{equation}
E_{ee}[\rho]= 
\frac{1}{2} \int \int \frac{\rho(\rvec) \rho(\rpvec)}{|\rvec-\rpvec|} \ \mathrm{d}\rvec \mathrm{d}\rpvec
+E_{nc}[\rho] = J[\rho]+E_{nc}[\rho].
\end{equation}
Here $J[\rho]$ signifies classical Coulomb repulsion whereas the last term is associated with the 
\emph{non-classical} contribution to electron repulsion, containing all the effects of exchange, correlation as
well as self-interaction correction.
 
Despite the charm and simplicity of HK theorem, very little progress could be made in terms of practical applications
to realistic atoms/molecules. \emph{All} this tells us is that, in principle, a unique mapping between ground-state density 
and energy exists. However, there is no guideline whatsoever, about the construction of this functional which 
delivers the ground-state energy. In so far as theoretical prediction of molecular properties are concerned 
through computational DFT, no conspicuous changes could be observed with the culmination of these theorems. Because
one is still left with the difficult problem of solving many-body system in presence of $v_{ext}(\rvec)$; calculations
are as hard as before the HK theory. The variational principle in second theorem also calls for caution. In any
real calculation, in absence of the exact functional, one is invariably left with no choice, but to use some 
approximate forms. The variational theorem, however, applies only to the case of \emph{exact} functionals, implying 
that in DFT world, energy delivered by a trial functional has absolutely no meaning. This is in sharp contrast to 
the conventional wave function-based, variational methods such as HF or CI, where the lower an energy E, better a
trial function approximates the true wave function.

A year later, in a ground breaking work, Kohn and Sham \cite{kohn65} proposed a route to approach the hitherto unknown
universal functional. The central part of their ingenious idea stems from the realization that the original many-body
problem of an interacting system could be replaced by an auxiliary, fictitious non-interacting system of particles. 
This ansatz, then, in principle, holds the promise for exact calculations of realistic systems using only an 
independent-particle picture of non-interacting fermions, which are \emph{exactly soluble} (in practice by numerical 
methods). The non-interacting reference system is constructed from a set of one-electron orbitals, facilitating the 
major portion of kinetic energy to be computed to a good accuracy (exact wave functions of non-interacting fermions 
are Slater determinants). The residual part of kinetic energy, which is usually fairly small, is then merged with the 
non-classical component of electron-electron repulsion, which is also unknown, 
\begin{eqnarray}
F[\rho] & = & T_s[\rho]+J[\rho]+E_{xc}[\rho] \nonumber \\
E_{xc}[\rho] & = & (T[\rho]-T_s[\rho]) + (E_{ee}[\rho]-J[\rho]) = T_c[\rho]+E_{nc}[\rho].
\end{eqnarray}
Here $T_s[\rho]$ corresponds to the exact kinetic energy of a hypothetical non-interacting system having the same
electron density as that of our real interacting system. The exchange-correlation (XC) functional $E_{xc}[\rho]$ now contains
everything that is unknown. Thus it includes not only the non-classical electrostatic effects of electron-electron 
repulsion, but also the difference of true kinetic energy $T_c[\rho]$ and $T_s[\rho]$. Now we are in a position to
write down the celebrated Kohn-Sham (KS) equation in its standard form, 
\begin{equation}
\left[ -\frac{1}{2} \nabla^2 +v_{eff}(\rvec) \right] \psi_i(\rvec) = \epsilon_i \psi_i(\rvec)
\end{equation}
where the ``effective" potential $v_{eff}(\rvec)$ includes the following terms, 
\begin{equation}
v_{eff}(\rvec)=v_{ext} (\rvec) + \int \frac{\rho(\rpvec)}{|\rvec-\rpvec|} \ \mathrm{d} \rpvec + v_{xc} (\rvec).
\end{equation}
Here $v_{eff}(\rvec)$ and $v_{ext}(\rvec)$ signify the effective and external potentials respectively. Literature
in DFT is very vast; many excellent books and review articles are available. Here we refer the readers to 
\cite{parr89,jones89,dreizler90,chong95,seminario96,joulbert98,dobson98,nagy98,kohn99,koch01,parr02,fiolhais03,
gidopoulos03,martin04,sholl09} for a more detailed and thorough exposition on the subject. 

So far, we have restricted our focus to the ground states. Now let us shift our attention to the calculation of excited 
states. It is well known that DFT has been one of the most powerful and successful tools in predicting numerous ground-state 
properties of many-electron systems such as atoms, molecules, solids, over the past four decades. However, mainly due 
to its inherent weaknesses, its extension to excited-state problems has been quite difficult and rather less 
straightforward, so much so that DFT is often dubbed as a \emph{ground-state theory}. Results on excited states 
remained very scarce until very recently.  Although considerable progress has been made by employing several different 
strategies to address the problem lately, there are many crucial unresolved issues as yet, which require further 
attention. 

At this point, however, it may be appropriate to discuss the major difficulties in dealing with excited states within 
the realm of Hohenberg-Kohn-Sham (HKS) DFT. As already mentioned in the beginning, very foundation of DFT relies on a
presupposition that the electron density alone is sufficient to describe \emph{all states} (both ground and excited) 
of a desired system. Indeed, $\rho(\rvec)$ contains all relevant informations on excited states as well besides
the ground state, but the problem is that no practical way to extract this information has been found out as yet. 
Moreover, there is no HK theorem for excited state parallel to ground state. This is presumably due to the fact that, 
for a general excited state, the wave function (in general, a complex quantity) can not be bypassed 
through the pure state density (a real quantity). Because from a hydrodynamical point of view, \emph{phase} part of the 
hydrodynamic function is \emph{constant} for ground and some excited states (the \emph{static} stationary states), 
but not so for a general excited state. Working completely in terms of single-particle density in contrast to the state 
function may be advantageous for ground states, but it is disadvantageous for excited states, because an individual 
excited state can not be characterized solely in terms of density. Besides, approximate functional forms of $T[\rho]$
and $E_{xc}[\rho]$, valid for both and ground states are unknown; certainly there are insurmountable difficulties 
in constructing \emph{exact} functionals for these. Note that, there is no reason that this functional will have same 
general form for both ground and an arbitrary excited state. Finally one also has to deal with the nagging and tedious
problem of ensuring both Hamiltonian and wave function orthogonalities, as in any standard variational calculation.  
Due to these reasons, excited states within DFT remains a very challenging and important area of research. 

The purpose of this chapter is to present a detailed account on an excited state density functional method, which has been
shown to be very promising for arbitrary excited states of atoms. This relies on a \emph{time-independent} SCF procedure
in contrast to the so-called \emph{time-dependent} DFT (TDDFT). Its success and usefulness has been well documented in 
a series of papers for diverse atomic excited states \cite{singh96,singh96a,roy97,roy97a,roy97b,roy98,singh98,singh99,
vikas00,roy02,roy04b,roy05b,roy07}. Section II gives a brief review of some of the most prominent DFT methods for 
excited states currently available in the literature. Section III presents the methodology and computational 
implementation used in our present work. Results from our calculation are discussed in Section IV, with reference to 
other theoretical methods as well as with experiments, wherever possible. Finally we conclude with a few remarks on 
the past, present and future of this method.

\section{A brief review of excited-state DFT}
In this section we will briefly mention some of the excited state DFT methods, with emphasis on those having 
realistic practical applications. However, before that, a very obvious question arises: Can the original KS equation 
(12) be excited? To answer this, we rewrite the expression for ``effective" potential in an atom in the following 
equivalent form,
\begin{equation}
v_{eff}(\rvec)=-\frac{Z}{r}+ \int \frac{\rho(\rpvec)}{|\rvec-\rpvec|} \ \mathrm{d} \rpvec + 
\frac{\delta E_{xc}[\rho]}{\delta \rho}.
\end{equation}
Here Z denotes nuclear charge on the atom. There is an inherent degeneracy in terms of electron spin. Moreover the 
other angular momentum quantum number ($l,m$) information required to characterize an individual excited state
is clearly missing. So it is not possible to select an excited state of a given space-spin symmetry corresponding 
to a particular electronic configuration. Next immediate question comes: can one possibly calculate an average
of a set of degenerate excited states? That depends on the form of XC functional used. As mentioned before, 
exact forms of XC functionals are unknown as yet and almost all existing functionals are designed for ground states 
only. The validity of these functionals for arbitrary excited states is unknown and in fact, common sense dictates 
that ground and excited state functional may not have same same; in all possibility they would be different.

Introduction of spin density, $\rho^S (\rvec) = \rho^{\alpha}(\rvec) - \rho^{\beta} (\rvec)$, made it possible for 
the development of more flexible potentials. Here $\rho^{\alpha}, \ \rho^{\beta}$ signify electron densities 
corresponding to the up, down spins respectively. It is easy to show that for different spin densities, 
$v_{eff}^{\alpha}(\rvec)$, $v_{eff}^{\beta}(\rvec)$ will be different; self-consistent solution of the
spin-polarized KS equation can be obtained. A thermodynamical version of the spin-density formalism has been
presented \cite{gunnarsson76}, which proves that standard DFT methods could be used for calculation of lowest
lying state of each spatial or spin irreducible representation of a given system, since, in a sense, these 
represent the ``ground state" in that particular symmetry. However there are inherent problems where the 
non-interacting case does not reduce to a single determinant. Many excited states can not be described by a single
determinant and intrinsically need a multi-configurational description. For example, consider the open-shell 
configuration $p^2$, as in the ground state of carbon atom, giving rise to three multiplets $^3P, ^1D, ^1S$. In 
accordance with the above discussion, one can obtain the energy of $^3P$ and $^1D$ terms, but not $^1S$. This can 
be understood from a consideration of the fact that, in absence of spin-orbit coupling, all valid states of an 
atom must be simultaneous eigenfunctions of not only the Hamiltonian operator $H$, but also angular momentum 
operators $L^2, L_z, S^2, S_z$, as well as parity operator $\Pi$, such that any given state is characterized by 
the following quantum numbers $L, M_L, S, M_S, \pi$, associated with these operators. In traditional wave functional
methods, these informations are carried by wave functions. But in DFT, the basic variable is electron density; so,
in principle, it should be $E_{xc}[\rho]$, which should contain this dependence on these above quantum numbers. 
On the other hand, most of the functionals in DFT depend solely in terms of charge or spin densities; therefore
clearly lacking any of these mentioned quantities needed for a complete description of the state of an atom. 
Stated otherwise, we are facing a very important conceptual problem here: how can one describe the state of a 
many-electron system, which are eigenfunctions of all these operators working entirely in terms of density with 
no access to the correct N-electron wave function and its associated symmetry characteristics?

This problem is partially resolved by making reference to Slater's transition state theory \cite{slater72,slater74}.
Following this prescription, an \emph{ad-hoc} approach to solve the multiplet problem was proposed, the so-called 
\emph{sum method} \cite{ziegler77,barth79}. Here the working equations are exactly like the KS equations, but 
density is assumed to correspond to a fictitious transition state where one or more orbitals are fractionally
occupied. These authors argue that energy of a term, not representable by a 
single determinant, but requires a linear combination of determinants, can not be computed by spin densities of 
the corresponding state functions, but can be written as a weighted sum of determinantal energies as follows:
\begin{equation}
E(M_j)= \sum_i A_{ji} E(D_i)
\end{equation}
Here $E(M_i)$, $E(D_i)$ refer to the energies of different multiplets and determinants respectively. Note that
the sum method has no firm theoretical justification; rather an empirical extension of the HF to the $X\alpha$ 
method. By using some elegant group theoretical method, a semi-automatic protocol has been developed to obtain 
the weights of various determinants \cite{daul94,stuckl97}. 

The above intuitive ideas was utilized on a rigorous foundation to formulate a scheme in which the density is a
sum of $M$ lowest-energy eigenstate densities with equal weightage \cite{theophilou79}. Here we discuss at some
length the \emph{ensemble density} or \emph{fractional occupation approach} to excited states 
\cite{gross88,gross88a,oliveira88}. This suggests to work with an ensemble of densities rather than a pure-state 
density and the subspace formulation \cite{theophilou79} can be considered a special case of it.

The generalized eigenvalue problem for a time-independent Hamiltonian $H$ with $M$ eigenvalues $E_1 \le E_2 \le 
\cdots  \le E_M$ for its $M$ low-lying states is:
\begin{equation}
H\Psi_k=E_k\Psi_k \ \ \ (k=1,2,\cdots,M)
\end{equation}
Applying the Rayleigh-Ritz variational principle, one can write the ensemble energy as:
\begin{equation}
\mbox{\boldmath $\cal{E}$}= \sum_{k=1}^M w_k E_k, \ \ \ 0 \leq w \leq 1/M, \ \ \ 1\leq g \leq M-1
\end{equation}
where $w_1 \geq w_2 \geq \cdots \geq w_M \geq 0$ are the weighting factors chosen such that: $w_1 = w_2 = \cdots = 
w_{M-g} = \frac{1-w}{M-g}$, $w_{M-g+1}=w_{M-g+2} = \cdots = w_M =w$. The limit $w=0$ corresponds to the 
eigenensemble of $M-g$ states ($w_1= w_2 = \cdots = w_{M-g} = \frac{1}{M-g}$ and $w_{M-g+1}= w_{M-g+2} = \cdots =
w_M=0)$, whereas $w=1/M$ leads to the eigenensemble of M states  ($w_1=w_2= \cdots = w_M = 1/M$).

The generalized HK theorem can be established; as well as the KS equation for ensembles following the variational 
principle in standard manner,
\begin{equation}
\left[ -\frac{1}{2} \nabla^2 + v_{KS} \right] \ u_i(\rvec) = \epsilon_i u_i (\rvec) 
\end{equation}
where the ensemble KS potential, given below,  
\begin{equation}
v_{KS}(\rvec; \rho_w) = v_{ext} (\rvec) + \int \frac{\rho_w(\rvec)}{|\rvec-\rpvec|} \ \mathrm{d}\rvec 
+ v_{xc} (\rvec; w,\rho_w)
\end{equation}
could be defined as a functional of the ensemble density as follows,
\begin{equation}
\rho^I_w (\rvec) = \frac{1-wg_I}{M_{I-1}} \sum_{m=1}^{M_I-g_I} \sum_j \lambda_{mj}\ |u_j(\rvec)|^2 
+ w \sum_{m=M_I-g_I+1}^{M_I} \sum_j \lambda_{mj}\ |u_j(\rvec)|^2.
\end{equation}
Here $g_I$ denotes degeneracy of the $I$th multiplet, $M_I=\sum_{i=1}^{I} g_i$ defines multiplicity of the 
ensemble, $\lambda_{mj}$ are occupation numbers, with $0 \leq w \leq 1/M_I$. Density matrix is defined as, 
\begin{equation}
P^{M,g}(w)= \sum_{m=1}^{M} w_m |\Psi_m \rangle \langle \Psi_m|.
\end{equation}
The XC potential $v_{xc}$ is the functional derivative of ensemble XC energy functional $E_{xc}$,
\begin{equation}
v_{xc}(\rvec; w, \rho) = \frac{\delta E_{xc}[\rho,w]}{\delta \rho (\rvec)}
\end{equation}
One can then express excitation energies in terms of one-electron energies $\epsilon_j$,
\begin{equation}
\overline{E}^I= \left. \frac{1}{g_I} \ \frac{d \mbox{\boldmath $\cal{E}$}^I(w)} {dw} \right|_{w=w_I} +
\ \sum_{i=2}^{I-1} \frac{1}{M_I} \ \left. \frac{d \mbox{\boldmath $\cal{E}$}^I(w)} {dw} \right|_{w=w_i}
\end{equation}
where
\begin{equation}
\frac{d \mbox{\boldmath $\cal{E}$}^I(w)} {dw} = \sum_{j=N+M_{I-1}}^{N-1+M_I} \epsilon_j - 
\frac{g_I}{M_{I-1}} \sum_{j=N}^{N-1+M_{I-1}} \epsilon_j + \left. \frac{\partial E_{xc}^I} {\partial w} \right|_{\rho_w}
\end{equation}
Clearly, excitation energies cannot be calculated as a difference of the one-electron energies; there is an extra quantity 
(the last term) that needs to be determined.

The two-particle density matrix of the ensemble is obtained as an weighted sum of two-particle density matrices of ground 
and excited states as follows,
\begin{equation}
\Gamma^{M,g,w}(\rvec_1,\rvec_2; \rpvec_1,\rpvec_2) = \sum_{m=1}^{M} w_m \Gamma^{m}(\rvec_1,\rvec_2; \rpvec_1,\rpvec_2)
\end{equation}
The total ensemble density takes the following form, 
\begin{equation}
\mbox{\boldmath $\cal{E}$}^{M,g}_w =\mathrm{Tr} \{P^{M,g}(w)H \} = 
\mathrm{Tr} \{P^{M,g}(T+E_{ee}) \} + \mathrm{Tr} \{P^{M,g}(w)V \} = F^{M,g}(w)+ \int \rho(\rvec) v_{ext}(\rvec) 
\mathrm{d} \rvec
\end{equation}
where $\rho(\rvec)$ is the ensemble density; $V= \sum_{i=1}^N v_{ext}(\rvec_i)$. The ensemble XC energy is given by,
\begin{equation}
E_{xc}^{M,g}[w,\rho]=F^{M,g}[w,\rho]-T_s^{M,g}[w,\rho] -J[\rho].
\end{equation}
Here the last two terms denote ensemble non-interacting kinetic and Coulomb energies.

Solution of the ensemble KS equation (18) requires knowledge of the ensemble XC potential, exact form of which remains
unknown and several approximations have been proposed. In \cite{gross88, gross88a, oliveira88}, excitation energies
of He were studied using the quasi-local density approximation \cite{kohn86}. First excitation energies of several
atoms \cite{nagy90} as well higher excitation energies \cite{nagy91} have been reported using the parameter-free 
exchange potential of G\'asp\'ar \cite{gaspar74}, which depends explicitly on spin orbitals. Several ground-state LDA
functionals have been employed for this purpose: \emph{viz.}, Gunnarsson-Lundqvist-Wilkins \cite{gunnarsson74},
von Barth-Hedin \cite{barth72}, Ceperley-Alder \cite{ceperley80}, local density approximations parametrized by Perdew
and Zunger \cite{perdew81} and Vosko \emph{et al.} \cite{vosko80}. It is found that, in general, spin-polarized 
calculations provide better results compared to the non-spin-polarized ones; however, in most cases, estimated
excitation energies are highly overestimated. Generally, these functionals provide results which are in close
agreement with each other. These references use minimum (0) and maximum values of the weighting
factor $w$. Any $w$ value satisfying the inequality in (18) is appropriate, provided that one uses the \emph{exact} XC 
energy. However since the latter is unknown, one has to take recourse to approximate functionals; thus different 
excitation energies are obtained with different $w$. This variation is studied in detail in \cite{nagy91}; in some 
occasions the change is small, while for others considerably large variation is observed. Simple local \emph{ensemble
potential} has been proposed \cite{nagy96} for this purpose as well,
\begin{equation}
v_x(\rho_w,w) = -3\alpha(w) \left( \frac{3\rho_w}{8\pi} \right)^{1/3}; \ \ \ \ 
E_x[\rho_w,w]=-\frac{9}{4} \left( \frac{3}{8\pi} \right)^{1/3} \alpha(w) \int \rho_{w}^{4/3} \mathrm{d} \rvec
\end{equation}
However, calculated excitation energies are still very far from the actual values. This leads us to the conclusion 
that like the ground-state DFT, search for accurate XC functional again remains one of the major bottlenecks in 
the success of ensemble or fractional occupation approach to excited-state energies and densities.

In another development \cite{fritsche86,cordes89}, KS equations were obtained by partitioning the wave function into
following two components,
\begin{equation}
\psi(\rvec_1,\rvec_2, \cdots ,\rvec_N) = \phi(\rvec_1, \rvec_2, \cdots ,\rvec_N) +
\tilde {\psi}(\rvec_1,\rvec_2, \cdots ,\rvec_N)   
\end{equation}
such that the two-particle density matrix becomes, 
\begin{equation}
\rho_2(\rpvec,\rvec)= \rho^0_2 (\rpvec,\rvec)+ \tilde{\rho}_2(\rpvec,\rvec)
\end{equation}
where
\begin{eqnarray}
\rho^0_2 (\rpvec,\rvec) & = & N(N-1) \int |\phi(\rpvec,\rvec,\rvec_3, \cdots ,\rvec_N|^2 
\ \mathrm{d}^4 \rvec_3 \ \mathrm{d}^4 \rvec_4 \ \cdots \ \mathrm{d}^4 \rvec_N  \nonumber \\
\tilde{\rho}_2 (\rpvec,\rvec) & = & N(N-1) \int [\phi^* \tilde{\psi} + \phi \tilde{\psi}^* + \tilde{\psi}\tilde{\psi}^* ] 
\ \mathrm{d}^4 \rvec_3 \ \mathrm{d}^4 \rvec_4 \ \cdots \ \mathrm{d}^4 \rvec_N  
\end{eqnarray}
A factor of 2 is included in $\rho^0_2$ and $\tilde{\rho}_2$. The symbol $\cdots \int \mathrm{d}^4 \rvec_j$
stands for real-space integration and spin summation for the jth particle. The spin-independent one-particle density
$\rho_s(\rvec)$ is,
\begin{equation}
\rho_s(\rvec) = \frac{1}{N-1} \sum_{s'} \int \rho^0_2(\rpvec,\rvec) \ \mathrm{d}\rpvec
\end{equation}
As a result,
\begin{equation}
\sum_{s'} \int \tilde{\rho}_2 (\rpvec,\rvec) \ \mathrm{d} \rpvec =0
\end{equation} 
The above properties uniquely define two components $\phi(\rvec_1,\rvec_2, \cdots ,\rvec_N)$ and $\tilde{\psi}
(\rvec_1, \rvec_2, \cdots ,\rvec_N)$ of any eigenstate $\psi(\rvec_1,\rvec_2, \cdots ,\rvec_N)$. Now, the variational
optimization involving the N-particle Hamiltonian yields the following KS type equation, 
\begin{equation}
\left[ \frac{1}{2} \nabla^2 + v_{ext}(\rvec) + v_H(\rvec) + v_{xc}^s (\rvec) \right] \ \psi_{is}(\rvec) 
= \epsilon_{is} \psi_{is}(\rvec)
\end{equation}
where the three $v$ terms denote external, Hartree and XC potentials respectively,
\begin{equation}
v^s_{xc} (\rvec) = - \sum_{s'} \int \frac{\rho_{s'}(\rpvec)}{|\rvec-\rpvec|} 
\left[ f_{s',s} (\rpvec,\rvec) + \frac{1}{2} \frac{\sum_{s''} \delta f_{s',s''} (\rpvec,\rvec) \rho_{s''}(\rvec)}
{\delta \rho_s(\rvec)} \right] \ \mathrm{d} \rpvec.
\end{equation}
Here $f_{s's}(\rpvec,\rvec) = 1-g_{s's}(\rpvec,\rvec)$ and $g_{s's}(\rpvec,\rvec)$ is the pair correlation 
function defined as $\rho_2^{s's}(\rpvec,\rvec) = \rho_{s'} (\rpvec) \rho_s(\rvec) g_{s's}(\rpvec,\rvec)$.
The method has produced reasonably good agreements with experimental as well as other density functional methods
for total ground-state energies of free atoms, ionization and affinity energies, etc. \cite{cordes89}. However, 
there are significant difficulties as far as practical computations are concerned for general excited states.

A configuration-interaction scheme restricted to single excitations (CIS) has been used in the realm of 
DFT for electronic excitations \cite{grimme96}. HF orbital energies in the matrix elements of CIS Hamiltonian are 
then replaced by the corresponding eigenvalues obtained from gradient-corrected KS calculations. Additionally it
requires three empirical parameters determined from a reference set to scale the Coulomb integrals and shifting 
the  diagonal CIS matrix elements. Even though this also suffers from a lack of a solid theoretical foundation, 
resultant excitation energies of molecules obtained by this method show fairly good agreement. This has also been 
extended to multi-reference CI schemes \cite{grimme99}.

Many other attempts have been made to calculate individual excited states. Some important ones are mentioned below. 
A time-independent quantal density functional theory (Q-DFT) of singly or multiply excited bound non-degenerate 
states have been proposed \cite{sahni01}. Existence of a variational KS DFT, with a minimum principle, for the 
self-consistent determination of an individual excited state energy and density has been established \cite{levy99}.
A perturbative approach is also suggested \cite{gorling96,filippi97}, where the non-interacting KS Hamiltonian serves
as the zeroth-order Hamiltonian. Two variants of perturbation theory (PT) were used: (a) the so-called \emph{standard
DF PT}, where zeroth-order Hamiltonian takes the form $H_0=T+V_{ext}+V_H+V_{xc}$, and the perturbation is given by
$H_1=V_{ee}-V_H-V_{xc}$, (b) \emph{the coupling-constant PT} is based on the adiabatic connection of the Hamiltonian where
a link is made between KS Hamiltonian and fully interacting Hamiltonian keeping ground-state density constant, 
independent of $\alpha$, such that, $H^{\alpha}=T+V_{ext}+V_H+V_{xc}+\alpha(V_{ee}-V_H-V_x)-V^{\alpha}_c.$ Here 
$V^{\alpha}_c$ is second order in $\alpha$ and equals the correlation potential when $\alpha=1$. Zeroth-order
Hamiltonian is again the KS Hamiltonian; the perturbing Hamiltonian contains a term $H_1^{(1)}=\alpha 
(V_{ee}-V_H-V_x)$, which is linear in perturbation parameter $\alpha$ and a component $-v^{\alpha}_c$, which contains 
second and higher order contributions. Accurate calculation of correlation energies of excited states has been 
proposed via a suitable multi-reference DFT method (such as MCSCF including the complete active space). They successfully
describe the \emph{non-dynamical} correlation; the fraction of dynamic correlation can be taken into account by DFT
\cite{sanfabian03}. Applicability of subspace DFT for atomic excited states have been studied \cite{tasnadi03}. 
Theories for individual excited states have been proposed elsewhere as well \cite{gorling99}. A localized HF-based 
DFT has been put forth for excitation energies of atoms, molecules \cite{vitale05}. This is based on separating the
electron-electron interaction energy of KS wave function of a given excited state as Coulomb and exchange energy
as follows \cite{sala02}. The former is given as, 
\begin{equation}
U=\sum_{\Gamma,a,\Lambda,b} f_a^{\Gamma}f_b^{\Lambda} \times
\sum_{\gamma,\lambda} \int \mathrm{d} \rpvec \mathrm{d} \rvec'' \ 2  \ 
\frac{\phi_a^{\Gamma,\gamma}(\rpvec) \phi_a^{\Gamma,\gamma}(\rpvec) 
\phi_b^{\Lambda,\lambda}(\rvec'') \phi_b^{\Lambda,\lambda} (\rvec'')} {|\rpvec-\rvec''|} =
\frac{1}{2} \int \mathrm{d}\rpvec \mathrm{d}\rvec'' \ 
\frac{\bar{\rho}(\rpvec) \bar{\rho}(\rvec'')} {|\rpvec-\rvec''|}.
\end{equation}
The totally symmetric part $\bar{\rho}(\rvec)$ of electron density being given by, 
\begin{equation}
\bar{\rho}(\rvec) = 2 \ \sum_{\Gamma,a,\gamma} f_a^{\Gamma} \phi_a^{\Gamma,\gamma} (\rvec) 
\phi_a^{\Gamma,\gamma} (\rvec) 
\end{equation}
Here $\phi_a^{\Gamma,\gamma}(\rvec)$ denotes the spatial part of orbital which belongs to the energy level $a$ of 
irreducible representation $\Gamma$ and which transforms under symmetry operations according to the symmetry 
partner $\gamma$ of $\Gamma$. Occupation number of energy level $a$ of $\Gamma$ is denoted by $f_a^{\Gamma}$. 
Summation indices $a,b$ run over all at least partially occupied, i.e., not completely unoccupied levels of $\Gamma$.
The exchange energy is then the reminder of electron-electron interaction energy of the KS wave function. The 
corresponding open-shell localized HF exchange potential is then expressed as $v_x^{OSLHF} (\rvec) = 
v_x^s (\rvec)+ v_x^{c} (\rvec)$, where the two terms represent a generalized Slater potential and a correction term
respectively. 

Recently an optimized effective potential approach and its exchange-only implementation for excited states has been 
reported \cite{glushkov07}. This uses a bifunctional DFT for excited states \cite{levy99,nagy01} that employs a 
simple method of taking orthogonality constraints into account (TOCIA) \cite{glushkov02,glushkov02a} for solving 
eigenvalue problems with restrictions. A $\Delta$SCF approach, or $\Delta$KS approach \cite{besley09}, wherein the 
excitation energy is simply the difference in energy between ground- and excited-state HF or KS calculation, has 
been employed as well, and found to be especially successful for core-excited states. 

So far all the methods we have discussed lie within the purview of \emph{time-independent} DFT. Now we move on to the
formalisms for excitation energies within the TDDFT framework. Many excellent articles and reviews are available on 
the subject \cite{casida95,bauernschmitt96,petersilka96,tozer98,gisbergen98,stratmann98,gorling99a,yabana99,heinze00,
furche01,grimme04,burke05}; here we mention only the essential details. Consider the unperturbed, ground-state of a 
many-electron system characterized by an external potential $v_0(\rvec)$, subject to a TD perturbation
$v_1(\rvec,t)$, such that at a later time, the external potential (a functional of the TD density), is given by  
$v_{ext}(\rvec,t)=v_0(\rvec)+v_1(\rvec,t)$. The density-density response function takes the form, 
\begin{equation}
\chi(\rvec,t; \rpvec,t') = \left. \frac{\delta \rho[v_{ext}](\rvec,t)}{\delta v_{ext}(\rpvec,t')} \right|_{v[\rho_0]}
\end{equation} 
where the functional derivative needs to be evaluated at the external potential corresponding to an unperturbed
ground-state density $\rho_0$. The first-order, linear density response to the perturbation $v_1(\rvec,t)$ is then
given by, 
\begin{equation}
\rho_1(\rvec,t) = \int \mathrm{d}t' \int \mathrm{d}\rpvec \chi(\rvec,t; \rpvec,t') \ v_1(\rpvec, t')
\end{equation}

Now, realizing that the Runge-Gross theorem also holds for non-interacting particles moving in an external 
potential $v_s(\rvec,t)$, one can write the KS response function of a non-interacting, unperturbed many-electron 
density $\rho_0$ as,
\begin{equation}
\chi_s(\rvec,t; \rpvec,t') = \left. \frac{\delta \rho[v_s](\rvec,t)} {\delta v_s(\rpvec,t')} \right|_{v_s[\rho_0]}
\end{equation} 
With this definition, $\chi_s(\rvec,t; \rpvec,t')$ is expressed in terms of static KS orbitals $\{\phi_k\}$,
\begin{equation}
\chi_s(\rvec, \rpvec;\omega) = \sum_{j,k} (f_k-f_j) 
\frac{\phi_k^*(\rvec) \phi_j(\rvec) \phi_j^*(\rpvec) \phi_k(\rpvec)}{\omega-(\epsilon_j-\epsilon_k)+i\delta }
\end{equation}
Here $f_k,f_j$ denote occupation numbers of KS orbitals; $\epsilon_j, \epsilon_k$ signify KS orbital energies; and
$\omega$ is the frequency obtained after applying a Fourier transform with respect to time. Summation includes both
occupied and unoccupied orbitals, plus the continuum states. Now, the first-order density change $\rho_1 (\rvec,t)$ in
terms of linear response of the non-interacting system to the effective perturbation $v_{s,1}(\rvec,t)$ can be written
in terms of frequency $\omega$ as, 
\begin{eqnarray}
\rho_1(\rvec, \omega) & = & \int \chi_s (\rvec, \rpvec; \omega) \ v_1 (\rpvec, \omega) \ \mathrm{d}\rpvec +
\\ \nonumber
& & \int \int \chi_s (\rvec, \rpvec; \omega) \times 
\left( \frac{1}{|\rpvec-\rvec''|}+f_{xc}[\rho_0](\rpvec,\rvec''; \omega)\right) \rho_1 (\rvec'',\omega) 
\ \mathrm{d}\rpvec \ \mathrm{d} \rvec''
\end{eqnarray}

It is established that the frequency-dependent linear response of a finite interacting system has discrete poles 
at the true excitation energies $\Omega_m=E_m -E_0$ of an unperturbed system. So the idea is to calculate the shifts
in KS orbital energy differences $\omega_{jk} = \epsilon_j-\epsilon_k$, which are poles of the KS response function.
True excitation energies ($\Omega$) are generally \emph{not} identical with the KS excitation energies $\omega_{jk}$. 
The exact density response $\rho_1$, however, has poles at true excitation energies $\omega= \Omega$. True excitation 
energies can then be described by those frequencies where the eigenvalues $\lambda(\omega)$ of the following equation, 
\begin{equation}
\int \mathrm{d}\rvec \int \mathrm{d} \rpvec \chi_s(\rvec'',\rvec; \omega) 
\left( \frac{1}{|\rvec-\rpvec|} +f_{xc} [\rho_0] (\rvec,\rpvec; \omega) \right)  \xi (\rpvec, \omega) =
\lambda(\omega) \ \xi (\rpvec, \omega)
\end{equation}
satisfy $\lambda(\Omega) = 1$. For practical purposes, one needs to expand $\Omega$ about one particular KS
energy difference $\omega_{\nu}= \omega_{jk}$:
\begin{eqnarray}
\chi_s(\rvec'',\rvec; \omega) & = & 2 \alpha_{\nu} \frac{\Phi_{\nu}(\rvec'') \Phi_{\nu}^*(\rvec)}
{\omega-\omega_{\nu}} + 2 \sum_{k \neq \nu} \alpha_k \ 
\frac{\Phi_k(\rvec'') \Phi_k^* (\rvec)}{\omega_{\nu}-\omega_k+i\delta } + \cdots \nonumber \\
f_{xc}[\rho_0](\rvec, \rpvec; \omega) & = & f_{xc} [\rho_0] (\rvec, \rpvec; \omega_{\nu}) \ 
\left. \frac{\mathrm{d}f_{xc}[\rho_0] (\rvec,\rpvec; \omega) }{\mathrm{d} \omega } 
\right|_{\omega_{\nu}} (\omega-\omega_{\nu}) + \cdots  \nonumber \\
\xi(\rvec'', \omega) & = & \xi (\rvec'', \omega_{\nu}) + 
\left. \frac{\mathrm{d}\xi(\rvec'',\omega)}{\mathrm{d} \omega} 
\right|_{\omega_{\nu}} (\omega-\omega_{\nu}) + \cdots \nonumber \\
\lambda(\omega) & = & \frac{A(\omega_{\nu})}{\omega-\omega_{\nu}} + B(\omega_{\nu}) + \cdots 
\end{eqnarray}
The index $\nu=(j,k)$ denotes a contraction implying a single-particle transition ($k\rightarrow j$), i.e., 
$\Phi_{\nu}(\rvec)=\Phi_k^*(\rvec)\Phi_j(\rvec)$ and $\alpha_{\nu}=n_k-n_j$. Assuming that the true excitation 
energy is not too far away from $\omega_{\nu}$ and inserting Laurent expansions for $\chi_s, f_{xc},\xi,\lambda$
into the above expressions, one finds that, 
\begin{eqnarray}
A(\omega_{\nu}) & = & M_{\nu \nu}(\omega_{\nu})   \\
B(\omega_{\nu}) & = & \left. \frac{\mathrm{d}M_{\nu \nu}}{\mathrm{d}\omega} \right|_{\omega_{\nu}} 
+ \frac{1}{M_{\nu \nu}(\omega_{\nu})} \ 
\sum_{k \neq \nu} \frac{M_{\nu k}(\omega_{\nu}) M_{k\nu}(\omega_{\nu})}
{\omega_{\nu} - \omega_k + i\delta }     \nonumber
\end{eqnarray}
where the matrix elements are given by, 
\begin{equation}
M_{k\nu}(\omega_{\nu})= 2\alpha_{\nu} \int \int \phi_k^*(\rvec) \left( \frac{1}{|\rvec-\rpvec|} + 
f_{xc}(\rvec, \rpvec; \omega) \right) \ \phi_{\nu}(\rpvec) \mathrm{d}\rvec \mathrm{d} \rpvec
\end{equation}
So, the condition $\lambda(\Omega)=1$ and its complex conjugate then, leads to, in lowest order, 
\begin{equation}
\Omega = \omega_{\nu} + \Re M_{\nu \nu} 
\end{equation}
Just like the time-independent case, now one has to approximate the TD XC potential. The simplest construction is
the adiabatic approximation, which makes use of ground-state XC potential, but replaces ground-state density
$\rho_0(\rvec)$ with the instantaneous TD density $\rho(\rvec,t)$, 
\begin{equation}
v_{xc}^{ad} ([\rho];\rvec,t) = \left. \frac{\delta E_{xc} [\rho_0(\rvec)]}{\delta \rho_0(\rvec)} 
\right|_{\rho_0(\rvec)=\rho(\rvec,t)}
\end{equation}
Within the adiabatic approximation, the XC kernel can be calculated from, 
\begin{equation}
f_{xc}^{ad} (\rvec,t; \rpvec, t') \equiv \frac{\delta v_{xc}([\rho_0];\rvec)}{\delta \rho_0 (\rpvec)} \ 
\delta (t-t')
\end{equation}

The kernel above is local in time, but not necessarily local in space. Clearly, this approximation completely neglects 
the frequency dependence arising from the XC vector potential; consequently the retardation and dissipation effects are
completely ignored in this picture. 
This has been widely used for single-particle excitation energies (see, for example, \cite{dreuw03,tozer03,jacquemin06,
adams07}, for some recent work) with good success, although it performs rather poorly for multiple excitations and
charge-transfer states. For explicit functionals of density, it is straightforward to calculate XC kernel. However, 
for orbital-dependent, such as meta-generalized gradient approximated (GGA) or hybrid functionals, it is not so and 
may be evaluated with the help of optimized effective potential or other simple, accurate approach. 

In practice, modern TDDFT excitation energies $\omega$ and corresponding response functions {\bf X}, {\bf Y} are 
generally obtained by solving a non-Hermitian eigenvalue equation, 
\begin{equation}
\left(  \begin{array}{cc}
\mathbf{A} & \mathbf{B} \\
\mathbf{B} & \mathbf{A} 
\end{array}  \right)
\left( \begin{array}{c} 
\mathbf{X} \\
\mathbf{Y} 
\end{array} \right) = \omega
\left(  \begin{array}{cc}
\mathbf{1} & \mathbf{0} \\
\mathbf{0} & \mathbf{-1} 
\end{array}  \right)
\left( \begin{array}{c} 
\mathbf{X} \\
\mathbf{Y} 
\end{array} \right)
\end{equation}
Here {\bf X,Y} are the excitation vectors representing excitation, deexcitation components of electronic density 
change, whereas the elements of {\bf A,B} are given by, 
\begin{equation}
A_{ai\sigma,bj\sigma'} = \delta_{ab} \delta_{ij} \delta_{\sigma \sigma'} (\epsilon_{a\sigma} - \epsilon_{i\sigma'}) 
+K_{ai\sigma,bj\sigma'}, \ \ \ \ \
B_{ai\sigma,bj\sigma'}= K_{ai\sigma,jb\sigma'} 
\end{equation}
where $\sigma, \sigma'$ denote spin indices, $\epsilon_{p\sigma}$ is the $p$th KS molecular orbital energy.
Indices $i,j,\cdots$ and $a,b,\cdots$ correspond to occupied, virtual orbitals. Matrix element
$K_{ai\sigma,bj\sigma'}$ is given by, 
\begin{equation}
K_{pq\sigma,rs\sigma'} = (pq\sigma|rs\sigma') -c_x \delta_{\sigma \sigma'} (pr\sigma | qs \sigma') +
f_{pq\sigma rs \sigma'}^{xc}.
\end{equation}
Here $p,q,\cdots$ indicate general MOs and $(pq\sigma|rs\sigma')$ identifies a two-electron repulsion integral in
the Mulliken notation, whereas $c_x$ is a mixing parameter of HF exchange integral in case of hybrid functionals. 
The last term, $f_{pq\sigma rs \sigma'}^{xc}$ represents a Hessian matrix element of the XC energy functional
$E_{xc}$ in terms of density, in the adiabatic approximation,
\begin{equation}
f_{\sigma \sigma'}^{xc}= \frac{\delta^2 E_{xc}}{\delta \rho_{\sigma}(\rvec_1) \delta \rho_{\sigma'} (\rvec_2)}
\end{equation}
Finally note that, if the orbitals are assumed real, following matrices can be defined,
\begin{equation}
\mathbf{F}= (\mathbf{A}-\mathbf{B})^{-1/2} (\mathbf{X}+\mathbf{Y}), \ \ \ \ 
\mathbf{\Omega}= (\mathbf{A}-\mathbf{B})^{1/2} (\mathbf{A}+\mathbf{B}) (\mathbf{A}-\mathbf{B})^{1/2},
\end{equation}
to express the problem in compact form, 
\begin{equation}
\mathbf{\Omega F}= \omega^2 \mathbf{F}
\end{equation}

Although, in general, the non-adiabatic correction is needed for even in low-frequency limit, it has been 
demonstrated that, at least for smaller systems, the largest source of error for accurate excitation energies, arises
from the approximation to static XC potential. This justifies validity and wellness of adiabatic approximation for 
low-lying excitations in atoms, molecules. Applicability and performance of other functionals such as adiabatic, 
non-empirical meta-GGA as well other adiabatic hybrid functionals have been reported lately \cite{tao08}. XC functionals
with varying fractions of HF exchange \cite{nakata06,nakata07}, self-interaction correction \cite{imamura07} or
other combinations \cite{imamura06}, ``short-range" corrected functionals \cite{song08}, methods such as CIS(D) which
use exact exchange \cite{asmuruf08} have been suggested for better representation. 

Numerous other variations of TDDFT for excited states have been proposed. In a resolution of identity (RI-J) 
approach to analytical TDDFT excited-state gradients \cite{rappoport05}, classical Coulomb energy and its derivatives 
are computed in an accelerated manner by expanding the density in an auxiliary basis. The Lagrangian of the excitation
energy is derived, which is stationary with respect to all electronic degrees of freedom. Now the excited-state
first-order properties are conveniently obtained because the Hellmann-Feynmann theorem holds. A state-specific 
scheme for TDDFT based on Davidson algorithm has been developed \cite{chiba06} to reduce the rank of response matrix
and efficient memory use, without loss of accuracy. In another work \cite{chiba07}, two-body fragment MO method (FMO2) 
was combined with TDDFT by dividing the system into fragments and electron density of each of these latter being
determined self-consistently. In another work \cite{guan06}, the excitation spectrum was calculated by means of
Tamm-Dancoff approximation and the spin-flip formalism \cite{wang04,wang05}. A double-hybrid DFT for excited states
\cite{grimme07} is also available, where a mixing of GGA XC with HF exchange and a perturbative second-order 
correlation part (obtained from KS GGA orbitals and eigenvalues) is advocated. TDDFT within the Tamm-Dancoff 
approximation is also implemented using a pseudospectral method to evaluate the two-electron repulsion integrals 
\cite{ko08}. On a separate work, a subspace formulation of TDDFT within the frozen-density embedding framework has 
been presented \cite{neugebauer07}. This allows to incorporate the couplings between electronic transitions on 
different subsystems which becomes very important in aggregates composed of several similar chromophores, e.g., in
biological or biomimetic light-harvesting systems. An occupation number averaging scheme \cite{hu07} for TDDFT 
response theory in frequency domain has been prescribed lately, where an average of excitation energies over the 
occupation number is adopted; this leads to equations of non-symmetric matrix form. Another work \cite{vahtras07}
combines generalized orbital expansion operators (designed to generate excited states having well-defined 
multiplicities) and the non-collinear formulation of DFT, for treatment of excited states. 

\section{The Work-Function Route to Excited states}
In this section, we present a simple DFT method for ground and arbitrary excited states of an atom. This has been 
tremendously successful for many excited states of many-electron atoms. The approach is simple, computationally 
efficient, and has been overwhelmingly successful for an enormous number of atomic states, such as singly, doubly,
triply excited states; low-, moderately high- and high-lying Rydberg states; valence as well as core excitation;
autoionizing resonances and satellite states; negative atoms as well. This is within a time-independent framework 
and results have been presented in the references 
\cite{singh96,singh96a,roy97,roy97a,roy97b,roy98,singh98,singh99,vikas00,roy02,roy04b,roy05b,roy07}. 

In this approach, a physical understanding of KS theory via quantum mechanical interpretation of electron-electron
interaction energy functional, $E_{ee}^{KS}[\rho]$, and its functional derivative (potential), 
$v_{ee}^{KS} (\rvec) = \delta E_{ee}^{KS} [\rho]/\delta \rho (\rvec)$, is established in terms of fields arising from 
source charge 
distributions (quantum mechanical expectations of Hermitian operators). Further, clear provisions are made to
distinguish Pauli-Coulomb correlation (due to Pauli exclusion principle and Coulomb repulsion) and kinetic energy 
correlation components of the energy functional and potential; each components arises from a separate field and source 
charge distribution \cite{harbola89,sahni90,holas95,sahni97}. It may be recalled  that $E_{ee}^{KS}[\rho]$ in KS 
theory represents Pauli and Coulomb correlations as well as correlation contributions to the kinetic energy. The
corresponding local potential $v_{ee}^{KS} (\rvec)$ (obtained as a functional derivative), consists of two 
separate contributions: (i) a \emph{purely} quantum mechanical (Pauli and Coulomb) electron-electron correlation
component $W_{ee}(\rvec)$, and (ii) a correlation kinetic energy component $W_{t_c}(\rvec)$. 

The interaction potential, $v_{ee}^{KS}(\rvec)$, is defined as the work done to bring an electron from infinity to
its position ar $\rvec$ against a field $\mbox{\boldmath $\cal{F}$}(\rvec)$,
\begin{equation}
v_{ee}^{KS} (\rvec) = \frac{\delta E_{ee}^{KS}[\rho]}{\delta \rho(\rvec)} = - \int_{\infty}^{\rvec} 
\mbox{\boldmath $\cal {F}$} (\rpvec) \cdot \  d\mathbf{l}'.
\end{equation}
The field is a sum of two fields: $\mbox{\boldmath $\cal {F}$}(\rvec)= \mbox{\boldmath $\cal {E}$}_{ee}(\rvec)+
Z_{t_c}(\rvec)$. The first term originates from Pauli and Coulomb correlations as its quantum mechanical 
source-charge distribution is the pair-correlation density $g(\rvec, \rpvec)$, while the second terms arises from a 
kinetic energy-density tensor $t_{\alpha \beta} (\rvec)$. The latter accounts for the difference of the fields derived 
from tensor for the interacting and KS non-interacting system. The electron-electron interaction potential, 
$v_{ee}^{KS}(\rvec)$, is 
expressed as a sum: $v_{ee}^{KS}(\rvec) = W_{ee}(\rvec) + W_{t_c}(\rvec)$, where, 
\begin{equation}
W_{ee}(\rvec) = - \int_{\infty}^{\rvec} \mbox{\boldmath $\cal {E}$}_{ee} (\rpvec) \cdot \  
d\mathbf{l}', \ \ \ \  \ \ \ \ \ \ \ \ \ 
W_{t_c}(\rvec) = - \int_{\infty}^{\rvec} Z_{t_c} (\rpvec) \cdot \  d\mathbf{l}' .
\end{equation}
The functional derivative in Eq.~(56) can be identified as the work done due to the fact that 
$\nabla v_{ee}^{KS} (\rvec) = - \mbox{\boldmath $\cal {F}$} (\rvec)$, so that the sum of two works $W_{ee}(\rvec)$ 
and $W_{t_c}(\rvec)$ 
is path-independent. The latter is rigorously valid provided the field $\mbox{\boldmath $\cal {F}$} (\rvec)$ is smooth, 
i.e., it is
continuous, differentiable and has continuous first derivative. It is also implicit that curl of the field vanishes, 
i.e., $\nabla \times \mbox{\boldmath $\cal {F}$} (\rvec) = 0$. Furthermore, for certain systems, such as 
closed-shell atoms, jellium 
metal clusters, jellium metal surfaces, open-shell atoms in central-field approximation, etc., the work done 
$W_{ee}(\rvec)$ and $W_{t_c}(\rvec)$ are separately path independent, i.e., 
$\nabla \times \mbox{\boldmath $\cal {E}$}_{ee}(\rvec) = \nabla \times Z_{t_c} (\rvec) = 0$.
 
Now, it is known that the pair-correlation density $g(\rvec, \rpvec)$ is not a static, but rather describes a 
\emph{dynamic} charge distribution, whose structure varies as a function of electron position. This dynamic nature is 
incorporated into the definition of local potential (through the force field 
$\mbox{\boldmath $\cal {E}$}_{ee}(\rvec)$) in which electrons move, via Coulomb's law as,
\begin{equation}
\mbox{\boldmath $\cal {E}$}_{ee}(\rvec) = \int \frac{g(\rvec,\rpvec)(\rvec-\rpvec)}{|\rvec-\rpvec|^3} \ 
\mathrm{d}\rpvec . 
\end{equation}
So, one can define the component $W_{ee}(\rvec)$ as work done to bring an electron from infinity to its position 
at $\rvec$ against this force field, as given in Eq.~(57). However, this can be further simplified by recognizing 
that pair-correlation density, $g(\rvec,\rpvec)$ can be expressed as a sum of density $\rho(\rpvec)$ and 
Fermi-Coulomb hole charge density $\rho_{xc}(\rvec, \rpvec)$: $g(\rvec,\rpvec)=\rho(\rpvec)+\rho_{xc}(\rvec, \rpvec)$.
The field $\mbox{\boldmath $\cal {E}$}_{ee}(\rvec)$ is then constituted of two fields, namely, the Hartree 
($\mbox{\boldmath $\cal {E}$}_H (\rvec))$ and
XC ($\mbox{\boldmath $\cal {E}$}_{xc}(\rvec)$) fields as: $\mbox{\boldmath $\cal {E}$}_{ee}(\rvec)= 
\mbox{\boldmath $\cal {E}$}_H(\rvec)+ \mbox{\boldmath $\cal {E}$}_{xc}(\rvec)$. These fields
are defined again as:
\begin{equation}
\mbox{\boldmath $\cal {E}$}_{H}(\rvec) = \int \frac{\rho(\rpvec)(\rvec-\rpvec)}{|\rvec-\rpvec|^3} 
\ \mathrm{d}\rpvec, \ \ \ \  \ \ \ \  
\mbox{\boldmath $\cal {E}$}_{xc}(\rvec) = \int \frac{\rho_{xc}(\rvec,\rpvec)(\rvec-\rpvec)}{|\rvec-\rpvec|^3} 
\ \mathrm{d}\rpvec . 
\end{equation}

The component $W_{ee}(\rvec)$ is a sum of works $W_H(\rvec)$ and $W_{xc}(\rvec)$, done to move an electron
in the corresponding Hartree and XC fields as $W_{ee}(\rvec)=W_H(\rvec)+W_{xc}(\rvec)$, with,
\begin{equation}
W_H(\rvec) = - \int_{\infty}^{\rvec} \mbox{\boldmath $\cal {E}$}_H (\rpvec) \cdot \  
d\mathbf{l}', \ \ \ \  \ \ \ \ \ \ \ \ \ 
W_{xc}(\rvec) = - \int_{\infty}^{\rvec} \mbox{\boldmath $\cal {E}$}_{xc} (\rpvec) \cdot \  d\mathbf{l}' .
\end{equation}
The work $W_H(\rvec)$ is path-independent, $\nabla \times {\cal E}_H(\rvec) = 0$, and also it is recognized as the
Hartree potential $v_H(\rvec)$ in DFT. The functional derivative of Coulomb self-energy functional $E_H[\rho]$ 
can be physically interpreted as work done in the field of electron density. The component $W_{ee}(\rvec)$ is 
then given as a sum of Hartree potential and the work done to move an electron in the field of quantum mechanical
Fermi-hole charge distribution: $W_{ee}(\rvec)=v_H(\rvec)+W_{xc}(\rvec)$. The latter is path independent for 
symmetrical density systems as mentioned previously, since $\nabla \times \mbox{\boldmath $\cal {E}$}_{xc}(\rvec) =0$ 
in all those cases. 
However, note that $\rho_{xc}(\rvec,\rpvec)$ that gives rise to the field $\mbox{\boldmath $\cal {E}$}_{xc}(\rvec)$ 
need not possess the same symmetry for arbitrary electron position. 

Finally, the KS electron-interaction energy $E_{ee}^{KS}[\rho]$ can also be expressed in terms of above fields
(and hence source charge distribution) as follows. The quantum mechanical electron-interaction energy is,
\begin{equation}
E_{ee}[\rho]= \int \mathrm{d} \rvec \ \rho(\rvec) \rvec \mathbf{\cdot}\  \mbox{\boldmath $\cal {E}$}_{ee} (\rvec) ,
\end{equation}
which can be further reduced to its Coulomb self-energy and XC components, 
\begin{equation}
E_{H}[\rho]= \int \mathrm{d} \rvec \ \rho(\rvec) \rvec \cdot \mbox{\boldmath $\cal {E}$}_{H} (\rvec), \ \ \ \ \ \ \ \
E_{xc}[\rho]= \int \mathrm{d} \rvec \ \rho(\rvec) \rvec \cdot \mbox{\boldmath $\cal {E}$}_{xc} (\rvec) ,
\end{equation}
and the correlation-kinetic energy component is,
\begin{equation}
T_c[\rho]= \frac{1}{2} \int \mathrm{d} \rvec \ \rho(\rvec) \rvec \cdot Z_{t_c} (\rvec).
\end{equation}

Such a description for XC potential in terms of the source charge distribution, gives a hope of writing KS 
equation of an interacting many-electron system which could, in principle, be applicable for both ground and 
excited states. Because, this procedure leads to a \emph{universal} prescription, independent of any state, as it 
does not have a definite functional form; it is completely and \emph{uniquely} determined by the electronic 
configuration of a particular state in question. Hence the applicability for ground as well as excited states; the same 
equation gives it all. Although the present method falls within the spirit of exchange energy as defined in Slater's
theory via Fermi-hole charge distribution, it is expected to offer improvement over the Hartree-Fock-Slater (equivalent
to LDA method in DFT) theory, because current scheme accounts for the dynamic nature of charge distribution. This
definition gives the expected falling off ($1/r$) of exchange potential at large $r$. Since at $r \rightarrow \infty$, the 
Coulomb-hole contributions to the interaction in Eq.~(62) is already zero, this implies that current method should 
give almost \emph{exact} results in the asymptotic region.  

Now, we proceed to some details of the actual numerical implementation. Note that the work $v_x(\rvec)$ against the 
force field due to a Fermi-hole charge can be determined exactly since the latter is known explicitly in terms of 
orbitals as,
\begin{equation}
\rho_x(\rvec,\rpvec)=-\frac{|\gamma (\rvec,\rpvec)|^2}{2 \rho(\rvec)}, \ \ \ \ \ \ \ \  \ \ \ \ \ \ \ \ \ 
\gamma(\rvec,\rpvec)=\sum_{\mathit{i}} \phi_{\mathit{i}}^\ast(\rvec) 
\phi_{\mathit{i}}(\rpvec).
\end{equation}
Here the terms have following meaning. $\gamma(\rvec,\rpvec)$ refers to the single-particle density matrix spherically 
averaged over electronic coordinates for a given orbital angular quantum number, 
$\phi_i(\rvec)= R_{nl}(r)\ Y_{lm}(\Omega)$ signifies the single-particle orbital, and $\rho(\rvec)$ is the total 
electron density expressed in terms of occupied orbitals, $\rho(\rvec)=\sum_{i} |\phi_i(\rvec)|^2.$
For spherically symmetric systems, exchange part in Eq.~(59) can be simplified as, 
\begin{equation}
\mbox{\boldmath $\cal{E}$}_{x,r} (r)=-\frac{1}{4\pi} \int \rho_x (\rvec,\rpvec) \ \frac{\partial}{\partial r} \ \ 
\frac{1}{|\rvec -\rpvec|}\ \mathrm{d}\rpvec \mathrm{d} \Omega_r.
\end{equation}
Now one can use the well-known expansion,
\begin{equation}
\frac{1}{|\rvec-\rpvec|}=4\pi \sum_{l'',m''} \frac{1}{2l''+1} Y^\ast_{l''m''}
(\Omega) \ Y_{l''m''}(\Omega') \ \frac{r_<^{l''}}{r_>^{l''+1}},
\end{equation}
to obtain
\begin{widetext}
\begin{eqnarray}
\mbox{\boldmath $\cal{E}$}_{x,r} (r) & = & \frac {1}{2\pi \rho(r)}
\int \sum_{n,l,m,n',l',m',l''} R_{nl}(r) R_{nl}(r') R_{n'l'}(r) R_{n'l'}(r') \left[ \frac{\partial}{\partial r} 
\frac {r_<^{l''}} {r_>^{l''+1}} \right] \nonumber \\
& & r'^2 \mathrm{d}r' \frac{(2\mathit{l}+\mathrm{1})}{(2\mathit{l'}+\mathrm{1})} 
\times C^2(\mathit{ll''l';m,m'-m,m'})\mathrm{C^2}(\mathit{ll''l'};\mathrm{000}),
\end{eqnarray}
\end{widetext}
where $R_{nl}(r)$ denotes radial part of the single-particle orbitals and C's are the Clebsch-Gordan coefficients 
\cite{rose57}. Now the exchange integral in Eq.~(60) can be written as an integral over radial coordinates only,
\begin{equation}
v_x (r) = - \int_{\infty}^{r} 
\mbox{\boldmath $\cal{E}$}_{x,r}(r') \ \mathrm{d}r'. 
\end{equation}
While the exchange potential $v_x(\rvec)$ can be accurately calculated through the procedure as delineated above, the 
correlation potential $v_c(\rvec)$ remains unknown and must be approximated for practical calculations. In the 
present work, two correlation functionals are used, (i) a simple, local, parametrized Wigner-type potential 
\cite{brual78} (ii) a slightly more complicated, generalized gradient-corrected correlation energy functional of 
Lee-Yang-Parr (LYP) \cite{lee88}. 

With this choice of $v_x(\rvec)$ and $v_c(\rvec)$, the following KS-type differential equation is solved 
self-consistently to produce a self-consistent set of orbitals, from which $\rho(\rvec)$ is constructed,
\begin{equation}
\left[ -\frac{1}{2} \nabla^2 +v_{es} (\rvec) +v_{xc}(\rvec) \right] \phi_i(\rvec) = \varepsilon_i \phi_i(\rvec),
\end{equation}
where $v_{es}(\rvec)$ represents the usual Hartree electrostatic potential including electron-nuclear 
attraction and inter-electronic Coulomb repulsion, whereas $v_{xc}(\rvec)= v_x(\rvec) +v_c(\rvec)$. Total energy
is then obtained as a sum of following terms in the usual manner,
\begin{equation}
T=-\frac{1}{2} \sum_i \int \phi_i^* (\rvec) \nabla^2 \phi_i (\rvec) \ \mathrm{d}\rvec, \ \ \ \ \ \  \ \ \ 
E_{es} = -Z \int \frac{\rho(\rvec)}{r} \ \mathrm{d}\rvec + 
\frac{1}{2} \int \int \frac{\rho(\rvec) \rho(\rpvec)}{|\rvec-\rpvec|} \ \mathrm{d}\rvec \ \mathrm{d}\rpvec .
\end{equation}
Two-electron Hartree and exchange energies can be simplified further,
\begin{eqnarray}
E_H & = & \frac{1}{2} \sum  \int \int R^2_{nl}(r) R^2_{n',l'}(r') \ \frac{r^{l''}_{<}}{r^{l''+1}_{>}} \ 
r^2 r'^2 \ \mathrm{d}r \ \mathrm{d}r' \nonumber \\
    &   & \times C(ll''l;m0m) \ C(ll''l;000) \ C(ll''l';m'0m') \ C(l'l''l'';000) \nonumber  \\
E_x & = & \sum  (\mathrm{pairs \ with \ parallel \ spin}) \int \int R_{nl}(r) R_{n',l'}(r) R_{nl}(r') R_{n',l'}(r') \ 
\frac{r^{l''}_{<}}{r^{l''+1}_{>}} \ r^2 r'^2 \ \mathrm{d}r \ \mathrm{d}r'  \nonumber \\
    &   & \times C^2(ll''l';m,m-m',m') \ C^2(ll''l';000)  \times \left( \frac{2l+1}{2l'+1} \right)
\end{eqnarray}

Now a few words should be mentioned regarding numerical solution of the KS equation for orbitals. In earlier stages
of the development of this method \cite{singh96,singh96a,roy97,roy97a,roy97b,roy98,singh98,singh99,vikas00}, a Numerov-type 
finite difference (FD) scheme was adopted for the discretization of spatial coordinates. It is, however, well-known that, due
to existence of Coulomb singularity at the origin and presence of long-range nature of the Coulomb potential, FD methods
require a large number of grid points to achieve decent accuracy even for ground-state calculations. Certainly excited 
(especially those higher-lying Rydberg ones) would need much more grid points to properly describe their long tail. Here, we 
describe the extension of the generalized pseudospectral (GPS) method for \emph{nonuniform} and optimal spatial discretization and 
solution of KS equation, Eq.~(69). This procedure has been demonstrated to be capable of providing high precision solution 
of eigenvalues and wave functions for a variety of \emph{singular} as well as non-singular potentials, like Hulthen, Yukawa,
Spiked harmonic oscillators, logarithmic, Hellmann potentials; very accurate results have also been obtained for static and 
dynamic calculation in Coulomb singular systems (like atoms, molecules) such as electronic structure, multi-photon processes in 
strong fields, Rydberg atom spectroscopy and dynamics, etc. \cite{roy02,roy02a,roy02b,roy04,roy04a,roy04b,roy05,roy05a,roy05b,
sen06,roy07,roy08,roy08a,roy08b}. In addition, the GPS method is computationally orders of magnitude faster than the 
equal-spacing FD methods. In what follows, we briefly outline the GPS procedure appropriate for our present DFT study. 
General discussion on the approach could be found in additional references \cite{gottlieb84,canuto88}.

The most important feature of this method is to approximate an \emph{exact} function $f(x)$ defined on the interval 
$[-1,1]$ by an Nth-order polynomial $f_N(x)$, 
\begin{equation}
f(x) \cong f_N(x) = \sum_{j=0}^{N} f(x_j)\ g_j(x),
\end{equation}
such that the approximation be \emph{exact} at \emph{collocation points} $x_j$,
\begin{equation}
f_N(x_j) = f(x_j).
\end{equation}
We chose to employ the Legendre pseudospectral method where $x_0=-1$, $x_N=1$, and $x_j (j=1,\ldots,N-1)$ are determined by 
roots of first derivative of the Legendre polynomial $P_N(x)$ with respect to x, i.e.,  $P'_N(x_j) = 0$.
In Eq.~(72), $g_j(x)$ are the cardinal functions satisfying a unique property $g_j(x_{j'}) = \delta_{j'j}$, and defined by,
\begin{equation}
g_j(x) = -\frac{1}{N(N+1)P_N(x_j)}\ \  \frac{(1-x^2)\ P'_N(x)}{x-x_j},
\end{equation}

The general eigenvalue problem for our radial KS-type equation can now be written as,
\begin{equation}
\hat{H}(r) \psi(r) =E \psi(r),
\end{equation}
with 
\begin{equation}
\hat{H}(r) =-\frac{1}{2} \ \ \frac{d^2}{dr^2} +V(r),
\end{equation}
For structure and dynamics calculations this involves Coulomb potential, which typically has a singularity problem at 
$r=0$, as well as the long-range $-1/r$ behavior. This usually requires a large number of grid points in the {\em equal-spacing} 
finite-difference methods, which are not feasible to extend to Rydberg state calculations. This can be overcome by first 
mapping the semi-infinite domain $r \in [0, \infty]$ into a finite domain $x \in [-1,1]$ by a mapping transformation 
$r=r(x)$ and then using the Legendre pseudospectral discretization technique. At this stage, following algebraic nonlinear 
mapping \cite{yao93,wang94} is used, 
\begin{equation}
r=r(x)=L\ \ \frac{1+x}{1-x+\alpha},
\end{equation}
where L and $\alpha=2L/r_{max}$ are the mapping parameters. Further, introducing, 
\begin{equation}
\psi(r(x))=\sqrt{r'(x)} f(x)
\end{equation}
and following a symmetrization procedure, a transformed Hamiltonian is obtained as,
\begin{equation}
\hat{H}(x)= -\frac{1}{2} \ \frac{1}{r'(x)}\ \frac{d^2}{dx^2} \ \frac{1}{r'(x)}
+ V(r(x))+V_m(x),
\end{equation}
where
\begin{equation}
V_m(x)=\frac {3(r'')^2-2r'''r'}{8(r')^4}.
\end{equation}
The advantage of this mapping scheme is that this leads to a \emph{symmetric} matrix eigenvalue problem. Note that for the 
mapping used here, $V_m(x)=0$. Therefore, discretizing our Hamiltonian by GPS method leads to the following set of coupled 
equations,
\begin{widetext}
\begin{equation}
\sum_{j=0}^N \left[ -\frac{1}{2} D^{(2)}_{j'j} + \delta_{j'j} \ V(r(x_j))
+\delta_{j'j}\ V_m(r(x_j))\right] A_j = EA_{j'},\ \ \ \ j=1,\ldots,N-1,
\end{equation}
\end{widetext}
\begin{equation}
A_j = r'(x_j)\ f(x_j)\ \left[ P_N(x_j)\right]^{-1} 
  =  \left[ r'(x_j)\right]^{1/2} \psi(r(x_j))\ \left[ P_N(x_j)\right]^{-1}.
\end{equation}
Here $D^{(2)}_{j'j}$ represents symmetrized second derivative of the cardinal function in respect to r,
\begin{equation}
D^{(2)}_{j'j} =  \left[r'(x_{j'}) \right]^{-1} d^{(2)}_{j'j} 
\left[r'(x_j)\right]^{-1}, 
\end{equation}
and 
\begin{eqnarray}
d^{(2)}_{j',j} & = & \frac{1}{r'(x)} \ \frac{(N+1)(N+2)} {6(1-x_j)^2} \ 
\frac{1}{r'(x)}, \ \ \ j=j', \nonumber \\
 & & \nonumber \\
& = & \frac{1}{r'(x_{j'})} \ \ \frac{1}{(x_j-x_{j'})^2} \ \frac{1}{r'(x_j)}, 
\ \ \ j\neq j'.
\end{eqnarray}

The orbitals $\{\phi_i(\rvec)\}$ obtained from self-consistent solution of KS equation (69) are used to construct various
determinants for a given electronic configuration of an atom, which, in turn, could be employed to calculate the associated
multiplets related to this configuration. Here we use Slater's diagonal sum rule for the multiplet energies \cite{slater60}. 
Similar strategy for multiplets has been adopted earlier \cite{ziegler77,wood80,lannoo81,dickson96,stuckl97,pollak97}. 

\begingroup
\squeezetable
\begin{table}
\caption {Comparison of singly excited-state energies of Li and Be (in a.u.) with literature data. Numbers in parentheses  
denote absolute percentage deviations with respect to reference values. Taken from ref.~\cite{roy02}.}
\begin{ruledtabular}
\begin{center}
\begin{tabular}{cccc|cccc}
      State&  $-E$(X)    &   $-E$(XC) &   $-E$(Literature) & State & $-E$(X)  & $-E$(XC) & $-E$(Literature)  \\ \hline
\underline{\textbf{Li}}    &              &           &         & \underline{\textbf{Be}} &  &  &  \\ 
1s$^2$3s $^2$S & 7.30966 & 7.35773 (0.05)& 7.35394\footnotemark[2]  &  
1s$^2$2s3s  $^3$S  &14.37798& 14.42917 (0.20) &14.42629\footnotemark[5] \\ 
                    & 7.31021\footnotemark[1]  &   &     & & & &   \\ 
1s$^2$5s $^2$S  & 7.25996 & 7.30466 (0.01)& 7.30339\footnotemark[2]  & 
1s$^2$2s5s  $^3$S  & 14.30562 & 14.34996 &               \\ 
1s$^2$2p $^2$P  & 7.36486 & 7.41204 (0.03)& 7.41016\footnotemark[3]  &  
1s$^2$2s2p  $^3$P &14.51068 &14.56660 (0.03)&14.56223\footnotemark[5]\\ 
                    & 7.36507\footnotemark[1]  &   &      &
                    & 14.51150\footnotemark[4]  &   &       \\ 
1s$^2$4p $^2$P  & 7.26859 & 7.31262 (0.01) & 7.31190\footnotemark[3]  & 
1s$^2$2s4p $^3$P & 14.31462 & 14.35910 &          \\ 
   & & &        &             & 14.31464\footnotemark[4]  &   &      \\ 
\end{tabular}  
\begin{tabbing}
$^{\mathrm{a}}\!$ HF result, ref.~\cite{goddard68}. \hspace{0.4in} \= $^{\mathrm{b}}\!$ Ref.~\cite{wang92}. \hspace{0.4in} \= 
$^{\mathrm{c}}\!$ Ref.~\cite{sims75}. \hspace{0.4in} \= $^{\mathrm{d}}\!$ HF result, ref.~\cite{weiss72}. \hspace{0.4in} \= 
$^{\mathrm{e}}\!$ Ref.~\cite{koyama86}.
\end{tabbing}
\end{center}
\vspace{-0.2in}
\end{ruledtabular}
\end{table}
\endgroup

\section{Results and Discussion}
At first, we give some sample results for singly excited 1s$^2$ns $^2$S, 1s$^2$np $^2$P states of Li, as well 1s$^2$2sns 
$^3$S Be, 1s$^2$2snp $^3$P states of Be, in Table I. Note that in this and all other 
following tables, we present only non-relativistic results; state energies are in atomic units, while excitation energies in 
eV. For all these calculations, a convergence criteria of $10^{-5}$ and $10^{-6}$, as well a radial grid of 500 points have
been used. In the literature generally excitation energies are reported, while individual state energies are given seldom. 
However, in a DFT study of low-lying singly excited states of some open-shell atoms (B, C, O, F, Na, Mg, Al, Si, P, Cl) 
\cite{singh98}, excitation energies from X-only and numerical HF methods are found to be in good agreement with each other. 
Surprisingly,
however, the two correlation energy functionals (Wigner and LYP) did not show any considerable improvements in excitation
energies although excited-state energies were dramatically improved. Therefore, in this work, we consider both the state
energies and excitation energies. In this table, two sets of calculations are performed; solution of Eq.~(69) with 
(i) $v_{xc}=v_x$ (exchange-only 
or E(X)) and (ii) $v_x+v_c$ (exchange plus correlation or E(XC)). These states as well the other ones in proceeding tables 
are of great significance in atomic physics; thus have been studied by both experimentalists and theoreticians by employing a
multitude of techniques and formalisms. Some prominent reference values are quoted for comparison, wherever possible. The
X-only results are fairly close to the HF values \cite{goddard68,weiss72}, errors ranging from 0.0057\% to as low as 0.0001\% 
for Be 1s$^2$2s4p $^3$P, indicating the accuracy in our calculation. The doublet $S$ states of Li are compared with the 
full-core-plus-correlation method with multi-configuration interaction wave functions \cite{wang92}, while doublet $P$ states 
with a combined configuration-interaction-Hylleraas method \cite{sims75}. For Be, literature is quite scanty and present 
density functional results match very closely with the multi-configuration calculations \cite{koyama86}. One finds some 
overestimation in total energy caused by the LYP correlation functional employed here; errors ranging from 0.052\%--0.003\%.
For a more detailed discussion, see \cite{roy02}. 

\begingroup
\squeezetable
\begin{table}
\caption {Calculated doubly excited-state (ns$^2$  $^1$S, np$^2$  $^1$D) energies of He (in a.u.) along with literature 
data for comparison. Numbers in parentheses denote absolute percentage errors with respect to literature data. Adopted from 
ref.~\cite{roy02}.}
\begin{ruledtabular}
\begin{center}
\begin{tabular}{lll|lll} 
      State&     $-E$(XC)  &   $-E$(Literature) &  State  &  $-E$(XC)   &  $-E$(Literature)  \\ \hline
2s$^2$  $^1$S    &  0.76637 (1.48)& 0.77787\footnotemark[1] & 2p$^2$ $^1$D  &  0.69272 (1.31)& 0.70195\footnotemark[3]\\
3s$^2$ $^1$S    &  0.34578 (2.19)& 0.35354\footnotemark[1] & 3p$^2$  $^1$D  &  0.31540 (0.04)& 0.31554\footnotemark[3]\\
4s$^2$  $^1$S    &  0.19659 (2.19)& 0.20099\footnotemark[1] & 4p$^2$  $^1$D  &  0.18095       &                        \\
5s$^2$  $^1$S    &  0.12754 (2.12)& 0.13030\footnotemark[2] & 5p$^2$  $^1$D  &  0.11610       &                        \\
6s$^2$  $^1$S    &  0.08808 (3.05)& 0.09085\footnotemark[2] & 6p$^2$  $^1$D  &  0.08115       &                        \\
7s$^2$  $^1$S    &  0.06524 (3.35)& 0.0675\footnotemark[2]  & 7p$^2$  $^1$D  &  0.05980       &                        \\
9s$^2$ $^1$S    &  0.03889 &                               & 9p$^2$  $^1$D  &  0.03604       &              \\
11s$^2$  $^1$S   &  0.02503 &                               & 11p$^2$  $^1$D &  0.02414 &              \\
13s$^2$  $^1$S   &  0.01811 &                               & 13p$^2$  $^1$D &  0.01728 &              \\
15s$^2$ $^1$S   &  0.01348 &                               & 15p$^2$ $^1$D &  0.01297 &              \\
17s$^2$  $^1$S   &  0.01132 &                               & 17p$^2$  $^1$D &  0.01010 &              \\
\end{tabular} 
\begin{tabbing}
$^{\mathrm{a}}\!$ Ref.~\cite{burgers95}. \hspace{0.6in} \= $^{\mathrm{b}}\!$ Ref.~\cite{koyama86}. \hspace{0.6in} \= 
$^{\mathrm{c}}\!$ Ref.~\cite{lindroth94}. 
\end{tabbing}
\end{center}
\vspace{-0.2in}
\end{ruledtabular}
\end{table}
\endgroup

Next, in Table II, some even-parity doubly excited states (ns$^2$ $^1$S, np$^2$ $^1$D, n=2--17) of He are presented. Many of 
these have been identified to be autoionizing in nature, e.g., ns$^2$ $^1$S. It is seen that the calculated energy values have 
never fallen below the quoted results. In the former case, DFT results are comparable to literature data for smaller $n$ and
tends to increase gradually with an increasing $n$, as evident from the absolute per cent deviations given in the parentheses.  
This could occur either because of the inadequate description of long-range nature of correlation potential employed or some
deficiencies in the 
work-function formalism itself. Finally we see that while accuracy of doubly excited state calculation is not as good as that
of singly excited state, error in the former still remains well within 3.6\%. More details on these could be found in
ref.~\cite{roy02}.  

\begingroup
\squeezetable
\begin{table}
\caption {Single and double excitation energies of He and Be (in a.u.) compared with literature data. Numbers in parentheses 
denote absolute percentage errors with respect to the best theoretical data available. Adopted from ref.~\cite{roy02}. } 
\begin{ruledtabular}
\begin{center}
\begin{tabular}{lllll}
   State&  Present Work  &$\Delta\epsilon_{KS}$& Other theory  & 
Experiment  \\  \hline
\underline{\textbf{Single excitation of He and Be}}    &    &  &    &  \\                  
He 1s2s $^3$S  & 0.72839 (0.02) & 0.7460\footnotemark[1] & 
0.72850\footnotemark[2] & 0.72833\footnotemark[3]   \\
He 1s2s $^1$S  & 0.75759 (0.02) &        & 
0.75775\footnotemark[2] & 0.75759\footnotemark[3]   \\
He 1s2p $^3$P  & 0.77041 (0.02) & 0.7772\footnotemark[1] & 
0.77056\footnotemark[2] & 0.77039\footnotemark[3]   \\
He 1s2p $^1$P  & 0.77971 (0.02) &        & 
0.77988\footnotemark[2] & 0.77972\footnotemark[3]   \\
He 1s3s $^3$S  & 0.83494 (0.01) & 0.8392\footnotemark[1] & 
0.83504\footnotemark[2] & 0.83486\footnotemark[3]   \\
He 1s3s $^1$S  & 0.84231 (0.02) &        & 
0.84245\footnotemark[2] & 0.84228\footnotemark[3]   \\
He 1s3p $^3$P  & 0.84548 (0.02) & 0.8476\footnotemark[1] & 
0.84564\footnotemark[2] & 0.84547\footnotemark[3]   \\
He 1s3p $^1$P  & 0.84841 (0.02) &        & 
0.84858\footnotemark[2] & 0.84841\footnotemark[3]   \\
Be 1s$^2$2s2p $^3$P  & 0.10089 & 0.1327\footnotemark[1] &    & 
0.100153\footnotemark[3]  \\
Be 1s$^2$2s3s $^3$S  & 0.23832(0.63) & 0.2444\footnotemark[1] & 
0.236823\footnotemark[4]   & 0.237304\footnotemark[3]  \\
Be 1s$^2$2s4p $^3$P  & 0.30839 & 0.3046\footnotemark[1] &    & 
0.300487\footnotemark[3]  \\
Be 1s$^2$2s5s $^3$S  & 0.31753 & 0.3153\footnotemark[1] &    & 
0.314429\footnotemark[3]  \\
\underline{\textbf{Double excitation of He}}    &    &        &   &       \\
He 2s$^2$ $^1$S & 2.13747(0.54) &               & 
2.1259\footnotemark[5],2.1285\footnotemark[6] &       \\
He 3s$^2$ $^1$S & 2.55806(0.61) &               & 
2.5425\footnotemark[7],2.5496\footnotemark[6] &       \\
He 4s$^2$ $^1$S & 2.70725(0.48) &               & 
2.6942\footnotemark[7],2.7017\footnotemark[6] &       \\
He 5s$^2$ $^1$S & 2.77630(0.10) &  & 2.7735\footnotemark[7]    &   \\
He 6s$^2$ $^1$S & 2.81576(0.10) &  & 2.8129\footnotemark[7]    &   \\
He 7s$^2$ $^1$S & 2.83860(0.09) &  & 2.8362\footnotemark[8]    &   \\
He 2p$^2$ $^1$D & 2.21120(0.68) &               & 
2.1961\footnotemark[8],2.2082\footnotemark[6] &       \\
He 3p$^2$ $^1$D & 2.58844(1.60) &            & 
2.5477\footnotemark[8],2.5595\footnotemark[6] &       \\
He 4p$^2$  $^1$D & 2.72289(1.08) &  & 2.6938\footnotemark[8]    &   \\
\end{tabular}
\begin{tabbing}
$^{\mathrm{a}}\!$ Ref.~\cite{savin98}. \hspace{0.1in} \= $^{\mathrm{b}}\!$ Ref.~\cite{drake93,drake94}. \hspace{0.1in} \= 
$^{\mathrm{c}}\!$ Ref.~\cite{bashkin75}. \hspace{0.1in} \= $^{\mathrm{d}}\!$ Ref.~\cite{begue98}. \hspace{0.1in} \= 
$^{\mathrm{e}}\!$ Ref.~\cite{ho81}. \hspace{0.1in} \= $^{\mathrm{f}}\!$ Ref.~\cite{herrick75}. \hspace{0.1in}  \= 
$^{\mathrm{g}}\!$ Ref.~\cite{koyama86}. \hspace{0.1in} \= $^{\mathrm{h}}\!$ Ref.~\cite{fukuda87}.
\end{tabbing}
\end{center}
\vspace{-0.2in}
\end{ruledtabular}
\end{table}
\endgroup

Now, single and double excitation energies of selected states of He, Be, are displayed in Table III, along with some reference 
values. These are estimated with respect to our calculated, non-relativistic ground-state energies of He, Be, i.e., 
$-$2.90384 and $-$14.66749 a.u. (as obtained from the same KS equation (69)). No experimental results could be found for
doubly excited resonances. In some occasions, our calculated excitation energies have fallen below the experimental results. 
This is not surprising keeping in mind that the present methodology is \emph{non-variational}. As a consequence, the variational
restriction on a particular excited state being the lowest of a given space-spin symmetry does not hold good. Here we also 
report the single-particle KS energies (obtained from the difference of KS eigenvalues) for single excitations in He, Be 
\cite{savin98}, which, of course, do not show the multiplet separation. The excitation energies from true KS potential 
for He, Be are clearly quite good. However, those from some other commonly used approximate exchange energy functionals 
(such as LDA) produce large errors in excitation energy \cite{savin98}. Also, it may be mentioned here that, for excitation
energies in Ne satellites \cite{roy98}, both LDA and one of the most commonly used gradient-corrected exchange functional 
\cite{becke88} have been found to be absolutely unsuitable for such studies, producing very large errors. Besides, present
results are not corrected for relativistic effects, which is included in the experimental values. Single excitations in He
show reasonably good agreement with both theory and experiment, while for Be the corresponding discrepancy is somehow larger. 
Nevertheless, 
the overall agreement between current results and literature data is quite satisfactory. Apart from the errors in XC 
potentials as discussed earlier, another possible source could be rooted in the \emph{single-determinantal} nature of our method.   
Assumption of spherical symmetry in dealing with the exchange potential could also bring some inaccuracies. Stated differently, 
the solenoidal component of the electric field $\mbox{\boldmath $\cal{E}$}_x(\rvec)$ may not be negligible compared to the 
irrotational component for these states, although this usually holds quite well for atoms \cite{sahni97}. As a further check, 
some representative radial expectation values for singly and doubly excited states of He as well as singly excited states of 
Li, Be have also been studied \cite{roy02}. These match quite well with the HF values \cite{fischer77}, once again reassuring 
the accuracy in our calculation.  


Now we move on to the triple excitations. For this purpose, we compare our DFT excitation energies for all the eight 
$2l2l'2l''$ ($n=2$ intrashell) triply excited states, \emph{viz.,} 2s$^2$2p $^2$P$^o$; 2s2p$^2$ $^2$D$^e$, $^4$P$^e$, 
$^2$P$^e$, $^2$S$^e$; and 2p$^3$ $^2$D$^o$, $^2$P$^o$, $^4$S$^o$ of selected members of Li-isoelectronic series, i.e., 
B$^{2+}$, N$^{4+}$ and F$^{6+}$ in Table IV. At this stage, it may be appropriate to illustrate the details of a multiplet calculation
from individual determinants by an example. For this, we consider all the four multiplets $^2$D, $^4$P, $^2$P, $^2$S 
associated with a 2s2p$^2$ configuration. This gives rise to 30 determinants which satisfy following relations (left and 
right correspond to the multiplet and determinantal energies), 
\begin{eqnarray}
^2D & = & (0^+1^+1^-)  \nonumber \\
^4P & = & (0^+1^+0^+)  \nonumber \\
^2D + ^4P + ^2P & = & (0^+1^+0^-)+ (0^+1^-0^+)  + (0^-1^+0^+)  \nonumber \\
^2D + ^4P + ^2P + ^2S & = & (0^+1^+-1^-)+ (0^+1^--1^+)  + (0^-1^+-1^+) + (0^+0^+0^-)
\end{eqnarray}
where the numbers denote $m_l$ values while ($+,-$) $m_s$ values. For more details, see \cite{roy97a}.

\begingroup
\squeezetable
\begin{table}
\caption {Comparison of calculated excitation energies (in eV) of the n=2 intrashell triply excited 
states of B$^{2+}$, N$^{4+}$ and F$^{6+}$  relative to the non-relativistic ground states of \cite{yan98}. GPS signifies 
present work. See ref.~\cite{roy05b} for details.}
\begin{ruledtabular}
\begin{center}
\begin{tabular}{lcccccc}
State & \multicolumn{2}{c}{B$^{2+}$} & \multicolumn{2}{c}{N$^{4+}$} & 
\multicolumn{2}{c}{F$^{6+}$} \\ 
\cline{2-3} \cline{4-5} \cline{6-7}
     &   GPS  & Ref. &  GPS  & Ref.  &  GPS &  Ref. \\  
\hline
2s$^2$2p $^2$P$^o$   & 436.588  & 436.07\footnotemark[1],436.59\footnotemark[2]
                     & 894.541  & 893.93\footnotemark[1],894.12\footnotemark[3]
                     &1514.229  & 1515.67\footnotemark[1],1514.90\footnotemark[3] \\
2s2p$^2$ $^4$P$^e$   & 436.917  & 436.69\footnotemark[1],436.89\footnotemark[2]
                     & 894.876  & 894.54\footnotemark[1],894.51\footnotemark[3]
                     &1514.474  & 1516.33\footnotemark[1],1515.16\footnotemark[3]  \\
2s2p$^2$ $^2$D$^e$   & 441.893  & 441.34\footnotemark[1],442.00\footnotemark[2]
                     & 902.655  & 901.93\footnotemark[1],902.32\footnotemark[3]
                     &1524.898  & 1526.43\footnotemark[1],1525.86\footnotemark[3]  \\
2p$^3$ $^4$S$^o$     & 443.852  & 443.86\footnotemark[1],444.15\footnotemark[2],
                     & 905.329  & 905.15\footnotemark[1],905.15\footnotemark[3],
                     &1528.187  & 1530.42\footnotemark[1],1529.35\footnotemark[3],  \\
                     &          & 443.63\footnotemark[4]
                     &          & 904.43\footnotemark[4]
                     &          & 1528.51\footnotemark[4]                          \\
2s2p$^2$ $^2$S$^e$   & 445.387  & 445.11\footnotemark[1],445.75\footnotemark[2]
                     & 907.930  & 907.41\footnotemark[1],907.87\footnotemark[3]
                     &1531.822  & 1533.61\footnotemark[1],1533.18\footnotemark[3]  \\
2s2p$^2$ $^2$P$^e$   & 445.814  & 445.35\footnotemark[1],446.21\footnotemark[2]
                     & 908.455  & 907.99\footnotemark[1],908.59\footnotemark[3]
                     &1532.717  & 1534.55\footnotemark[1],1534.17\footnotemark[3]  \\
2p$^3$ $^2$D$^o$     & 446.173  & 446.02\footnotemark[1],446.58\footnotemark[2]
                     & 909.089  & 909.02\footnotemark[1],909.37\footnotemark[3]
                     &1533.816  & 1536.01\footnotemark[1],1535.35\footnotemark[3]  \\
2p$^3$ $^2$P$^o$     & 450.088  & 450.04\footnotemark[1],450.65\footnotemark[2]
                     & 915.023  & 915.00\footnotemark[1],915.47\footnotemark[3]
                     &1541.946  & 1543.95\footnotemark[1],1543.46\footnotemark[3]  \\
\end{tabular}
\begin{tabbing}
$^{\mathrm{a}}\!$ Ref.~\cite{safronova98}. \hspace{0.4in} \= $^{\mathrm{b}}\!$ Ref.~\cite{conneely02}. \hspace{0.4in} \= 
$^{\mathrm{c}}\!$ Ref.~\cite{conneely04}. \hspace{0.4in} \= $^{\mathrm{d}}\!$ Ref.~\cite{davis90}.
\end{tabbing}
\end{center}
\vspace{-0.2in}
\end{ruledtabular}
\end{table}
\endgroup

In this case, we no more report individual state energies as these are very difficult 
to compare directly; instead only excitation energies are reported. To put our results in proper perspective, all triple
excitation energies in this table are computed with respect to the accurate non-relativistic ground state 
of \cite{yan98}. No experimental results have been reported as yet in the literature and appropriate theoretical results 
are quoted here. All these states are autoionizing except the 2p$^3$ $^4$S$^o$, which is bound, metastable against 
auto-ionization by conservation of parity and angular momentum. These are studied through a multi-configuration-interaction
type formalism within a Rayleigh-Ritz variational principle \cite{davis90}. Recently a perturbation theory method 
(1/Z expansion) \cite{safronova98} as well a truncated diagonalization method \cite{conneely02,conneely04} have been employed
to determine the position of all these states. It is gratifying that our current positions for all these 8 states for these 
3 ions follow the same orderings as in \cite{safronova98,conneely02,conneely04}, which clearly demonstrates the reliability 
in our calculation. All these excitation energies show excellent agreement with the literature data, with a maximum 
discrepancy of 0.125\%; for the three ions the deviation ranges are 0.0005--0.125\%, 0.007--0.049\% and 0.044--0.099\% 
respectively. Both underestimation and over-estimation is observed in excitation energies. 

\begingroup
\squeezetable
\begin{table}
\caption {2s$^2$ns $^2$S$^e$ and 2s$^2$np $^2$P$^o$ resonances of Li. State energies and 
excitation energies, relative to the ground state of \cite{chung91}. GPS signifies present work. See \cite{roy04b}
for details.}
\begin{ruledtabular}
\begin{center}
\begin{tabular}{lllllllll}
n   & \multicolumn{4}{c}{$\langle$A,ns$\rangle$ $^2$S$^e$}& \multicolumn{4}{c}
{$\langle$A,np$\rangle$ $^2$P$^o$} \\ 
   & \multicolumn{2}{c}{$-$E(a.u.)} & \multicolumn{2}{c}{Excitation energy(eV)} 
   & \multicolumn{2}{c}{$-$E(a.u.)} & \multicolumn{2}{c}{Excitation energy(eV)} \\ 
\cline{2-3} \cline{4-5} \cline{6-7} \cline{8-9}
   & GPS  & Ref. & GPS & Ref.  
   & GPS  & Ref. & GPS & Ref.  \\
 2 &     &   &    &  
   & 2.2448  & 2.2503\footnotemark[1],2.247\footnotemark[2], & 142.385 
   & 142.255\footnotemark[1],142.439\footnotemark[3],142.12\footnotemark[4], \\
   &    &    &    &   
   &    & 2.2428\footnotemark[3] &   
   & 142.310\footnotemark[5],142.35\footnotemark[6],142.33\footnotemark[7]  \\
 3 & 1.9871  & 2.0048\footnotemark[3]$^,$\footnotemark[8], &  149.396 
   & 148.632\footnotemark[9],148.822\footnotemark[5],  
   & 1.9740  & 1.9935\footnotemark[10],1.991\footnotemark[2], & 149.753
   & 149.241\footnotemark[10],149.07\footnotemark[11],  \\
   &  & 2.0102\footnotemark[12] & & 148.788\footnotemark[12],148.914\footnotemark[3]
   &  & 1.9879\footnotemark[3]  & & 149.222\footnotemark[5],149.374\footnotemark[3] \\
 6 & 1.9094  & 1.9165\footnotemark[3] & 151.510  
   & 150.855\footnotemark[9],151.317\footnotemark[3],
   & 1.9075  & 1.9214\footnotemark[10],1.9145\footnotemark[3] & 151.562 
   & 151.203\footnotemark[10],150.88\footnotemark[11], \\
   &   &   &   &  151.025\footnotemark[5] &  &  &  
   & 151.371\footnotemark[3],151.057\footnotemark[5]  \\
 8 & 1.9004  & 1.9072\footnotemark[3] & 151.755  
   & 151.092\footnotemark[9],151.570\footnotemark[3],
   & 1.8996  & 1.9068\footnotemark[3] & 151.777 
   & 151.11\footnotemark[11],151.581\footnotemark[3], \\
   &  &   &  & 151.263\footnotemark[5]  &  &  &  & 151.285\footnotemark[5] \\
10 & 1.8965  & 1.9037\footnotemark[3] & 151.861  
   & 151.190\footnotemark[9],151.665\footnotemark[3]
   & 1.8961  & 1.9034\footnotemark[3] & 151.872
   & 151.20\footnotemark[11],151.673\footnotemark[3]  \\
12 & 1.8944  & 1.9018\footnotemark[3] & 151.918  
   & 151.247\footnotemark[9],151.717\footnotemark[3]
   & 1.8943  &    & 151.921 &     \\
16 & 1.8925  &    & 151.970 & 151.296\footnotemark[9]
   & 1.8925  &    & 151.970 &     \\
20 & 1.8917  &    & 151.992 & 151.318\footnotemark[9]  
   & 1.8918  &    & 151.989 &     \\     
22 & 1.8914  &    & 152.000 & 151.325\footnotemark[9]
   & 1.8916  &    & 151.994 &      \\     
\end{tabular}
\begin{tabbing}
$^{\mathrm{a}}\!$ Ref.~\cite{chung95}. \hspace{0.2in} \= $^{\mathrm{b}}\!$ Ref.~\cite{madsen00}. \hspace{0.2in} \= 
$^{\mathrm{c}}\!$ Ref.~\cite{conneely02}. \hspace{0.2in} \= $^{\mathrm{d}}\!$ Ref.~\cite{diehl96}. \hspace{0.2in} \= 
$^{\mathrm{e}}\!$ Ref.~\cite{berrington98}. \hspace{0.2in} $^{\mathrm{f}}\!$ Ref.~\cite{azuma95}.  \\ 
$^{\mathrm{g}}\!$ Ref.~\cite{kiernan95}. \hspace{0.2in} \= $^{\mathrm{h}}\!$ Ref.~\cite{morishita03}. \hspace{0.2in} \= 
$^{\mathrm{i}}\!$ Ref.~\cite{zhou00}. \hspace{0.2in} \= $^{\mathrm{j}}\!$ Ref.~\cite{chung96}. \hspace{0.2in} \= 
$^{\mathrm{k}}\!$ Ref.~\cite{voky98}. \hspace{0.2in} \= $^{\mathrm{l}}\!$ Ref.~\cite{zhang98}.
\end{tabbing}
\end{center}
\vspace{-0.2in}
\end{ruledtabular}
\end{table}
\endgroup

\begingroup
\squeezetable
\begin{table}
\caption {Selected $3l3l'nl''$ term energies (in a.u.) and excitation energies (in eV) of Li, relative to 
the ground state of \cite{chung91}. Adopted from ref.~\cite{roy04b}.}
\begin{ruledtabular}
\begin{tabular}{lllllllll}
State & $-$E & Exc. energy & State & $-$E & Exc. energy & 
State & $-$E & Exc. energy \\  
\hline
3s$^2$4s  $^2$S$^e$   &  0.90054  &  178.959   &
3p$^3$  $^4$S$^o$     &  1.00055  &  176.238   & 
3p$^2$4s  $^4$P$^e$   &  0.89860  &  179.012   \\ 
3s$^2$6s  $^2$S$^e$   &  0.85729  &  180.136   &
3s3p4s  $^4$P$^o$     &  0.93313  &  178.072   & 
3p$^2$6s  $^4$P$^e$   &  0.85744  &  180.132   \\ 
3s$^2$3p  $^2$P$^o$   &  1.01210\footnotemark[1]  &  
175.924\footnotemark[2]$^,$\footnotemark[3]   &
3s3p5s  $^4$P$^o$     &  0.90193  &  178.921   & 
3p$^2$4s  $^2$D$^e$   &  0.87282  &  179.713   \\ 
3s$^2$6p  $^2$P$^o$   &  0.85558  &  180.182   &
3s3p5p  $^4$D$^e$     &  0.89924  &  178.994   & 
3p$^2$4p  $^4$D$^o$   &  0.89642  &  179.071   \\ 
3s$^2$3d  $^2$D$^e$   &  0.97108  &  177.040   & 
3s3p6p  $^4$D$^e$     &  0.88767  &  179.309   & 
3p$^2$5p  $^4$D$^o$   &  0.86846  &  179.832   \\ 
3s$^2$6d  $^2$D$^e$   &  0.85352  &  180.238   & 
3s3p5d  $^4$F$^o$     &  0.89400  &  179.137   & 
3p$^2$5p  $^2$F$^o$   &  0.84343  &  180.513   \\ 
3s3p$^2$  $^4$P$^e$   &  1.02288\footnotemark[4]  &  175.630   & 
3s3p6d  $^4$F$^o$     &  0.88517  &  179.377   & 
3p$^2$6p  $^2$F$^o$   &  0.83225  &  180.817   \\ 
\end{tabular}
\end{ruledtabular}
\footnotetext[1]{Reference theoretical values are:  1.043414 a.u. \cite{diehl97}, 1.043 a.u. \cite{madsen00} and 
1.040985 a.u. \cite{piangos03}.}
\footnotetext[2]{Reference experimental results are: 175.25 eV \cite{azuma97} and 175.165$\pm$0.050 eV \cite{diehl97}.}
\footnotetext[3]{Reference theoretical values are: 174.11 eV \cite{diehl97}, 174.14 eV \cite{azuma97} and
175.15 eV \cite{piangos03}.}
\footnotetext[4]{Reference theoretical value is: 1.0393859 a.u. \cite{piangos03}.}
\end{table}
\endgroup

Now we present results for triply excited hollow resonances, $2l2l'nl'' (n \geq 2)$ in Li. In \cite{roy04b}, 12 such resonance 
series, {\em viz.,} 2s$^2$ns $^2$S$^e$, 2s$^2$np $^2$P$^o$, 2s$^2$nd $^2$D$^e$, 2p$^2$ns $^2$D$^e$,$^4$P$^e$, 2s2pns $^4$P$^o$, 
2s2pnp $^4$D$^e$, 2p$^2$np $^2$F$^o$,$^4$D$^o$, 2p$^2$nd $^2$G$^e$, $^4$F$^e$ and 2s2pnd $^4$F$^o$, covering a total of about 
270 low-, moderately high- and high-lying states (with n as high as 25) were studied in some detail. These represent the 
model case of a highly correlated, multi-excited three electron system in presence of a nucleus, and hence a four-body
Coulombic problem. These are often termed as \emph{hollow} states, as all three electrons reside in higher shells leaving the
K shell empty. They have many fascinating properties, as well are very difficult for both theory and experiments. For example, 
these are difficult to produce from ground state by single photon absorption or electron impact excitation; also they have
proximity to more than one thresholds; moreover there are infinite open channels associated with these resonance. Table V
gives some representative state energies and excitation energies of even-parity 2s$^2$ns $^2$S$^e$ and odd-parity 2s$^2$np 
$^2$P$^o$ resonances in Li (n=2--22). The latter is calculated relative to the accurate ground state of Li, using full core 
plus correlation within a multi-configuration
interaction wave function \cite{chung91}. The reference energy value $-$7.47805953 a.u. is to be compared with our present 
value of $-$7.4782839 a.u. In literature, these states are conveniently classified using our independent particle model 
classification 
\cite{conneely00, conneely02} where the six core Li$^+$ n=2 intrashell doubly excited states, {\em viz.,} 2s$^2$ 
$^1$S$^e$, 2s2p $^3$P$^o$, 2p$^2$ $^3$P$^e$, 2p$^2$ $^1$D$^e$, 2s2p $^1$P$^o$ and 2p$^2$ $^1$S$^e$ are denoted by A, B, C, D, E 
and F respectively. For the former, no experimental results are available as yet. Lower members of the former series have been
studied in considerable detail by a variety of techniques, such as a hyper-spherical coordinate method \cite{morishita03}, a
combination of saddle point and complex coordinate rotation \cite{zhang98}. Of late, an eigenphase derivative technique 
in conjunction with a quantum defect theory \cite{zhou00} reported the low and high resonances up to $n=22$, whereas the same
up to $n=12$ were done by a truncated diagonalization method \cite{conneely02}. Our DFT results are in good agreement with
these references; state energies lie about 0.36--0.88\% above \cite{conneely02}, whereas the excitation energies are higher
by 0.41--0.51\% from those of \cite{zhou00}. The $\langle A,ns \rangle$ $^2$P$^o$ resonances are the most widely studied
series in Li, both theoretically and experimentally. Position of the lowest state in this series has been experimentally 
measured at
142.33 eV \cite{kiernan95}, 142.35 eV \cite{azuma95}. These are generally supported by theoretical calculations, e.g., 
combined saddle-point and complex coordinate rotation approach \cite{chung95}, a complex scaling method having correlated
basis functions constructed from B-splines \cite{madsen00}, an R-matrix theory \cite{berrington98}, etc. Present
excitation energy is only 0.02\% above the experimental results. Other members of the series with $n=3-7$ are in reasonably 
good agreement with complex coordinate rotation calculations \cite{chung96}. Our state energies are underestimated by
0.24--0.98\% with respect to those of \cite{chung95,chung96}, leading to higher excitation energies (deviations with respect 
to \cite{berrington98,voky98} being only 0.05--0.32\% and 0.43--0.46\% for $n=2-9$ and $n=3-10$ respectively). For further 
discussion on these and other hollow resonances see \cite{roy04b}.  

Table VI extends the method for some higher lying triply excited hollow resonances of Li having both K and L shells empty, 
the so-called \emph{doubly hollow} states, {\em viz.,} $3l3l'$n$l''$(3$\le$n$\le$6) ($^2$S$^e$, $^2$P$^o$, $^2$D$^e$, 
$^2$F$^o$, $^4$S$^o$, $^4$P$^{e,o}$, $^4$D$^{e,o}$, $^4$F$^o$). For $2l2l'nl''$ resonances, several accurate, reliable 
experimental and theoretical results are available; however, the same for $3l3l'nl''$ resonances, are very limited mainly
because of the greater challenges encountered. These have very distinctive features: (a) they are weak (by about an order 
of magnitude compared to the lower hollow states), broad and having much larger widths \cite{azuma97}. The principal 
difficulties with these higher hollow states at larger photon energies are mainly due to a very rapid increase in the density 
of possible triply excited and other lower states of same symmetry, as well as of the number of open channels available, 
giving rise to very strong and quite complicated electron correlation effects. Nevertheless some attempts have been made, 
which are quoted here. The energies and decay rates of N$^{4+}$ and N$^{2+}$ $3l3l'$3$l''$ have been studied using a CI 
approach \cite{vacek92}. Positions and widths of N$^{4+}$ (3,3,3) $^2$S$^{e,o}$ states are investigated by a space partition 
as well as a stabilization procedure both of which use the L$^2$  discretization \cite{bachau96}. A large scale state specific 
theory calculations for 11 n=3 resonances of He$^-$ has been suggested \cite{nicolaides01}. Critical issues in the theory and 
computation of the lowest three n=3 intrashell states, {\em viz}., 3s$^2$3p $^2$P$^o$, 3s3p$^2$ $^4$P$^e$ and 3s3p$^2$ 
$^2$D$^e$ of Z=2--7 in the light of state specific theory, has been published \cite{piangos03}. Energies, widths and Auger 
branching ratios for eight He$^-$ $3l3l'$3$l''$ states are calculated by complex rotation method \cite{piangos03}. A
semi-quantitative analysis of the angular correlation of 64 n=3 intrashell states of a model three-electron atom confined on 
the surface of a sphere were presented recently \cite{morishita01}. The only result available for such triply photo-excited 
(3,3,3) KL hollow state for Li are the 3s$^2$3p $^2$P$^o$ and 3s3p$^2$ $^4$P$^e$, both theoretically, whereas only the former 
experimentally. The former's position has been measured at 175.25 and 175.165 eV by synchrotron radiation measurement 
\cite{azuma97} 
and photo-ion spectroscopy \cite{diehl97}. In a saddle-point calculation with R-matrix approximation \cite{diehl97}, a 
570-term 25 angular component wave function gives an energy of $-1.043414$ a.u., and position at 174.11 eV. This is in 
reasonable agreement with the complex rotation calculation of $-1.043$ a.u., \cite{madsen00}, and the state specific result 
\cite{piangos03} of $-1$.0409856 a.u., as well as, with the multi-configuration Dirac-Fock \cite{azuma97} excitation energy of 
174.14 eV. DFT energy value of $-1.01210$ a.u., gives its position at 175.940 eV, (about 0.67 eV above the experimental value 
of \cite{azuma97}) and matches well with the state specific result of 175.15 eV \cite{piangos03}. Calculated 3s3p$^2$ 
$^4$P$^e$ state energy of $-1$.02288 a.u., matches closely with the state specific result of $-1$.0393859 a.u. \cite{piangos03}. 
Leaving aside a few of those as mentioned above, most of these can not be compared directly due to lack of any reference
values and we expect that these results may be useful in future studies of these resonances. Note that our result gives 
3s3p$^2$ $^4$P$^e$ as the lowest n=3 resonance rather than the 3s$^2$3p $^2$P$^o$, the former lying 0.0108 a.u., 
below the latter which coincides with the ordering found in other calculation such as complex rotation for He$^-$ 
\cite{chung01} and CI calculation for N$^{4+}$ \cite{vacek92}. However this disagrees with the state specific calculation of 
\cite{piangos03}, where the ordering is reversed and separation for Li being about 0.0016 a.u. Clearly, more accurate
calculation with better correlation functionals would be required to achieve such smaller separations (of the order of 
1$\times$10$^{-3}$ a.u.) within this DFT formalism to reach a more authentic conclusion. Now Fig. 1 depicts the radial 
densities for some representative (a) $2l2l'$n$l''$ and (b) $3l3l'$n$l''$ hollow states; as expected, they show the 
characteristic shell structures (superpositions of orbital radial densities).

\begin{figure}
\centering
\begin{minipage}[c]{0.40\textwidth}
\centering
\includegraphics[scale=0.45]{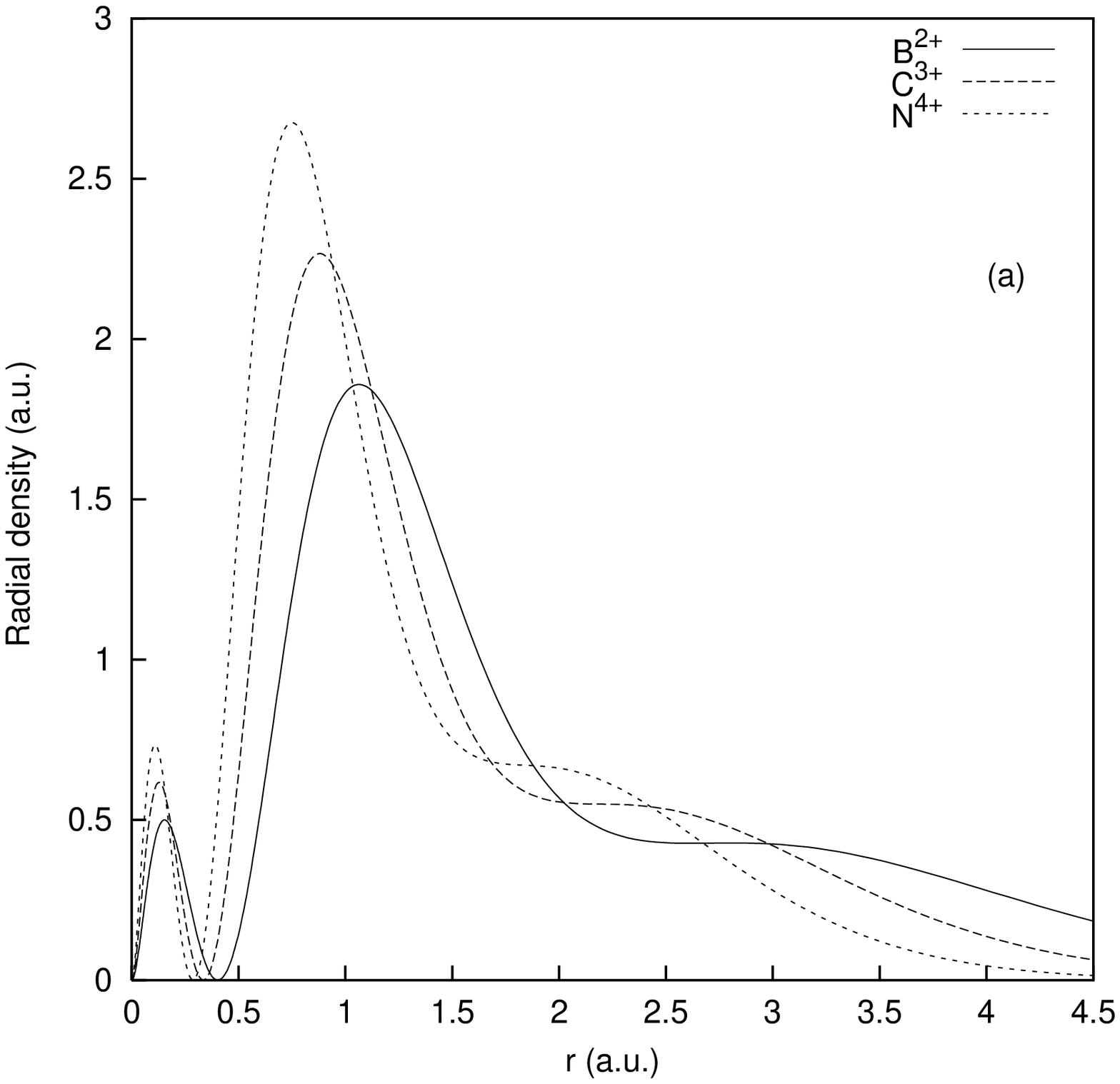}
\end{minipage}%
\hspace{0.6in}
\begin{minipage}[c]{0.40\textwidth}
\centering
\includegraphics[scale=0.45]{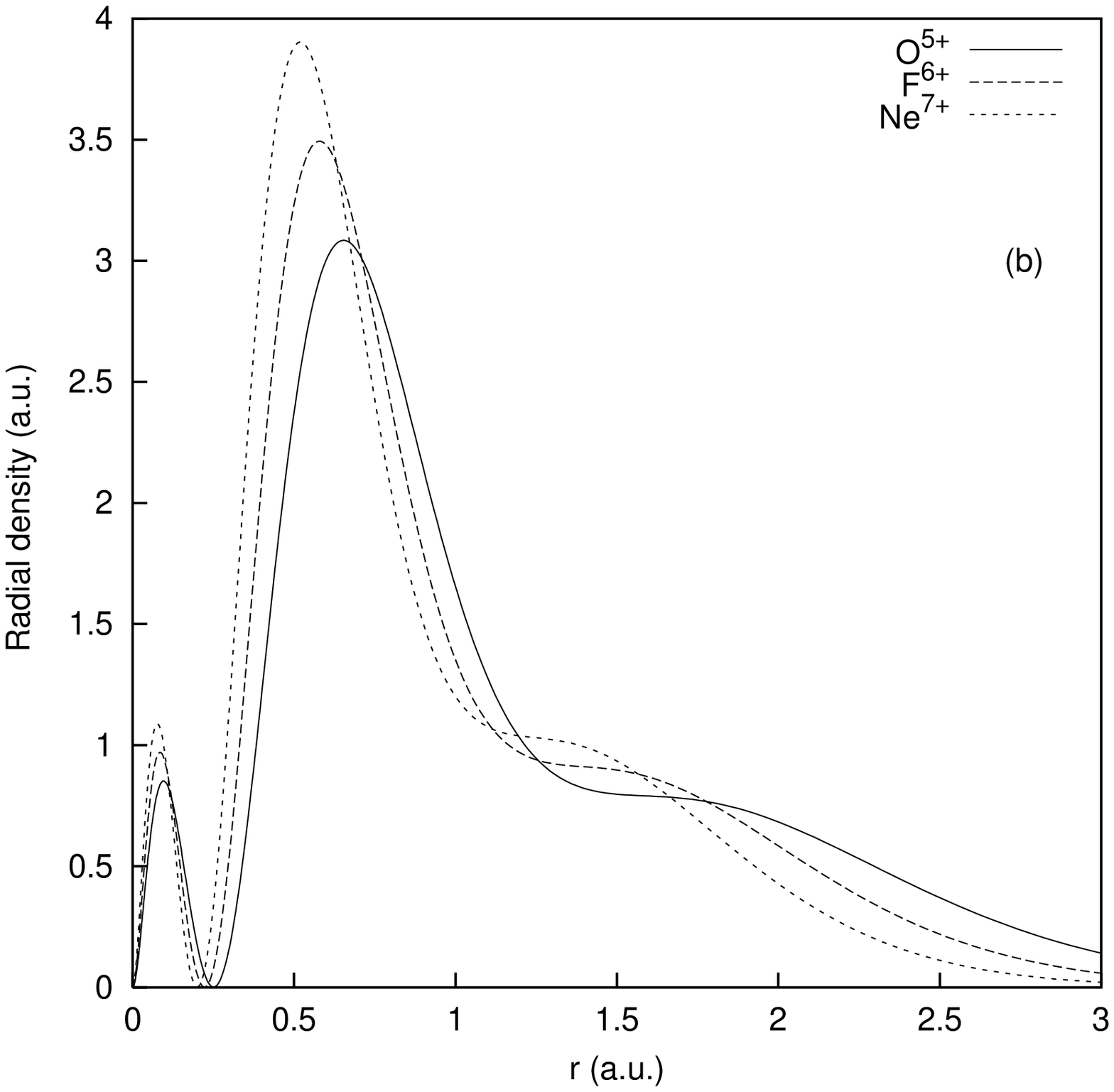}
\end{minipage}%
\caption{The radial densities (a.u.) of 2s$^2$3s $^2$S$^e$ states for (a) B$^{2+}$,
C$^{3+}$, N$^{4+}$ and (b) O$^{5+}$, F$^{6+}$, Ne$^{7+}$ respectively. Taken from ref.~\cite{roy05b}.}
\end{figure}

Now Table VII reports DFT energies for ground and excited states of some negative atoms, namely, Li$^-$ and Be$^-$. For the  
former, 4 states are considered, \emph{viz.,} [He]2s$^2$ $^1$S$^e$, 1s2s2p$^2$ $^5$P$^e$, [He]2p$^3$ $^5$S$^o$, 
1s2s2p3p $^5$P$^e$; while for the latter 3 states, i.e, [He]2s2p$^2$ $^4$P$^e$, [He]2p$^3$ $^4$S$^o$, 1s2s2p$^3$ $^6$S$^o$. 
Two sets of energies are reported \emph{viz.,} X-only (non-correlated) and XC (correlated, with LYP functional). Excepting the 
core-excited even-parity 1s2s2p3p $^5$P$^e$ of Li$^-$ (reported lately), the rest 4 states have been investigated quite 
extensively. Comparisons with literature data are made, wherever possible.  Per cent deviations are given in parentheses; 
for X-only case these are relative to the lone literature results in column 4; for XC case, these are with respect to the 
recent variational Monte Carlo (VMC) \cite{galvez06} values except for the 4th state of Li$^-$, where such a result is
unavailable; and this is given in reference to the saddle-point calculation of \cite{wang06}. Our X-only ground state of 
Li$^-$ is higher from the accurate HF calculation of \cite{fischer93} by a marginal 0.0004 a.u. The XC energy shows 
fairly good agreement (slightly above) with the accurate correlated MCHF-$n$ expansion considering all expansions 
\cite{fischer93}, as well as VMC method \cite{galvez06}. These seem to be in considerable disagreement with the earlier 
result of \cite{weiss68}. X-only results for the core-excited high-spin even-parity $^5$P$^e$ and odd-parity $^5$S$^o$ 
states of Li$^-$ also show excellent agreement with the HF energy \cite{galvez06}, while XC energies match well with 
literature values, such as VMC \cite{galvez06}, CI \cite{bunge80}, variational multi-configuration calculation 
\cite{yang95}, saddle-point \cite{wang06}, MCHF \cite{fischer90}, etc. Note that XC energies for these two states are lower 
than all of these reference results by 0.171 and 0.137\%, respectively giving maximum deviations in our calculation. As
already verified, the X-only results are practically of HF quality; hence this overestimation is probably caused by the 
approximate correlation potential used. Note that while ours is a single determinantal method, some of these correlated 
calculations are highly elaborate and extensive; for example, in \cite{yang95}, a 45 angular component 1004-term wave function
was used, \cite{bunge80} used a 320-term CI and \cite{wang06} used 7-50 angular spin components with 541--1298 linear parameters 
for the former. Nevertheless, as clearly seen, our results exhibit a rather small discrepancy from these references. 
Even-parity $^5$P$^e$ state has been considered rather lately \cite{wang06}. Current energy value shows very good matching 
with the reference, where the authors used a saddle-point restricted variational method with accurate multi-configurational 
wave functions built from STO basis sets. Be$^-$ does not have a [He]2s$^2$2p $^2$P ground state; three metastable bound 
states found in the discrete spectrum are given here. [He]2s2p$^2$ $^4$P$^e$, [He]2p$^3$ $^4$S$^o$ and 1s2s2p$^3$ $^6$S$^o$ 
states lie below the Be 1s$^2$2s2p $^3$P, 1s$^2$2p$^2$ $^3$P and 1s2s2p$^2$ $^5$P excited states. X-only energies are 
again in excellent agreement with the HF results \cite{beck81}. A decent number of sophisticated accurate theoretical results 
are found in literature for the correlated case. Notable amongst them are the CI calculation of \cite{beck81} including 
single, double, triple, quadrupole sub-shell excitations. First two states have also been studied through a method of 
full core plus correlation and restricted variation approach \cite{hsu95}. XC energies, in this case, show better agreements 
with literature results than those for Li$^-$; however, as in Li$^-$ again falling below the reference values for all three 
states. Also a combined Rayleigh-Ritz and a method of restricted variation exists for the $^6$S$^o$ state \cite{hsu95a}.
The transition wave lengths of Li$^-$ 1s2s2p$^2$ $^5$P$^e$ $\rightarrow$ 1s2p$^3$ $^5$S$^o$ and Be$^-$ [He]2s2p$^2$ $^4$P$^e$
$\rightarrow$ [He] 2p$^3$ $^4$S$^o$ have also been found to be in good agreement with literature results \cite{roy07}. 

\begingroup
\squeezetable
\begin{table}
\caption {Calculated ground and excited states of Li$^-$, Be$^-$ along with literature data. 
Numbers in the parentheses denote absolute per cent deviations. Adopted from \cite{roy07}.}
\begin{ruledtabular}
\begin{tabular}{llllll}
Ion     &  State  & \multicolumn{4}{c}{$-$E(a.u.)}                  \\
\cline{3-6} 
        &         & \multicolumn{2}{c}{X-only}  &  \multicolumn{2}{c}{XC}       \\
\cline{3-4}  \cline{5-6}
        &                       &  This work &  Ref.     &   This work  &  Ref.  \\
\hline 
Li$^-$ & [He]2s$^2$ $^1$S$^e$   & 7.4278(0.005)   & 7.4282\footnotemark[1]     
                                & 7.4984(0.009)   & 7.4553\footnotemark[2],7.5008\footnotemark[1],
                                                    7.4991\footnotemark[3]        \\ 
       & 1s2s2p$^2$ $^5$P$^e$   & 5.3640(0.006)   & 5.3643\footnotemark[3] 
                                & 5.3925(0.171)   & 5.3866\footnotemark[4]$^,$\footnotemark[5],
                                                    5.3833\footnotemark[3],5.3865\footnotemark[6],
                                                    5.3863\footnotemark[11] \\
       & 1s2p$^3$ $^5$S$^o$     & 5.2223(0.004)   & 5.2225\footnotemark[3]
                                & 5.2608(0.137)   & 5.2561\footnotemark[4]$^,$\footnotemark[5],
                                                    5.2536\footnotemark[3],5.2560\footnotemark[6],
                                                    5.2558\footnotemark[11] \\
        & 1s2s2p3p $^5$P$^e$    & 5.3289          &   
                                & 5.3683(0.007)   & 5.3679\footnotemark[5]                      \\
Be$^-$ & [He]2s2p$^2$ $^4$P$^e$ & 14.5078(0.008)  & 14.5090\footnotemark[7] 
                                & 14.5806(0.062)  & 14.5779\footnotemark[8],14.5716\footnotemark[3],
                                                    14.5708\footnotemark[7],14.5769\footnotemark[10]\\
        & [He]2p$^3$ $^4$S$^o$  & 14.3272(0.002)  & 14.3275\footnotemark[7]
                                & 14.4081(0.049)  & 14.4063\footnotemark[8],14.4010\footnotemark[3],
                                                    14.4002\footnotemark[7]  \\
        & 1s2s2p$^3$ $^6$S$^o$  & 10.4279(0.009)  & 10.4288\footnotemark[7]
                                & 10.4758(0.092)  & 10.4662\footnotemark[3], 10.4615\footnotemark[7],
                                                    10.4711\footnotemark[9]  \\
\end{tabular}
\end{ruledtabular}
\begin{tabbing}
$^{\mathrm{a}}$Ref. \cite{fischer93}.  \hspace{0.3in}  \= $^{\mathrm{b}}${Ref. \cite{weiss68}.} \hspace{0.3in} \=
$^{\mathrm{c}}${Ref. \cite{galvez06}.} \hspace{0.3in} \= $^{\mathrm{d}}${Ref. \cite{yang95}.}  \hspace{0.3in} \=
$^{\mathrm{e}}${Ref. \cite{wang06}.}   \hspace{0.3in}  \= $^{\mathrm{f}}${Ref. \cite{bunge80}.} \\ 
$^{\mathrm{g}}${Ref. \cite{beck81}.}   \hspace{0.3in} \= $^{\mathrm{h}}${Ref. \cite{hsu95}.}  \hspace{0.3in} \= 
$^{\mathrm{i}}${Ref. \cite{hsu95a}.}   \hspace{0.3in} \= $^{\mathrm{j}}${Ref. \cite{bunge86}.} \hspace{0.3in} \= 
$^{\mathrm{k}}${Ref. \cite{fischer90}.} 
\end{tabbing}
\end{table}
\endgroup

Before passing, a few remarks should be made on the present approach. Although DFT has enjoyed remarkable success for studying 
properties of atoms, molecules, solids, clusters in ground states, the same for excited states has been less conspicuous, 
partly because of complete abandoning of the state-function concept. Other major problems are cumbersome wave function
and Hamiltonian orthogonality requirements between a given excited state and other lower states of same space-spin symmetry, 
as well as the unavailability of universal XC energy density functional. Despite all these difficulties, numerous attractive
and elegant attempts have been made over the years to tackle these issues. However, until only very recently, 
most of the proposed methods have been either found to be computationally rather difficult to implement 
or producing large errors in excitation energies, except TDDFT where decent excitation energies are obtained. Moreover, it is 
not a straightforward task to extract the radial density.
Also, most of these methods have dealt with lower and singly excited states; multiple and higher excitations, especially
the Rydberg series as studied here, have not been reported so far by any other DFT approach except the current method. 
Also note that while the
recently popular TDDFT route provides accurate excitation energies efficiently, extraction of individual state
energies as well as densities, expectation values are not easy. The present scheme however, offers a simple, attractive
way to produce energies, excitation energies as well as densities and expectation values with very good accuracy. 
In the current approach, all the well-known problems of DFT have essentially been bypassed by bringing the traditional 
wave function concept 
within DFT, so that the atomic orbital and electronic configuration pictures are retained. Fermi-hole charge distribution
and hence local exchange potential is precisely in terms of orbitals, which will be different for ground and excited states. 
Not being explicitly dependent on any functional form, the exchange potential is \emph{universal}; it is fixed by the given
electronic configuration of a particular state. Thus the same KS equation is now valid for both ground and excited states, 
obviating the necessity for obtaining the exchange potential as a functional derivative of exchange energy density functional. 
Instead it is now directly obtained from Fermi hole charge distribution. Due to the locality of XC potential, SCF
solution of KS equation is computationally much easier than the HF equations, which involve a nonlocal integral operator.
Yet, as demonstrated above, our X-only results are practically of HF quality and with correlation included, results go
much beyond the HF level. Unlike many other sophisticated quantum mechanical methods, the present methodology does not involve
basis set dependence, continuum mixing or explicit r$_{12}$ dependence; it works essentially in the single determinantal 
framework.     

In the HKS DFT, all many-body effects are incorporated into a local multiplicative potential, obtained as a
functional derivative $\delta E_{xc}[\rho]/\delta \rho$ within the variational principle. Although the exact form of this
$E_{xc}[\rho]$ is unknown, good approximations exist; however, with these approximations, bounds of total energy are no longer
rigorous. Therefore the work-function prescription is not derived from the variational principle for energy, in the sense
that it is not expressible as $\delta E_{xc}[\rho]/\delta \rho$, but is based on a \emph{physical} interpretation for the local
many body potential that an electron moves in an electrostatic potential arising from the Fermi-hole charge distribution. So
even though a KS-type equation is solved with the work-function potential, this procedure is not subject to a variational
bound. Thus the variational restriction on the excited state being the \emph{lowest} state of a particular space-spin 
symmetry is not applicable here. Furthermore, although the existence of a local effective potential is guaranteed in KS DFT,
no mathematical proof for the existence of such a potential for excited states has been known. Therefore, a key assumption is
that excited states can also be described by a local potential. This is based on the fact that the physical argument used for
the construction of ground state potential, can also be equally applied for excited states. Further discussion on the method
and its application could be found in \cite{singh96,singh96a,roy97,roy97a,roy97b,roy98,singh98,singh99,
vikas00,roy02,roy04b,roy05b,roy07}. 

\section{Conclusion}
A simple DFT methodology has been presented for accurate, reliable, efficient calculation of ground and excited states of 
neutral, 
positive, negative atoms. Nonrelativistic energies, excitation energies, radial densities, radial expectation values, transition
wave lengths are reported and compared with the best theoretical and experimental results available till date. The work-function
exchange in conjunction with a GPS scheme for the solution of resulting KS equation makes it a simple and computationally efficient
route for these important challenging systems. The accuracy achieved within this single determinantal framework is quite comparable 
to those from more elaborate and extensive calculations available in the literature. Success and usefulness of the method has been
clearly demonstrated for a wide variety of excitations from single to multiple and low to very high Rydberg resonances as well as the
satellite states, hollow, doubly hollow states etc. Computed quantities show excellent agreement with literature results. Almost all 
of these systems are highly correlated. Since the exchange potential is treated quite accurately (almost as good as HF), a major 
source of error in the present work is certainly due to the inefficiency of LYP potential in incorporating the delicate and intricate
correlation effects, which could be further improved or replaced by more accurate energy density functionals for better accuracy. 
The assumption of spherical symmetry in calculating the exchange potential could also account for partial errors as well. In other
words, the rotational component of electric field may not have insignificant contribution compared to the irrotational component 
for these states, in general, although this usually holds true. To summarize, this work presented a current account of a simple
general and efficient DFT-based method for accurate and faithful description of multiply excited atomic systems. 

\section{acknowledgements}
I thank Prof.~F.~Columbus, for the kind invitation to present some of my recent works in this exciting book. I gratefully acknowledge
the support of Prof.~B.~M.~Deb in numerous ways. It is in his laboratory I got introduced to this fascinating area of DFT, and the
earlier developments of this work took place there, when I was working as a graduate student. I thank Prof.~P.~Panigrahi for critical
reading of the manuscript. It is a pleasure to thank my current colleagues at IISER-Kolkata, for their kind support. 

\bibliography{refn.bib}

\begin{thebibliography}{186}
\expandafter\ifx\csname natexlab\endcsname\relax\def\natexlab#1{#1}\fi
\expandafter\ifx\csname bibnamefont\endcsname\relax
  \def\bibnamefont#1{#1}\fi
\expandafter\ifx\csname bibfnamefont\endcsname\relax
  \def\bibfnamefont#1{#1}\fi
\expandafter\ifx\csname citenamefont\endcsname\relax
  \def\citenamefont#1{#1}\fi
\expandafter\ifx\csname url\endcsname\relax
  \def\url#1{\texttt{#1}}\fi
\expandafter\ifx\csname urlprefix\endcsname\relax\def\urlprefix{URL }\fi
\providecommand{\bibinfo}[2]{#2}
\providecommand{\eprint}[2][]{\url{#2}}

\bibitem[{\citenamefont{Thomas}(1927)}]{thomas27}
\bibinfo{author}{\bibfnamefont{L.~H.} \bibnamefont{Thomas}},
  \bibinfo{journal}{Proc.~Camb~Phil.~Soc.} \textbf{\bibinfo{volume}{23}},
  \bibinfo{pages}{542} (\bibinfo{year}{1927}).

\bibitem[{\citenamefont{Fermi}(1927)}]{fermi27}
\bibinfo{author}{\bibfnamefont{E.}~\bibnamefont{Fermi}},
  \bibinfo{journal}{Rend.~Accad.~Lincei} \textbf{\bibinfo{volume}{6}},
  \bibinfo{pages}{602} (\bibinfo{year}{1927}).

\bibitem[{\citenamefont{Dirac}(1930)}]{dirac30}
\bibinfo{author}{\bibfnamefont{P.~A.~M.} \bibnamefont{Dirac}},
  \bibinfo{journal}{Proc.~Camb.~Phil.~Soc.} \textbf{\bibinfo{volume}{26}},
  \bibinfo{pages}{376} (\bibinfo{year}{1930}).

\bibitem[{\citenamefont{Hohenberg and Kohn}(1964)}]{hohenberg64}
\bibinfo{author}{\bibfnamefont{P.}~\bibnamefont{Hohenberg}} \bibnamefont{and}
  \bibinfo{author}{\bibfnamefont{W.}~\bibnamefont{Kohn}},
  \bibinfo{journal}{Phys.~Rev.} \textbf{\bibinfo{volume}{136}},
  \bibinfo{pages}{B864} (\bibinfo{year}{1964}).

\bibitem[{\citenamefont{Kohn and Sham}(1965)}]{kohn65}
\bibinfo{author}{\bibfnamefont{W.}~\bibnamefont{Kohn}} \bibnamefont{and}
  \bibinfo{author}{\bibfnamefont{L.~J.} \bibnamefont{Sham}},
  \bibinfo{journal}{Phys.~Rev.} \textbf{\bibinfo{volume}{140}},
  \bibinfo{pages}{A1133} (\bibinfo{year}{1965}).

\bibitem[{\citenamefont{Parr and Yang}(1989)}]{parr89}
\bibinfo{author}{\bibfnamefont{R.~G.} \bibnamefont{Parr}} \bibnamefont{and}
  \bibinfo{author}{\bibfnamefont{W.}~\bibnamefont{Yang}},
  \emph{\bibinfo{title}{Density Functional Theory of Atoms and Molecules}}
  (\bibinfo{publisher}{{Oxford University Press}}, \bibinfo{address}{{New
  York}}, \bibinfo{year}{1989}).

\bibitem[{\citenamefont{Jones and Gunnarsson}(1989)}]{jones89}
\bibinfo{author}{\bibfnamefont{R.~O.} \bibnamefont{Jones}} \bibnamefont{and}
  \bibinfo{author}{\bibfnamefont{O.}~\bibnamefont{Gunnarsson}},
  \bibinfo{journal}{Rev.~Mod.~Phys.} \textbf{\bibinfo{volume}{61}},
  \bibinfo{pages}{689} (\bibinfo{year}{1989}).

\bibitem[{\citenamefont{{R.~M.~Dreizler and E.~K.~U.~Gross
  (Eds)}}(1990)}]{dreizler90}
\bibinfo{author}{\bibnamefont{{R.~M.~Dreizler and E.~K.~U.~Gross (Eds)}}},
  \emph{\bibinfo{title}{Density Functional Theory: An Approach to the Quantum
  Many-Body Problem}} (\bibinfo{publisher}{{Springer-Verlag}},
  \bibinfo{address}{{Berlin}}, \bibinfo{year}{1990}).

\bibitem[{\citenamefont{{D.~P.~Chong (Eds)}}(1995)}]{chong95}
\bibinfo{author}{\bibnamefont{{D.~P.~Chong (Eds)}}},
  \emph{\bibinfo{title}{Recent Advances in Density Functional Methods}}
  (\bibinfo{publisher}{{World Scientific}}, \bibinfo{address}{{Singapore}},
  \bibinfo{year}{1995}).

\bibitem[{\citenamefont{{J.~M.~Seminario (Eds)}}(1996)}]{seminario96}
\bibinfo{author}{\bibnamefont{{J.~M.~Seminario (Eds)}}},
  \emph{\bibinfo{title}{Recent Developments and Applications of Modern DFT}}
  (\bibinfo{publisher}{{Elsevier}}, \bibinfo{address}{{Amsterdam}},
  \bibinfo{year}{1996}).

\bibitem[{\citenamefont{{D.~Joulbert (Eds)}}(1998)}]{joulbert98}
\bibinfo{author}{\bibnamefont{{D.~Joulbert (Eds)}}},
  \emph{\bibinfo{title}{Density Functionals: Theory and Applications}}
  (\bibinfo{publisher}{{Springer}}, \bibinfo{address}{{Berlin}},
  \bibinfo{year}{1998}).

\bibitem[{\citenamefont{{J.~F.~Dobson and G.~Vignale and M.~P.~Das
  (Eds)}}(1998)}]{dobson98}
\bibinfo{author}{\bibnamefont{{J.~F.~Dobson and G.~Vignale and M.~P.~Das
  (Eds)}}}, \emph{\bibinfo{title}{Density Functional Theory: Recent Progress
  and New Directions}} (\bibinfo{publisher}{{Plenum}}, \bibinfo{address}{{New
  York}}, \bibinfo{year}{1998}).

\bibitem[{\citenamefont{{\'A.~Nagy}}(1998)}]{nagy98}
\bibinfo{author}{\bibnamefont{{\'A.~Nagy}}}, \bibinfo{journal}{Phys.~Rep.}
  \textbf{\bibinfo{volume}{1}}, \bibinfo{pages}{298} (\bibinfo{year}{1998}).

\bibitem[{\citenamefont{Kohn}(1999)}]{kohn99}
\bibinfo{author}{\bibfnamefont{W.}~\bibnamefont{Kohn}},
  \bibinfo{journal}{Rev.~Mod.~Phys.} \textbf{\bibinfo{volume}{71}},
  \bibinfo{pages}{1253} (\bibinfo{year}{1999}).

\bibitem[{\citenamefont{Koch and Holthausen}(2001)}]{koch01}
\bibinfo{author}{\bibfnamefont{W.}~\bibnamefont{Koch}} \bibnamefont{and}
  \bibinfo{author}{\bibfnamefont{M.~C.} \bibnamefont{Holthausen}},
  \emph{\bibinfo{title}{A Chemist's guide to Density Functional Theory}}
  (\bibinfo{publisher}{{John Wiley}}, \bibinfo{address}{{New York}},
  \bibinfo{year}{2001}).

\bibitem[{\citenamefont{{R.~G.~Parr and K.~D.~Sen (Eds)}}(2002)}]{parr02}
\bibinfo{author}{\bibnamefont{{R.~G.~Parr and K.~D.~Sen (Eds)}}},
  \emph{\bibinfo{title}{Reviews of Modern Quantum Chemistry: A Celebration of
  the Contributions of Robert G. Parr}} (\bibinfo{publisher}{{World
  Scientific}}, \bibinfo{address}{{Singapore}}, \bibinfo{year}{2002}).

\bibitem[{\citenamefont{{C.~Fiolhais and F.~Nogueira and M.~Marques
  (Eds)}}(2003)}]{fiolhais03}
\bibinfo{author}{\bibnamefont{{C.~Fiolhais and F.~Nogueira and M.~Marques
  (Eds)}}}, \emph{\bibinfo{title}{A Primer in Density Functional Theory}}
  (\bibinfo{publisher}{{Springer}}, \bibinfo{address}{{Berlin}},
  \bibinfo{year}{2003}).

\bibitem[{\citenamefont{Gidopoulos and Wilson}(2003)}]{gidopoulos03}
\bibinfo{author}{\bibfnamefont{N.~I.} \bibnamefont{Gidopoulos}}
  \bibnamefont{and} \bibinfo{author}{\bibfnamefont{S.}~\bibnamefont{Wilson}},
  \emph{\bibinfo{title}{The Fundamentals of Electron Density, Density Matrix
  and Density Functional Theory in Atoms, Molecules and the Solid State}}
  (\bibinfo{publisher}{{Springer}}, \bibinfo{address}{{Berlin}},
  \bibinfo{year}{2003}).

\bibitem[{\citenamefont{Martin}(2004)}]{martin04}
\bibinfo{author}{\bibfnamefont{R.~M.} \bibnamefont{Martin}},
  \emph{\bibinfo{title}{Electronic Structure: Basic Theory and Practical
  Methods}} (\bibinfo{publisher}{{Cambridge University Press}},
  \bibinfo{address}{{Cambridge}}, \bibinfo{year}{2004}).

\bibitem[{\citenamefont{Sholl and Steckel}(2009)}]{sholl09}
\bibinfo{author}{\bibfnamefont{D.~S.} \bibnamefont{Sholl}} \bibnamefont{and}
  \bibinfo{author}{\bibfnamefont{J.~A.} \bibnamefont{Steckel}},
  \emph{\bibinfo{title}{Density functional Theory: A Practical Introduction}}
  (\bibinfo{publisher}{{John-Wiley}}, \bibinfo{address}{{Hoboken, NJ}},
  \bibinfo{year}{2009}).

\bibitem[{\citenamefont{Singh and Deb}(1996{\natexlab{a}})}]{singh96}
\bibinfo{author}{\bibfnamefont{R.}~\bibnamefont{Singh}} \bibnamefont{and}
  \bibinfo{author}{\bibfnamefont{B.~M.} \bibnamefont{Deb}},
  \bibinfo{journal}{J.~Mol.~Struct. (Theochem)} \textbf{\bibinfo{volume}{361}},
  \bibinfo{pages}{1321} (\bibinfo{year}{1996}{\natexlab{a}}).

\bibitem[{\citenamefont{Singh and Deb}(1996{\natexlab{b}})}]{singh96a}
\bibinfo{author}{\bibfnamefont{R.}~\bibnamefont{Singh}} \bibnamefont{and}
  \bibinfo{author}{\bibfnamefont{B.~M.} \bibnamefont{Deb}},
  \bibinfo{journal}{J.~Chem.~Phys.} \textbf{\bibinfo{volume}{104}},
  \bibinfo{pages}{5892} (\bibinfo{year}{1996}{\natexlab{b}}).

\bibitem[{\citenamefont{Roy et~al.}(1997{\natexlab{a}})\citenamefont{Roy,
  Singh, and Deb}}]{roy97}
\bibinfo{author}{\bibfnamefont{A.~K.} \bibnamefont{Roy}},
  \bibinfo{author}{\bibfnamefont{R.}~\bibnamefont{Singh}}, \bibnamefont{and}
  \bibinfo{author}{\bibfnamefont{B.~M.} \bibnamefont{Deb}},
  \bibinfo{journal}{J.~Phys.~B} \textbf{\bibinfo{volume}{30}},
  \bibinfo{pages}{4763} (\bibinfo{year}{1997}{\natexlab{a}}).

\bibitem[{\citenamefont{Roy et~al.}(1997{\natexlab{b}})\citenamefont{Roy,
  Singh, and Deb}}]{roy97a}
\bibinfo{author}{\bibfnamefont{A.~K.} \bibnamefont{Roy}},
  \bibinfo{author}{\bibfnamefont{R.}~\bibnamefont{Singh}}, \bibnamefont{and}
  \bibinfo{author}{\bibfnamefont{B.~M.} \bibnamefont{Deb}},
  \bibinfo{journal}{Int.~J.~Quant.~Chem.} \textbf{\bibinfo{volume}{65}},
  \bibinfo{pages}{317} (\bibinfo{year}{1997}{\natexlab{b}}).

\bibitem[{\citenamefont{Roy and Deb}(1997)}]{roy97b}
\bibinfo{author}{\bibfnamefont{A.~K.} \bibnamefont{Roy}} \bibnamefont{and}
  \bibinfo{author}{\bibfnamefont{B.~M.} \bibnamefont{Deb}},
  \bibinfo{journal}{Phys.~Lett.~A} \textbf{\bibinfo{volume}{234}},
  \bibinfo{pages}{465} (\bibinfo{year}{1997}).

\bibitem[{\citenamefont{Roy and Deb}(1998)}]{roy98}
\bibinfo{author}{\bibfnamefont{A.~K.} \bibnamefont{Roy}} \bibnamefont{and}
  \bibinfo{author}{\bibfnamefont{B.~M.} \bibnamefont{Deb}},
  \bibinfo{journal}{Chem.~Phys.~Lett.} \textbf{\bibinfo{volume}{292}},
  \bibinfo{pages}{461} (\bibinfo{year}{1998}).

\bibitem[{\citenamefont{Singh et~al.}(1998)\citenamefont{Singh, Roy, and
  Deb}}]{singh98}
\bibinfo{author}{\bibfnamefont{R.}~\bibnamefont{Singh}},
  \bibinfo{author}{\bibfnamefont{A.~K.} \bibnamefont{Roy}}, \bibnamefont{and}
  \bibinfo{author}{\bibfnamefont{B.~M.} \bibnamefont{Deb}},
  \bibinfo{journal}{Chem.~Phys.~Lett.} \textbf{\bibinfo{volume}{296}},
  \bibinfo{pages}{530} (\bibinfo{year}{1998}).

\bibitem[{\citenamefont{Singh and Deb}(1999)}]{singh99}
\bibinfo{author}{\bibfnamefont{R.}~\bibnamefont{Singh}} \bibnamefont{and}
  \bibinfo{author}{\bibfnamefont{B.~M.} \bibnamefont{Deb}},
  \bibinfo{journal}{Phys.~Rep.} \textbf{\bibinfo{volume}{311}},
  \bibinfo{pages}{47} (\bibinfo{year}{1999}).

\bibitem[{\citenamefont{Vikas et~al.}(2000)\citenamefont{Vikas, Roy, and
  Deb}}]{vikas00}
\bibinfo{author}{\bibnamefont{Vikas}}, \bibinfo{author}{\bibfnamefont{A.~K.}
  \bibnamefont{Roy}}, \bibnamefont{and} \bibinfo{author}{\bibfnamefont{B.~M.}
  \bibnamefont{Deb}}, \bibinfo{journal}{Ind.~J.~Chem.~Sec.~A (Special Issue)}
  \textbf{\bibinfo{volume}{39}}, \bibinfo{pages}{32} (\bibinfo{year}{2000}).

\bibitem[{\citenamefont{Roy and Chu}(2002{\natexlab{a}})}]{roy02}
\bibinfo{author}{\bibfnamefont{A.~K.} \bibnamefont{Roy}} \bibnamefont{and}
  \bibinfo{author}{\bibfnamefont{S.~I.} \bibnamefont{Chu}},
  \bibinfo{journal}{Phys.~Rev.~A} \textbf{\bibinfo{volume}{65}},
  \bibinfo{pages}{052508} (\bibinfo{year}{2002}{\natexlab{a}}).

\bibitem[{\citenamefont{Roy}(2004{\natexlab{a}})}]{roy04b}
\bibinfo{author}{\bibfnamefont{A.~K.} \bibnamefont{Roy}},
  \bibinfo{journal}{J.~Phys.~B} \textbf{\bibinfo{volume}{37}},
  \bibinfo{pages}{4369} (\bibinfo{year}{2004}{\natexlab{a}}).

\bibitem[{\citenamefont{Roy}(2005{\natexlab{a}})}]{roy05b}
\bibinfo{author}{\bibfnamefont{A.~K.} \bibnamefont{Roy}},
  \bibinfo{journal}{J.~Phys.~B.} \textbf{\bibinfo{volume}{38}},
  \bibinfo{pages}{1591} (\bibinfo{year}{2005}{\natexlab{a}}).

\bibitem[{\citenamefont{Roy and Jalbout}(2007)}]{roy07}
\bibinfo{author}{\bibfnamefont{A.~K.} \bibnamefont{Roy}} \bibnamefont{and}
  \bibinfo{author}{\bibfnamefont{A.~F.} \bibnamefont{Jalbout}},
  \bibinfo{journal}{Chem.~Phys.~Lett.} \textbf{\bibinfo{volume}{445}},
  \bibinfo{pages}{355} (\bibinfo{year}{2007}).

\bibitem[{\citenamefont{Gunnarsson and Lundqvist}(1976)}]{gunnarsson76}
\bibinfo{author}{\bibfnamefont{O.}~\bibnamefont{Gunnarsson}} \bibnamefont{and}
  \bibinfo{author}{\bibfnamefont{B.~I.} \bibnamefont{Lundqvist}},
  \bibinfo{journal}{Phys.~Rev.~B} \textbf{\bibinfo{volume}{13}},
  \bibinfo{pages}{4274} (\bibinfo{year}{1976}).

\bibitem[{\citenamefont{Slater}(1972)}]{slater72}
\bibinfo{author}{\bibfnamefont{J.~C.} \bibnamefont{Slater}},
  \bibinfo{journal}{Adv.~Quant.~Chem.} \textbf{\bibinfo{volume}{6}},
  \bibinfo{pages}{1} (\bibinfo{year}{1972}).

\bibitem[{\citenamefont{Slater}(1974)}]{slater74}
\bibinfo{author}{\bibfnamefont{J.~C.} \bibnamefont{Slater}},
  \emph{\bibinfo{title}{The Self-Consistent Field for Molecules and Solids,
  Vol. IV}} (\bibinfo{publisher}{{McGraw-Hill}}, \bibinfo{address}{{New York}},
  \bibinfo{year}{1974}).

\bibitem[{\citenamefont{Ziegler et~al.}(1977)\citenamefont{Ziegler, Rauk, and
  Baerends}}]{ziegler77}
\bibinfo{author}{\bibfnamefont{T.}~\bibnamefont{Ziegler}},
  \bibinfo{author}{\bibfnamefont{A.}~\bibnamefont{Rauk}}, \bibnamefont{and}
  \bibinfo{author}{\bibfnamefont{E.~J.} \bibnamefont{Baerends}},
  \bibinfo{journal}{Theor.~Chim.~Acta} \textbf{\bibinfo{volume}{43}},
  \bibinfo{pages}{261} (\bibinfo{year}{1977}).

\bibitem[{\citenamefont{von Barth}(1979)}]{barth79}
\bibinfo{author}{\bibfnamefont{U.}~\bibnamefont{von Barth}},
  \bibinfo{journal}{Phys.~Rev.~A} \textbf{\bibinfo{volume}{20}},
  \bibinfo{pages}{1693} (\bibinfo{year}{1979}).

\bibitem[{\citenamefont{Daul}(1994)}]{daul94}
\bibinfo{author}{\bibfnamefont{C.}~\bibnamefont{Daul}},
  \bibinfo{journal}{Int.~J.~Quant.~Chem.} \textbf{\bibinfo{volume}{52}},
  \bibinfo{pages}{867} (\bibinfo{year}{1994}).

\bibitem[{\citenamefont{{A.~C.~St\"uckl}
  et~al.}(1997)\citenamefont{{A.~C.~St\"uckl}, Daul, and
  {H.~U.~G\"udel}}}]{stuckl97}
\bibinfo{author}{\bibnamefont{{A.~C.~St\"uckl}}},
  \bibinfo{author}{\bibfnamefont{C.~A.} \bibnamefont{Daul}}, \bibnamefont{and}
  \bibinfo{author}{\bibnamefont{{H.~U.~G\"udel}}},
  \bibinfo{journal}{Int.~J.~Quant.~Chem.} \textbf{\bibinfo{volume}{61}},
  \bibinfo{pages}{579} (\bibinfo{year}{1997}).

\bibitem[{\citenamefont{Theophilou}(1979)}]{theophilou79}
\bibinfo{author}{\bibfnamefont{A.~K.} \bibnamefont{Theophilou}},
  \bibinfo{journal}{J.~Phys.~C} \textbf{\bibinfo{volume}{12}},
  \bibinfo{pages}{5419} (\bibinfo{year}{1979}).

\bibitem[{\citenamefont{Gross et~al.}(1988{\natexlab{a}})\citenamefont{Gross,
  Oliveira, and Kohn}}]{gross88}
\bibinfo{author}{\bibfnamefont{E.~K.~U.} \bibnamefont{Gross}},
  \bibinfo{author}{\bibfnamefont{L.~N.} \bibnamefont{Oliveira}},
  \bibnamefont{and} \bibinfo{author}{\bibfnamefont{W.}~\bibnamefont{Kohn}},
  \bibinfo{journal}{Phys.~Rev.~A} \textbf{\bibinfo{volume}{37}},
  \bibinfo{pages}{2805} (\bibinfo{year}{1988}{\natexlab{a}}).

\bibitem[{\citenamefont{Gross et~al.}(1988{\natexlab{b}})\citenamefont{Gross,
  Oliveira, and Kohn}}]{gross88a}
\bibinfo{author}{\bibfnamefont{E.~K.~U.} \bibnamefont{Gross}},
  \bibinfo{author}{\bibfnamefont{L.~N.} \bibnamefont{Oliveira}},
  \bibnamefont{and} \bibinfo{author}{\bibfnamefont{W.}~\bibnamefont{Kohn}},
  \bibinfo{journal}{Phys.~Rev.~A} \textbf{\bibinfo{volume}{37}},
  \bibinfo{pages}{2809} (\bibinfo{year}{1988}{\natexlab{b}}).

\bibitem[{\citenamefont{Oliveira et~al.}(1988)\citenamefont{Oliveira, Gross,
  and Kohn}}]{oliveira88}
\bibinfo{author}{\bibfnamefont{L.~N.} \bibnamefont{Oliveira}},
  \bibinfo{author}{\bibfnamefont{E.~K.~U.} \bibnamefont{Gross}},
  \bibnamefont{and} \bibinfo{author}{\bibfnamefont{W.}~\bibnamefont{Kohn}},
  \bibinfo{journal}{Phys.~Rev.~A} \textbf{\bibinfo{volume}{37}},
  \bibinfo{pages}{2821} (\bibinfo{year}{1988}).

\bibitem[{\citenamefont{Kohn}(1986)}]{kohn86}
\bibinfo{author}{\bibfnamefont{W.}~\bibnamefont{Kohn}},
  \bibinfo{journal}{Phys.~Rev.~A} \textbf{\bibinfo{volume}{34}},
  \bibinfo{pages}{737} (\bibinfo{year}{1986}).

\bibitem[{\citenamefont{{\'A.~Nagy}}(1990)}]{nagy90}
\bibinfo{author}{\bibnamefont{{\'A.~Nagy}}}, \bibinfo{journal}{Phys.~Rev.~A}
  \textbf{\bibinfo{volume}{42}}, \bibinfo{pages}{4388} (\bibinfo{year}{1990}).

\bibitem[{\citenamefont{{\'A.~Nagy}}(1991)}]{nagy91}
\bibinfo{author}{\bibnamefont{{\'A.~Nagy}}}, \bibinfo{journal}{J.~Phys.~B}
  \textbf{\bibinfo{volume}{24}}, \bibinfo{pages}{4691} (\bibinfo{year}{1991}).

\bibitem[{\citenamefont{{R.~G\'asp\'ar}}(1974)}]{gaspar74}
\bibinfo{author}{\bibnamefont{{R.~G\'asp\'ar}}}, \bibinfo{journal}{Acta
  Phys.~Hung.} \textbf{\bibinfo{volume}{35}}, \bibinfo{pages}{213}
  (\bibinfo{year}{1974}).

\bibitem[{\citenamefont{Gunnarsson et~al.}(1974)\citenamefont{Gunnarsson,
  Lundqvist, and Wilkins}}]{gunnarsson74}
\bibinfo{author}{\bibfnamefont{O.}~\bibnamefont{Gunnarsson}},
  \bibinfo{author}{\bibfnamefont{B.~I.} \bibnamefont{Lundqvist}},
  \bibnamefont{and} \bibinfo{author}{\bibfnamefont{J.~W.}
  \bibnamefont{Wilkins}}, \bibinfo{journal}{Phys.~Rev.~B}
  \textbf{\bibinfo{volume}{10}}, \bibinfo{pages}{1319} (\bibinfo{year}{1974}).

\bibitem[{\citenamefont{von Barth and Hedin}(1972)}]{barth72}
\bibinfo{author}{\bibfnamefont{U.}~\bibnamefont{von Barth}} \bibnamefont{and}
  \bibinfo{author}{\bibfnamefont{L.}~\bibnamefont{Hedin}},
  \bibinfo{journal}{J.~Phys.~C} \textbf{\bibinfo{volume}{5}},
  \bibinfo{pages}{1629} (\bibinfo{year}{1972}).

\bibitem[{\citenamefont{Ceperley and Alder}(1980)}]{ceperley80}
\bibinfo{author}{\bibfnamefont{D.~M.} \bibnamefont{Ceperley}} \bibnamefont{and}
  \bibinfo{author}{\bibfnamefont{B.~J.} \bibnamefont{Alder}},
  \bibinfo{journal}{Phys.~Rev.~Lett.} \textbf{\bibinfo{volume}{45}},
  \bibinfo{pages}{566} (\bibinfo{year}{1980}).

\bibitem[{\citenamefont{Perdew and Zunger}(1981)}]{perdew81}
\bibinfo{author}{\bibfnamefont{J.~P.} \bibnamefont{Perdew}} \bibnamefont{and}
  \bibinfo{author}{\bibfnamefont{A.}~\bibnamefont{Zunger}},
  \bibinfo{journal}{Phys.~Rev.~B} \textbf{\bibinfo{volume}{23}},
  \bibinfo{pages}{5048} (\bibinfo{year}{1981}).

\bibitem[{\citenamefont{Vosko et~al.}(1980)\citenamefont{Vosko, Wilk, and
  Nusair}}]{vosko80}
\bibinfo{author}{\bibfnamefont{S.~H.} \bibnamefont{Vosko}},
  \bibinfo{author}{\bibfnamefont{L.}~\bibnamefont{Wilk}}, \bibnamefont{and}
  \bibinfo{author}{\bibfnamefont{M.}~\bibnamefont{Nusair}},
  \bibinfo{journal}{Can.~J.~Phys.} \textbf{\bibinfo{volume}{58}},
  \bibinfo{pages}{1200} (\bibinfo{year}{1980}).

\bibitem[{\citenamefont{{\'A.~Nagy}}(1996)}]{nagy96}
\bibinfo{author}{\bibnamefont{{\'A.~Nagy}}}, \bibinfo{journal}{J.~Phys.~B}
  \textbf{\bibinfo{volume}{29}}, \bibinfo{pages}{389} (\bibinfo{year}{1996}).

\bibitem[{\citenamefont{Fritsche}(1986)}]{fritsche86}
\bibinfo{author}{\bibfnamefont{L.}~\bibnamefont{Fritsche}},
  \bibinfo{journal}{Phys.~Rev.~B} \textbf{\bibinfo{volume}{33}},
  \bibinfo{pages}{3976} (\bibinfo{year}{1986}).

\bibitem[{\citenamefont{Cordes and Fritsche}(1989)}]{cordes89}
\bibinfo{author}{\bibfnamefont{J.}~\bibnamefont{Cordes}} \bibnamefont{and}
  \bibinfo{author}{\bibfnamefont{L.}~\bibnamefont{Fritsche}},
  \bibinfo{journal}{Z.~Phys.~D} \textbf{\bibinfo{volume}{13}},
  \bibinfo{pages}{345} (\bibinfo{year}{1989}).

\bibitem[{\citenamefont{Grimme}(1996)}]{grimme96}
\bibinfo{author}{\bibfnamefont{S.}~\bibnamefont{Grimme}},
  \bibinfo{journal}{Chem.~Phys.~Lett.} \textbf{\bibinfo{volume}{259}},
  \bibinfo{pages}{128} (\bibinfo{year}{1996}).

\bibitem[{\citenamefont{Grimme and Waletzke}(1999)}]{grimme99}
\bibinfo{author}{\bibfnamefont{S.}~\bibnamefont{Grimme}} \bibnamefont{and}
  \bibinfo{author}{\bibfnamefont{M.}~\bibnamefont{Waletzke}},
  \bibinfo{journal}{J.~Chem.~Phys.} \textbf{\bibinfo{volume}{111}},
  \bibinfo{pages}{5645} (\bibinfo{year}{1999}).

\bibitem[{\citenamefont{Sahni et~al.}(2001)\citenamefont{Sahni, Massa, Singh,
  and Slamet}}]{sahni01}
\bibinfo{author}{\bibfnamefont{V.}~\bibnamefont{Sahni}},
  \bibinfo{author}{\bibfnamefont{L.}~\bibnamefont{Massa}},
  \bibinfo{author}{\bibfnamefont{R.}~\bibnamefont{Singh}}, \bibnamefont{and}
  \bibinfo{author}{\bibfnamefont{M.}~\bibnamefont{Slamet}},
  \bibinfo{journal}{Phys.~Rev.~Lett.} \textbf{\bibinfo{volume}{87}},
  \bibinfo{pages}{113002} (\bibinfo{year}{2001}).

\bibitem[{\citenamefont{{M.~Levy and \'A.~Nagy}}(1999)}]{levy99}
\bibinfo{author}{\bibnamefont{{M.~Levy and \'A.~Nagy}}},
  \bibinfo{journal}{Phys.~Rev.~Lett.} \textbf{\bibinfo{volume}{83}},
  \bibinfo{pages}{4361} (\bibinfo{year}{1999}).

\bibitem[{\citenamefont{{A.~G\"orling}}(1996)}]{gorling96}
\bibinfo{author}{\bibnamefont{{A.~G\"orling}}}, \bibinfo{journal}{Phys.~Rev.~A}
  \textbf{\bibinfo{volume}{54}}, \bibinfo{pages}{3912} (\bibinfo{year}{1996}).

\bibitem[{\citenamefont{Filippi et~al.}(1997)\citenamefont{Filippi, Umrigar,
  and Gonze}}]{filippi97}
\bibinfo{author}{\bibfnamefont{C.}~\bibnamefont{Filippi}},
  \bibinfo{author}{\bibfnamefont{C.~J.} \bibnamefont{Umrigar}},
  \bibnamefont{and} \bibinfo{author}{\bibfnamefont{X.}~\bibnamefont{Gonze}},
  \bibinfo{journal}{J.~Chem.~Phys.} \textbf{\bibinfo{volume}{107}},
  \bibinfo{pages}{9994} (\bibinfo{year}{1997}).

\bibitem[{\citenamefont{San-Fabi\'an and Pastor-Abia}(2003)}]{sanfabian03}
\bibinfo{author}{\bibfnamefont{E.}~\bibnamefont{San-Fabi\'an}}
  \bibnamefont{and}
  \bibinfo{author}{\bibfnamefont{L.}~\bibnamefont{Pastor-Abia}},
  \bibinfo{journal}{Int.~J.~Quant.~Chem.} \textbf{\bibinfo{volume}{91}},
  \bibinfo{pages}{451} (\bibinfo{year}{2003}).

\bibitem[{\citenamefont{{F.Tasn\'adi and \'A.~Nagy}}(2003)}]{tasnadi03}
\bibinfo{author}{\bibnamefont{{F.Tasn\'adi and \'A.~Nagy}}},
  \bibinfo{journal}{Int.~J.~Quant.~Chem.} \textbf{\bibinfo{volume}{92}},
  \bibinfo{pages}{234} (\bibinfo{year}{2003}).

\bibitem[{\citenamefont{{A.~G\"orling}}(1999)}]{gorling99}
\bibinfo{author}{\bibnamefont{{A.~G\"orling}}}, \bibinfo{journal}{Phys.~Rev.~A}
  \textbf{\bibinfo{volume}{59}}, \bibinfo{pages}{3359} (\bibinfo{year}{1999}).

\bibitem[{\citenamefont{Vitale et~al.}(2005)\citenamefont{Vitale, Sala, and
  {A.~G\"orling}}}]{vitale05}
\bibinfo{author}{\bibfnamefont{V.}~\bibnamefont{Vitale}},
  \bibinfo{author}{\bibfnamefont{F.~D.} \bibnamefont{Sala}}, \bibnamefont{and}
  \bibinfo{author}{\bibnamefont{{A.~G\"orling}}},
  \bibinfo{journal}{J.~Chem.~Phys.} \textbf{\bibinfo{volume}{122}},
  \bibinfo{pages}{244102} (\bibinfo{year}{2005}).

\bibitem[{\citenamefont{Sala and {A.~G\"orling}}(2002)}]{sala02}
\bibinfo{author}{\bibfnamefont{F.~D.} \bibnamefont{Sala}} \bibnamefont{and}
  \bibinfo{author}{\bibnamefont{{A.~G\"orling}}},
  \bibinfo{journal}{J.~Chem.~Phys.} \textbf{\bibinfo{volume}{118}},
  \bibinfo{pages}{10439} (\bibinfo{year}{2002}).

\bibitem[{\citenamefont{Glushkov and Levy}(2005)}]{glushkov07}
\bibinfo{author}{\bibfnamefont{V.~N.} \bibnamefont{Glushkov}} \bibnamefont{and}
  \bibinfo{author}{\bibfnamefont{M.}~\bibnamefont{Levy}},
  \bibinfo{journal}{J.~Chem.~Phys.} \textbf{\bibinfo{volume}{126}},
  \bibinfo{pages}{174106} (\bibinfo{year}{2005}).

\bibitem[{\citenamefont{{\'A.~Nagy and M.~Levy}}(2001)}]{nagy01}
\bibinfo{author}{\bibnamefont{{\'A.~Nagy and M.~Levy}}},
  \bibinfo{journal}{Phys.~Rev.~A} \textbf{\bibinfo{volume}{63}},
  \bibinfo{pages}{052502} (\bibinfo{year}{2001}).

\bibitem[{\citenamefont{Glushkov}(2002{\natexlab{a}})}]{glushkov02}
\bibinfo{author}{\bibfnamefont{V.~N.} \bibnamefont{Glushkov}},
  \bibinfo{journal}{J.~Math.~Chem.} \textbf{\bibinfo{volume}{31}},
  \bibinfo{pages}{91} (\bibinfo{year}{2002}{\natexlab{a}}).

\bibitem[{\citenamefont{Glushkov}(2002{\natexlab{b}})}]{glushkov02a}
\bibinfo{author}{\bibfnamefont{V.~N.} \bibnamefont{Glushkov}},
  \bibinfo{journal}{Opt.~Spectrosc.} \textbf{\bibinfo{volume}{93}},
  \bibinfo{pages}{11} (\bibinfo{year}{2002}{\natexlab{b}}).

\bibitem[{\citenamefont{Besley et~al.}(2009)\citenamefont{Besley, Gilbert, and
  Gill}}]{besley09}
\bibinfo{author}{\bibfnamefont{N.~A.} \bibnamefont{Besley}},
  \bibinfo{author}{\bibfnamefont{A.~T.~B.} \bibnamefont{Gilbert}},
  \bibnamefont{and} \bibinfo{author}{\bibfnamefont{P.~M.~W.}
  \bibnamefont{Gill}}, \bibinfo{journal}{J.~Chem.~Phys.}
  \textbf{\bibinfo{volume}{130}}, \bibinfo{pages}{124308}
  (\bibinfo{year}{2009}).

\bibitem[{\citenamefont{Casida}(1995)}]{casida95}
\bibinfo{author}{\bibfnamefont{M.~E.} \bibnamefont{Casida}}, in
  \emph{\bibinfo{booktitle}{Recent {A}dvances in {D}ensity {F}unctional
  {M}ethods}}, edited by \bibinfo{editor}{\bibfnamefont{D.~P.}
  \bibnamefont{Chong}} (\bibinfo{publisher}{{World Scientific}},
  \bibinfo{address}{Singapore}, \bibinfo{year}{1995}), p. \bibinfo{pages}{155}.

\bibitem[{\citenamefont{Bauernschmitt and Ahlrichs}(1996)}]{bauernschmitt96}
\bibinfo{author}{\bibfnamefont{R.}~\bibnamefont{Bauernschmitt}}
  \bibnamefont{and} \bibinfo{author}{\bibfnamefont{R.}~\bibnamefont{Ahlrichs}},
  \bibinfo{journal}{Chem.~Phys.~Lett.} \textbf{\bibinfo{volume}{256}},
  \bibinfo{pages}{454} (\bibinfo{year}{1996}).

\bibitem[{\citenamefont{Petersilka et~al.}(1996)\citenamefont{Petersilka,
  Gossmann, and Gross}}]{petersilka96}
\bibinfo{author}{\bibfnamefont{M.}~\bibnamefont{Petersilka}},
  \bibinfo{author}{\bibfnamefont{U.~J.} \bibnamefont{Gossmann}},
  \bibnamefont{and} \bibinfo{author}{\bibfnamefont{E.~K.~U.}
  \bibnamefont{Gross}}, \bibinfo{journal}{Phys.~Rev.~Lett.}
  \textbf{\bibinfo{volume}{76}}, \bibinfo{pages}{1212} (\bibinfo{year}{1996}).

\bibitem[{\citenamefont{Tozer and Handy}(1998)}]{tozer98}
\bibinfo{author}{\bibfnamefont{D.~J.} \bibnamefont{Tozer}} \bibnamefont{and}
  \bibinfo{author}{\bibfnamefont{N.~C.} \bibnamefont{Handy}},
  \bibinfo{journal}{J.~Chem.~Phys.} \textbf{\bibinfo{volume}{109}},
  \bibinfo{pages}{10180} (\bibinfo{year}{1998}).

\bibitem[{\citenamefont{van Gisbergen et~al.}(1998)\citenamefont{van Gisbergen,
  Kootstra, Schipper, Gritsenko, Snijders, and Baerends}}]{gisbergen98}
\bibinfo{author}{\bibfnamefont{S.~J.~A.} \bibnamefont{van Gisbergen}},
  \bibinfo{author}{\bibfnamefont{F.}~\bibnamefont{Kootstra}},
  \bibinfo{author}{\bibfnamefont{P.~R.~T.} \bibnamefont{Schipper}},
  \bibinfo{author}{\bibfnamefont{O.~V.} \bibnamefont{Gritsenko}},
  \bibinfo{author}{\bibfnamefont{J.~G.} \bibnamefont{Snijders}},
  \bibnamefont{and} \bibinfo{author}{\bibfnamefont{E.~J.}
  \bibnamefont{Baerends}}, \bibinfo{journal}{Phys.~Rev.~A}
  \textbf{\bibinfo{volume}{57}}, \bibinfo{pages}{2556} (\bibinfo{year}{1998}).

\bibitem[{\citenamefont{Stratmann et~al.}(1998)\citenamefont{Stratmann,
  Scuseria, and Frisch}}]{stratmann98}
\bibinfo{author}{\bibfnamefont{R.~E.} \bibnamefont{Stratmann}},
  \bibinfo{author}{\bibfnamefont{G.~E.} \bibnamefont{Scuseria}},
  \bibnamefont{and} \bibinfo{author}{\bibfnamefont{M.~J.}
  \bibnamefont{Frisch}}, \bibinfo{journal}{J.~Chem.~Phys.}
  \textbf{\bibinfo{volume}{109}}, \bibinfo{pages}{8218} (\bibinfo{year}{1998}).

\bibitem[{\citenamefont{{A.~G\"orling}
  et~al.}(1999)\citenamefont{{A.~G\"orling}, Heinze, Ruzankin, Staufer, and
  {N.~R\"osch}}}]{gorling99a}
\bibinfo{author}{\bibnamefont{{A.~G\"orling}}},
  \bibinfo{author}{\bibfnamefont{H.~H.} \bibnamefont{Heinze}},
  \bibinfo{author}{\bibfnamefont{S.~P.} \bibnamefont{Ruzankin}},
  \bibinfo{author}{\bibfnamefont{M.}~\bibnamefont{Staufer}}, \bibnamefont{and}
  \bibinfo{author}{\bibnamefont{{N.~R\"osch}}},
  \bibinfo{journal}{J.~Chem.~Phys.} \textbf{\bibinfo{volume}{110}},
  \bibinfo{pages}{2785} (\bibinfo{year}{1999}).

\bibitem[{\citenamefont{Yabana and Bertsch}(1999)}]{yabana99}
\bibinfo{author}{\bibfnamefont{K.}~\bibnamefont{Yabana}} \bibnamefont{and}
  \bibinfo{author}{\bibfnamefont{G.~F.} \bibnamefont{Bertsch}},
  \bibinfo{journal}{Int.~J.~Quant.~Chem.} \textbf{\bibinfo{volume}{75}},
  \bibinfo{pages}{55} (\bibinfo{year}{1999}).

\bibitem[{\citenamefont{Heinze et~al.}(2000)\citenamefont{Heinze,
  {A.~G\"orling}, and {N.~R\"osch}}}]{heinze00}
\bibinfo{author}{\bibfnamefont{H.~H.} \bibnamefont{Heinze}},
  \bibinfo{author}{\bibnamefont{{A.~G\"orling}}}, \bibnamefont{and}
  \bibinfo{author}{\bibnamefont{{N.~R\"osch}}},
  \bibinfo{journal}{J.~Chem.~Phys.} \textbf{\bibinfo{volume}{113}},
  \bibinfo{pages}{2088} (\bibinfo{year}{2000}).

\bibitem[{\citenamefont{Furche}(2001)}]{furche01}
\bibinfo{author}{\bibfnamefont{F.}~\bibnamefont{Furche}},
  \bibinfo{journal}{J.~Chem.~Phys.} \textbf{\bibinfo{volume}{114}},
  \bibinfo{pages}{5982} (\bibinfo{year}{2001}).

\bibitem[{\citenamefont{Grimme}(2004)}]{grimme04}
\bibinfo{author}{\bibfnamefont{S.}~\bibnamefont{Grimme}}, in
  \emph{\bibinfo{booktitle}{Reviews in {C}omputational {C}hemistry, {V}ol.
  20}}, edited by \bibinfo{editor}{\bibfnamefont{K.~B.}
  \bibnamefont{Lipkowitz}} \bibnamefont{and}
  \bibinfo{editor}{\bibfnamefont{D.~B.} \bibnamefont{Boyd}}
  (\bibinfo{publisher}{{Wiley-VCH}}, \bibinfo{address}{New York},
  \bibinfo{year}{2004}), p. \bibinfo{pages}{153}.

\bibitem[{\citenamefont{Burke et~al.}(2005)\citenamefont{Burke, Werschnik, and
  Gross}}]{burke05}
\bibinfo{author}{\bibfnamefont{K.}~\bibnamefont{Burke}},
  \bibinfo{author}{\bibfnamefont{J.}~\bibnamefont{Werschnik}},
  \bibnamefont{and} \bibinfo{author}{\bibfnamefont{E.~K.~U.}
  \bibnamefont{Gross}}, \bibinfo{journal}{J.~Chem.~Phys.}
  \textbf{\bibinfo{volume}{123}}, \bibinfo{pages}{062206}
  (\bibinfo{year}{2005}).

\bibitem[{\citenamefont{Dreuw et~al.}(2003)\citenamefont{Dreuw, Weisman, and
  Head-Gordan}}]{dreuw03}
\bibinfo{author}{\bibfnamefont{A.}~\bibnamefont{Dreuw}},
  \bibinfo{author}{\bibfnamefont{J.~L.} \bibnamefont{Weisman}},
  \bibnamefont{and}
  \bibinfo{author}{\bibfnamefont{M.}~\bibnamefont{Head-Gordan}},
  \bibinfo{journal}{J.~Chem.~Phys.} \textbf{\bibinfo{volume}{119}},
  \bibinfo{pages}{2943} (\bibinfo{year}{2003}).

\bibitem[{\citenamefont{Tozer}(2003)}]{tozer03}
\bibinfo{author}{\bibfnamefont{D.~J.} \bibnamefont{Tozer}},
  \bibinfo{journal}{J.~Chem.~Phys.} \textbf{\bibinfo{volume}{119}},
  \bibinfo{pages}{12697} (\bibinfo{year}{2003}).

\bibitem[{\citenamefont{Jacquemin et~al.}(2006)\citenamefont{Jacquemin, Bouhy,
  and Perpete}}]{jacquemin06}
\bibinfo{author}{\bibfnamefont{D.}~\bibnamefont{Jacquemin}},
  \bibinfo{author}{\bibfnamefont{M.}~\bibnamefont{Bouhy}}, \bibnamefont{and}
  \bibinfo{author}{\bibfnamefont{E.~A.} \bibnamefont{Perpete}},
  \bibinfo{journal}{J.~Chem.~Phys.} \textbf{\bibinfo{volume}{124}},
  \bibinfo{pages}{204321} (\bibinfo{year}{2006}).

\bibitem[{\citenamefont{Adams et~al.}(2007)\citenamefont{Adams, Captain, Hall,
  Trufan, and Yang}}]{adams07}
\bibinfo{author}{\bibfnamefont{R.~D.} \bibnamefont{Adams}},
  \bibinfo{author}{\bibfnamefont{B.}~\bibnamefont{Captain}},
  \bibinfo{author}{\bibfnamefont{M.~B.} \bibnamefont{Hall}},
  \bibinfo{author}{\bibfnamefont{E.}~\bibnamefont{Trufan}}, \bibnamefont{and}
  \bibinfo{author}{\bibfnamefont{X.~Z.} \bibnamefont{Yang}},
  \bibinfo{journal}{J.~Am.~Chem.~Soc} \textbf{\bibinfo{volume}{129}},
  \bibinfo{pages}{12328} (\bibinfo{year}{2007}).

\bibitem[{\citenamefont{Tao et~al.}(2008)\citenamefont{Tao, Tretiak, and
  Zhu}}]{tao08}
\bibinfo{author}{\bibfnamefont{J.}~\bibnamefont{Tao}},
  \bibinfo{author}{\bibfnamefont{S.}~\bibnamefont{Tretiak}}, \bibnamefont{and}
  \bibinfo{author}{\bibfnamefont{J.-X.} \bibnamefont{Zhu}},
  \bibinfo{journal}{J.~Chem.~Phys.} \textbf{\bibinfo{volume}{128}},
  \bibinfo{pages}{084110} (\bibinfo{year}{2008}).

\bibitem[{\citenamefont{Nakata et~al.}(2006)\citenamefont{Nakata, Imamura, and
  Nakai}}]{nakata06}
\bibinfo{author}{\bibfnamefont{A.}~\bibnamefont{Nakata}},
  \bibinfo{author}{\bibfnamefont{Y.}~\bibnamefont{Imamura}}, \bibnamefont{and}
  \bibinfo{author}{\bibfnamefont{H.}~\bibnamefont{Nakai}},
  \bibinfo{journal}{J.~Chem.~Phys.} \textbf{\bibinfo{volume}{125}},
  \bibinfo{pages}{064109} (\bibinfo{year}{2006}).

\bibitem[{\citenamefont{Nakata et~al.}(2007)\citenamefont{Nakata, Imamura, and
  Nakai}}]{nakata07}
\bibinfo{author}{\bibfnamefont{A.}~\bibnamefont{Nakata}},
  \bibinfo{author}{\bibfnamefont{Y.}~\bibnamefont{Imamura}}, \bibnamefont{and}
  \bibinfo{author}{\bibfnamefont{H.}~\bibnamefont{Nakai}},
  \bibinfo{journal}{J.~Chem.~Thoery Comput.} \textbf{\bibinfo{volume}{3}},
  \bibinfo{pages}{1295} (\bibinfo{year}{2007}).

\bibitem[{\citenamefont{Imamura and Nakai}(2007)}]{imamura07}
\bibinfo{author}{\bibfnamefont{Y.}~\bibnamefont{Imamura}} \bibnamefont{and}
  \bibinfo{author}{\bibfnamefont{H.}~\bibnamefont{Nakai}},
  \bibinfo{journal}{Int.~J.~Quant.~Chem.} \textbf{\bibinfo{volume}{107}},
  \bibinfo{pages}{23} (\bibinfo{year}{2007}).

\bibitem[{\citenamefont{Imamura and Nakai}(2006)}]{imamura06}
\bibinfo{author}{\bibfnamefont{Y.}~\bibnamefont{Imamura}} \bibnamefont{and}
  \bibinfo{author}{\bibfnamefont{H.}~\bibnamefont{Nakai}},
  \bibinfo{journal}{Chem.~Phys.~Lett.} \textbf{\bibinfo{volume}{419}},
  \bibinfo{pages}{297} (\bibinfo{year}{2006}).

\bibitem[{\citenamefont{Song et~al.}(2008)\citenamefont{Song, Watson, Nakata,
  and Hirao}}]{song08}
\bibinfo{author}{\bibfnamefont{J.-W.} \bibnamefont{Song}},
  \bibinfo{author}{\bibfnamefont{M.~A.} \bibnamefont{Watson}},
  \bibinfo{author}{\bibfnamefont{A.}~\bibnamefont{Nakata}}, \bibnamefont{and}
  \bibinfo{author}{\bibfnamefont{K.}~\bibnamefont{Hirao}},
  \bibinfo{journal}{J.~Chem.~Phys.} \textbf{\bibinfo{volume}{129}},
  \bibinfo{pages}{184113} (\bibinfo{year}{2008}).

\bibitem[{\citenamefont{Asmuruf and Besley}(2008)}]{asmuruf08}
\bibinfo{author}{\bibfnamefont{F.~A.} \bibnamefont{Asmuruf}} \bibnamefont{and}
  \bibinfo{author}{\bibfnamefont{N.~A.} \bibnamefont{Besley}},
  \bibinfo{journal}{Chem.~Phys.~Lett.} \textbf{\bibinfo{volume}{463}},
  \bibinfo{pages}{267} (\bibinfo{year}{2008}).

\bibitem[{\citenamefont{Rappoport and Furche}(2005)}]{rappoport05}
\bibinfo{author}{\bibfnamefont{D.}~\bibnamefont{Rappoport}} \bibnamefont{and}
  \bibinfo{author}{\bibfnamefont{F.}~\bibnamefont{Furche}},
  \bibinfo{journal}{J.~Chem.~Phys.} \textbf{\bibinfo{volume}{122}},
  \bibinfo{pages}{064105} (\bibinfo{year}{2005}).

\bibitem[{\citenamefont{Chiba et~al.}(2006)\citenamefont{Chiba, Tsuneda, and
  Hirao}}]{chiba06}
\bibinfo{author}{\bibfnamefont{M.}~\bibnamefont{Chiba}},
  \bibinfo{author}{\bibfnamefont{T.}~\bibnamefont{Tsuneda}}, \bibnamefont{and}
  \bibinfo{author}{\bibfnamefont{K.}~\bibnamefont{Hirao}},
  \bibinfo{journal}{Chem.~Phys.~Lett.} \textbf{\bibinfo{volume}{420}},
  \bibinfo{pages}{391} (\bibinfo{year}{2006}).

\bibitem[{\citenamefont{Chiba et~al.}(2007)\citenamefont{Chiba, Fedorov, and
  Kitaura}}]{chiba07}
\bibinfo{author}{\bibfnamefont{M.}~\bibnamefont{Chiba}},
  \bibinfo{author}{\bibfnamefont{D.~G.} \bibnamefont{Fedorov}},
  \bibnamefont{and} \bibinfo{author}{\bibfnamefont{K.}~\bibnamefont{Kitaura}},
  \bibinfo{journal}{J.~Chem.~Phys.} \textbf{\bibinfo{volume}{127}},
  \bibinfo{pages}{104108} (\bibinfo{year}{2007}).

\bibitem[{\citenamefont{Guan et~al.}(2006)\citenamefont{Guan, Wang, Ziegler,
  and Cox}}]{guan06}
\bibinfo{author}{\bibfnamefont{J.}~\bibnamefont{Guan}},
  \bibinfo{author}{\bibfnamefont{F.}~\bibnamefont{Wang}},
  \bibinfo{author}{\bibfnamefont{T.}~\bibnamefont{Ziegler}}, \bibnamefont{and}
  \bibinfo{author}{\bibfnamefont{H.}~\bibnamefont{Cox}},
  \bibinfo{journal}{J.~Chem.~Phys.} \textbf{\bibinfo{volume}{125}},
  \bibinfo{pages}{044314} (\bibinfo{year}{2006}).

\bibitem[{\citenamefont{Wang and Ziegler}(2004)}]{wang04}
\bibinfo{author}{\bibfnamefont{F.}~\bibnamefont{Wang}} \bibnamefont{and}
  \bibinfo{author}{\bibfnamefont{T.}~\bibnamefont{Ziegler}},
  \bibinfo{journal}{J.~Chem.~Phys.} \textbf{\bibinfo{volume}{121}},
  \bibinfo{pages}{12191} (\bibinfo{year}{2004}).

\bibitem[{\citenamefont{Wang and Ziegler}(2005)}]{wang05}
\bibinfo{author}{\bibfnamefont{F.}~\bibnamefont{Wang}} \bibnamefont{and}
  \bibinfo{author}{\bibfnamefont{T.}~\bibnamefont{Ziegler}},
  \bibinfo{journal}{J.~Chem.~Phys.} \textbf{\bibinfo{volume}{122}},
  \bibinfo{pages}{204103} (\bibinfo{year}{2005}).

\bibitem[{\citenamefont{Grimme and Neese}(2007)}]{grimme07}
\bibinfo{author}{\bibfnamefont{S.}~\bibnamefont{Grimme}} \bibnamefont{and}
  \bibinfo{author}{\bibfnamefont{F.}~\bibnamefont{Neese}},
  \bibinfo{journal}{J.~Chem.~Phys.} \textbf{\bibinfo{volume}{127}},
  \bibinfo{pages}{154116} (\bibinfo{year}{2007}).

\bibitem[{\citenamefont{Ko et~al.}(2008)\citenamefont{Ko, Malick, Braden,
  Friesner, and {T.~J.~Mart\'inez}}}]{ko08}
\bibinfo{author}{\bibfnamefont{C.}~\bibnamefont{Ko}},
  \bibinfo{author}{\bibfnamefont{D.~K.} \bibnamefont{Malick}},
  \bibinfo{author}{\bibfnamefont{D.~A.} \bibnamefont{Braden}},
  \bibinfo{author}{\bibfnamefont{R.~A.} \bibnamefont{Friesner}},
  \bibnamefont{and} \bibinfo{author}{\bibnamefont{{T.~J.~Mart\'inez}}},
  \bibinfo{journal}{J.~Chem.~Phys.} \textbf{\bibinfo{volume}{128}},
  \bibinfo{pages}{104103} (\bibinfo{year}{2008}).

\bibitem[{\citenamefont{Neugebauer}(2007)}]{neugebauer07}
\bibinfo{author}{\bibfnamefont{J.}~\bibnamefont{Neugebauer}},
  \bibinfo{journal}{J.~Chem.~Phys.} \textbf{\bibinfo{volume}{126}},
  \bibinfo{pages}{134116} (\bibinfo{year}{2007}).

\bibitem[{\citenamefont{Hu and Sugino}(2007)}]{hu07}
\bibinfo{author}{\bibfnamefont{C.}~\bibnamefont{Hu}} \bibnamefont{and}
  \bibinfo{author}{\bibfnamefont{O.}~\bibnamefont{Sugino}},
  \bibinfo{journal}{J.~Chem.~Phys.} \textbf{\bibinfo{volume}{126}},
  \bibinfo{pages}{074112} (\bibinfo{year}{2007}).

\bibitem[{\citenamefont{Vahtras and Rinkevicius}(2007)}]{vahtras07}
\bibinfo{author}{\bibfnamefont{O.}~\bibnamefont{Vahtras}} \bibnamefont{and}
  \bibinfo{author}{\bibfnamefont{Z.}~\bibnamefont{Rinkevicius}},
  \bibinfo{journal}{J.~Chem.~Phys.} \textbf{\bibinfo{volume}{126}},
  \bibinfo{pages}{114101} (\bibinfo{year}{2007}).

\bibitem[{\citenamefont{Harbola and Sahni}(1989)}]{harbola89}
\bibinfo{author}{\bibfnamefont{M.}~\bibnamefont{Harbola}} \bibnamefont{and}
  \bibinfo{author}{\bibfnamefont{V.}~\bibnamefont{Sahni}},
  \bibinfo{journal}{Phys.~Rev.~Lett.} \textbf{\bibinfo{volume}{62}},
  \bibinfo{pages}{489} (\bibinfo{year}{1989}).

\bibitem[{\citenamefont{Sahni and Harbola}(1990)}]{sahni90}
\bibinfo{author}{\bibfnamefont{V.}~\bibnamefont{Sahni}} \bibnamefont{and}
  \bibinfo{author}{\bibfnamefont{M.}~\bibnamefont{Harbola}},
  \bibinfo{journal}{Int.~J.~Quant.~Chem.~Symp.} \textbf{\bibinfo{volume}{24}},
  \bibinfo{pages}{569} (\bibinfo{year}{1990}).

\bibitem[{\citenamefont{Holas and March}(1995)}]{holas95}
\bibinfo{author}{\bibfnamefont{A.}~\bibnamefont{Holas}} \bibnamefont{and}
  \bibinfo{author}{\bibfnamefont{N.~H.} \bibnamefont{March}},
  \bibinfo{journal}{Phys.~Rev.~A} \textbf{\bibinfo{volume}{51}},
  \bibinfo{pages}{2040} (\bibinfo{year}{1995}).

\bibitem[{\citenamefont{Sahni}(1997)}]{sahni97}
\bibinfo{author}{\bibfnamefont{V.}~\bibnamefont{Sahni}},
  \bibinfo{journal}{Phys.~Rev.~A} \textbf{\bibinfo{volume}{55}},
  \bibinfo{pages}{1846} (\bibinfo{year}{1997}).

\bibitem[{\citenamefont{Rose}(1957)}]{rose57}
\bibinfo{author}{\bibfnamefont{M.~E.} \bibnamefont{Rose}},
  \emph{\bibinfo{title}{Elementary Theory of Angular Momentum}}
  (\bibinfo{publisher}{{Wiley}}, \bibinfo{address}{{New York}},
  \bibinfo{year}{1957}).

\bibitem[{\citenamefont{Brual and Rothstein}(1978)}]{brual78}
\bibinfo{author}{\bibfnamefont{G.}~\bibnamefont{Brual}} \bibnamefont{and}
  \bibinfo{author}{\bibfnamefont{S.~M.} \bibnamefont{Rothstein}},
  \bibinfo{journal}{J.~Chem.~Phys.} \textbf{\bibinfo{volume}{69}},
  \bibinfo{pages}{1177} (\bibinfo{year}{1978}).

\bibitem[{\citenamefont{Lee et~al.}(1988)\citenamefont{Lee, Wang, and
  Parr}}]{lee88}
\bibinfo{author}{\bibfnamefont{C.}~\bibnamefont{Lee}},
  \bibinfo{author}{\bibfnamefont{Y.}~\bibnamefont{Wang}}, \bibnamefont{and}
  \bibinfo{author}{\bibfnamefont{R.~G.} \bibnamefont{Parr}},
  \bibinfo{journal}{Phys.~Rev.~B} \textbf{\bibinfo{volume}{37}},
  \bibinfo{pages}{785} (\bibinfo{year}{1988}).

\bibitem[{\citenamefont{Roy and Chu}(2002{\natexlab{b}})}]{roy02a}
\bibinfo{author}{\bibfnamefont{A.~K.} \bibnamefont{Roy}} \bibnamefont{and}
  \bibinfo{author}{\bibfnamefont{S.~I.} \bibnamefont{Chu}},
  \bibinfo{journal}{Phys.~Rev.~A} \textbf{\bibinfo{volume}{65}},
  \bibinfo{pages}{043402} (\bibinfo{year}{2002}{\natexlab{b}}).

\bibitem[{\citenamefont{Roy and Chu}(2002{\natexlab{c}})}]{roy02b}
\bibinfo{author}{\bibfnamefont{A.~K.} \bibnamefont{Roy}} \bibnamefont{and}
  \bibinfo{author}{\bibfnamefont{S.~I.} \bibnamefont{Chu}},
  \bibinfo{journal}{J.~Phys.~B} \textbf{\bibinfo{volume}{35}},
  \bibinfo{pages}{2075} (\bibinfo{year}{2002}{\natexlab{c}}).

\bibitem[{\citenamefont{Roy}(2004{\natexlab{b}})}]{roy04}
\bibinfo{author}{\bibfnamefont{A.~K.} \bibnamefont{Roy}},
  \bibinfo{journal}{Phys.~Lett.~A} \textbf{\bibinfo{volume}{321}},
  \bibinfo{pages}{231} (\bibinfo{year}{2004}{\natexlab{b}}).

\bibitem[{\citenamefont{Roy}(2004{\natexlab{c}})}]{roy04a}
\bibinfo{author}{\bibfnamefont{A.~K.} \bibnamefont{Roy}},
  \bibinfo{journal}{J.~Phys.~G} \textbf{\bibinfo{volume}{30}},
  \bibinfo{pages}{269} (\bibinfo{year}{2004}{\natexlab{c}}).

\bibitem[{\citenamefont{Roy}(2005{\natexlab{b}})}]{roy05}
\bibinfo{author}{\bibfnamefont{A.~K.} \bibnamefont{Roy}},
  \bibinfo{journal}{Pramana-J.~Phys.} \textbf{\bibinfo{volume}{38}},
  \bibinfo{pages}{2189} (\bibinfo{year}{2005}{\natexlab{b}}).

\bibitem[{\citenamefont{Roy}(2005{\natexlab{c}})}]{roy05a}
\bibinfo{author}{\bibfnamefont{A.~K.} \bibnamefont{Roy}},
  \bibinfo{journal}{Int.~J.~Quant.~Chem.} \textbf{\bibinfo{volume}{104}},
  \bibinfo{pages}{861} (\bibinfo{year}{2005}{\natexlab{c}}).

\bibitem[{\citenamefont{Sen and Roy}(2006)}]{sen06}
\bibinfo{author}{\bibfnamefont{K.~D.} \bibnamefont{Sen}} \bibnamefont{and}
  \bibinfo{author}{\bibfnamefont{A.~K.} \bibnamefont{Roy}},
  \bibinfo{journal}{Phys.~Lett.~A.} \textbf{\bibinfo{volume}{357}},
  \bibinfo{pages}{112} (\bibinfo{year}{2006}).

\bibitem[{\citenamefont{Roy et~al.}(2008{\natexlab{a}})\citenamefont{Roy,
  Jalbout, and Proynov}}]{roy08}
\bibinfo{author}{\bibfnamefont{A.~K.} \bibnamefont{Roy}},
  \bibinfo{author}{\bibfnamefont{A.~F.} \bibnamefont{Jalbout}},
  \bibnamefont{and} \bibinfo{author}{\bibfnamefont{E.~I.}
  \bibnamefont{Proynov}}, \bibinfo{journal}{J.~Math.~Chem.}
  \textbf{\bibinfo{volume}{44}}, \bibinfo{pages}{260}
  (\bibinfo{year}{2008}{\natexlab{a}}).

\bibitem[{\citenamefont{Roy et~al.}(2008{\natexlab{b}})\citenamefont{Roy,
  Jalbout, and Proynov}}]{roy08a}
\bibinfo{author}{\bibfnamefont{A.~K.} \bibnamefont{Roy}},
  \bibinfo{author}{\bibfnamefont{A.~F.} \bibnamefont{Jalbout}},
  \bibnamefont{and} \bibinfo{author}{\bibfnamefont{E.~I.}
  \bibnamefont{Proynov}}, \bibinfo{journal}{Int.~J.~Quant.~Chem.}
  \textbf{\bibinfo{volume}{108}}, \bibinfo{pages}{827}
  (\bibinfo{year}{2008}{\natexlab{b}}).

\bibitem[{\citenamefont{Roy and Jalbout}(2008)}]{roy08b}
\bibinfo{author}{\bibfnamefont{A.~K.} \bibnamefont{Roy}} \bibnamefont{and}
  \bibinfo{author}{\bibfnamefont{A.~F.} \bibnamefont{Jalbout}},
  \bibinfo{journal}{J.~Mol.~Struct: Theochem} \textbf{\bibinfo{volume}{853}},
  \bibinfo{pages}{27} (\bibinfo{year}{2008}).

\bibitem[{\citenamefont{Gottlieb et~al.}(1984)\citenamefont{Gottlieb, Yousuff,
  and Orszag}}]{gottlieb84}
\bibinfo{author}{\bibfnamefont{D.}~\bibnamefont{Gottlieb}},
  \bibinfo{author}{\bibfnamefont{M.}~\bibnamefont{Yousuff}}, \bibnamefont{and}
  \bibinfo{author}{\bibfnamefont{S.~A.} \bibnamefont{Orszag}}, in
  \emph{\bibinfo{booktitle}{Spectral {M}ethods for {P}artial {D}ifferential
  {E}quations}}, edited by \bibinfo{editor}{\bibfnamefont{R.~G.}
  \bibnamefont{Voigt}},
  \bibinfo{editor}{\bibfnamefont{D.}~\bibnamefont{Gottlieb}}, \bibnamefont{and}
  \bibinfo{editor}{\bibfnamefont{M.~Y.} \bibnamefont{Hussaini}}
  (\bibinfo{publisher}{{SIAM}}, \bibinfo{address}{Philadelphia},
  \bibinfo{year}{1984}).

\bibitem[{\citenamefont{Canuto et~al.}(1988)\citenamefont{Canuto, Hussaini,
  Quarteroni, and Zang}}]{canuto88}
\bibinfo{author}{\bibfnamefont{C.}~\bibnamefont{Canuto}},
  \bibinfo{author}{\bibfnamefont{M.~Y.} \bibnamefont{Hussaini}},
  \bibinfo{author}{\bibfnamefont{A.}~\bibnamefont{Quarteroni}},
  \bibnamefont{and} \bibinfo{author}{\bibfnamefont{T.~A.} \bibnamefont{Zang}},
  \emph{\bibinfo{title}{Spectral {M}ethods in {F}luid {D}ynamics}}
  (\bibinfo{publisher}{{Springer}}, \bibinfo{address}{{Berlin}},
  \bibinfo{year}{1988}).

\bibitem[{\citenamefont{Yao and Chu}(1993)}]{yao93}
\bibinfo{author}{\bibfnamefont{G.}~\bibnamefont{Yao}} \bibnamefont{and}
  \bibinfo{author}{\bibfnamefont{S.~I.} \bibnamefont{Chu}},
  \bibinfo{journal}{Chem.~Phys.~Lett.} \textbf{\bibinfo{volume}{204}},
  \bibinfo{pages}{381} (\bibinfo{year}{1993}).

\bibitem[{\citenamefont{Wang et~al.}(1994)\citenamefont{Wang, Chu, and
  Laughlin}}]{wang94}
\bibinfo{author}{\bibfnamefont{J.}~\bibnamefont{Wang}},
  \bibinfo{author}{\bibfnamefont{S.~I.} \bibnamefont{Chu}}, \bibnamefont{and}
  \bibinfo{author}{\bibfnamefont{C.}~\bibnamefont{Laughlin}},
  \bibinfo{journal}{Phys.~Rev.~A} \textbf{\bibinfo{volume}{50}},
  \bibinfo{pages}{3028} (\bibinfo{year}{1994}).

\bibitem[{\citenamefont{Slater}(1960)}]{slater60}
\bibinfo{author}{\bibfnamefont{J.~C.} \bibnamefont{Slater}},
  \emph{\bibinfo{title}{Quantum Theory of Atomic Structure, Vol.II}}
  (\bibinfo{publisher}{{McGraw-Hill}}, \bibinfo{address}{{New York}},
  \bibinfo{year}{1960}).

\bibitem[{\citenamefont{Wood}(1980)}]{wood80}
\bibinfo{author}{\bibfnamefont{J.~H.} \bibnamefont{Wood}},
  \bibinfo{journal}{J.~Phys.~B} \textbf{\bibinfo{volume}{13}},
  \bibinfo{pages}{1} (\bibinfo{year}{1980}).

\bibitem[{\citenamefont{Lannoo et~al.}(1981)\citenamefont{Lannoo, Baraff, and
  {M.~Schl\"uter}}}]{lannoo81}
\bibinfo{author}{\bibfnamefont{M.}~\bibnamefont{Lannoo}},
  \bibinfo{author}{\bibfnamefont{G.~A.} \bibnamefont{Baraff}},
  \bibnamefont{and} \bibinfo{author}{\bibnamefont{{M.~Schl\"uter}}},
  \bibinfo{journal}{Phys.~Rev.~B} \textbf{\bibinfo{volume}{24}},
  \bibinfo{pages}{943} (\bibinfo{year}{1981}).

\bibitem[{\citenamefont{Dickson and Ziegler}(1996)}]{dickson96}
\bibinfo{author}{\bibfnamefont{R.~M.} \bibnamefont{Dickson}} \bibnamefont{and}
  \bibinfo{author}{\bibfnamefont{T.}~\bibnamefont{Ziegler}},
  \bibinfo{journal}{Int.~J.~Quant.~Chem.} \textbf{\bibinfo{volume}{58}},
  \bibinfo{pages}{681} (\bibinfo{year}{1996}).

\bibitem[{\citenamefont{Pollak et~al.}(1997)\citenamefont{Pollak, Rosa, and
  Baerends}}]{pollak97}
\bibinfo{author}{\bibfnamefont{C.}~\bibnamefont{Pollak}},
  \bibinfo{author}{\bibfnamefont{A.}~\bibnamefont{Rosa}}, \bibnamefont{and}
  \bibinfo{author}{\bibfnamefont{E.~J.} \bibnamefont{Baerends}},
  \bibinfo{journal}{J.~Am.~Chem.~Soc.} \textbf{\bibinfo{volume}{119}},
  \bibinfo{pages}{7324} (\bibinfo{year}{1997}).

\bibitem[{\citenamefont{{W.~A.~Goddard III}}(1968)}]{goddard68}
\bibinfo{author}{\bibnamefont{{W.~A.~Goddard III}}},
  \bibinfo{journal}{Phys.~Rev.} \textbf{\bibinfo{volume}{176}},
  \bibinfo{pages}{106} (\bibinfo{year}{1968}).

\bibitem[{\citenamefont{Wang et~al.}(1992)\citenamefont{Wang, Zh, and
  Chung}}]{wang92}
\bibinfo{author}{\bibfnamefont{Z.-W.} \bibnamefont{Wang}},
  \bibinfo{author}{\bibfnamefont{X.-W.} \bibnamefont{Zh}}, \bibnamefont{and}
  \bibinfo{author}{\bibfnamefont{K.~T.} \bibnamefont{Chung}},
  \bibinfo{journal}{Phys.~Rev.~A} \textbf{\bibinfo{volume}{46}},
  \bibinfo{pages}{6914} (\bibinfo{year}{1992}).

\bibitem[{\citenamefont{Sims and Hagstrom}(1975)}]{sims75}
\bibinfo{author}{\bibfnamefont{J.~S.} \bibnamefont{Sims}} \bibnamefont{and}
  \bibinfo{author}{\bibfnamefont{S.~A.} \bibnamefont{Hagstrom}},
  \bibinfo{journal}{Phys.~Rev.~A} \textbf{\bibinfo{volume}{11}},
  \bibinfo{pages}{418} (\bibinfo{year}{1975}).

\bibitem[{\citenamefont{Weiss}(1972)}]{weiss72}
\bibinfo{author}{\bibfnamefont{A.~W.} \bibnamefont{Weiss}},
  \bibinfo{journal}{Phys.~Rev.~A} \textbf{\bibinfo{volume}{6}},
  \bibinfo{pages}{1261} (\bibinfo{year}{1972}).

\bibitem[{\citenamefont{Koyama et~al.}(1986)\citenamefont{Koyama, Fukuda,
  Motoyama, and Matsuzawa}}]{koyama86}
\bibinfo{author}{\bibfnamefont{N.}~\bibnamefont{Koyama}},
  \bibinfo{author}{\bibfnamefont{H.}~\bibnamefont{Fukuda}},
  \bibinfo{author}{\bibfnamefont{T.}~\bibnamefont{Motoyama}}, \bibnamefont{and}
  \bibinfo{author}{\bibfnamefont{M.~J.} \bibnamefont{Matsuzawa}},
  \bibinfo{journal}{J.~Phys.~B} \textbf{\bibinfo{volume}{19}},
  \bibinfo{pages}{L331} (\bibinfo{year}{1986}).

\bibitem[{\citenamefont{{A.~B\"urgers}
  et~al.}(1995)\citenamefont{{A.~B\"urgers}, Wintgen, and Rost}}]{burgers95}
\bibinfo{author}{\bibnamefont{{A.~B\"urgers}}},
  \bibinfo{author}{\bibfnamefont{D.}~\bibnamefont{Wintgen}}, \bibnamefont{and}
  \bibinfo{author}{\bibfnamefont{J.-M.} \bibnamefont{Rost}},
  \bibinfo{journal}{J.~Phys.~B} \textbf{\bibinfo{volume}{28}},
  \bibinfo{pages}{3163} (\bibinfo{year}{1995}).

\bibitem[{\citenamefont{Lindroth}(1994)}]{lindroth94}
\bibinfo{author}{\bibfnamefont{E.}~\bibnamefont{Lindroth}},
  \bibinfo{journal}{Phys.~Rev.~A} \textbf{\bibinfo{volume}{49}},
  \bibinfo{pages}{4473} (\bibinfo{year}{1994}).

\bibitem[{\citenamefont{Savin et~al.}(1998)\citenamefont{Savin, Umrigar, and
  Gonze}}]{savin98}
\bibinfo{author}{\bibfnamefont{A.}~\bibnamefont{Savin}},
  \bibinfo{author}{\bibfnamefont{C.~J.} \bibnamefont{Umrigar}},
  \bibnamefont{and} \bibinfo{author}{\bibfnamefont{X.}~\bibnamefont{Gonze}},
  \bibinfo{journal}{Chem.~Phys.~Lett.} \textbf{\bibinfo{volume}{288}},
  \bibinfo{pages}{391} (\bibinfo{year}{1998}).

\bibitem[{\citenamefont{Drake}(1993)}]{drake93}
\bibinfo{author}{\bibfnamefont{G.~W.~F.} \bibnamefont{Drake}}, in
  \emph{\bibinfo{booktitle}{Casimir Forces: Theory and Esperiments on Atomic
  Systems}}, edited by \bibinfo{editor}{\bibfnamefont{F.~S.}
  \bibnamefont{Levin}} \bibnamefont{and} \bibinfo{editor}{\bibfnamefont{D.~A.}
  \bibnamefont{Micha}} (\bibinfo{publisher}{{Plenum}}, \bibinfo{address}{New
  York}, \bibinfo{year}{1993}).

\bibitem[{\citenamefont{Drake and Yan}(1994)}]{drake94}
\bibinfo{author}{\bibfnamefont{G.~F.} \bibnamefont{Drake}} \bibnamefont{and}
  \bibinfo{author}{\bibfnamefont{Z.~C.} \bibnamefont{Yan}},
  \bibinfo{journal}{Chem.~Phys.~Lett.} \textbf{\bibinfo{volume}{229}},
  \bibinfo{pages}{486} (\bibinfo{year}{1994}).

\bibitem[{\citenamefont{Bashkin and {J.~O.~Stoner Jr.}}(1975)}]{bashkin75}
\bibinfo{author}{\bibfnamefont{S.}~\bibnamefont{Bashkin}} \bibnamefont{and}
  \bibinfo{author}{\bibnamefont{{J.~O.~Stoner Jr.}}},
  \emph{\bibinfo{title}{Atomic Energy Levels and Grotrian Diagrams, Vol. I}}
  (\bibinfo{publisher}{{North-Holland}}, \bibinfo{address}{{Amsterdam}},
  \bibinfo{year}{1975}).

\bibitem[{\citenamefont{Begue et~al.}(1998)\citenamefont{Begue, Merawa, and
  Pouchan}}]{begue98}
\bibinfo{author}{\bibfnamefont{D.}~\bibnamefont{Begue}},
  \bibinfo{author}{\bibfnamefont{M.}~\bibnamefont{Merawa}}, \bibnamefont{and}
  \bibinfo{author}{\bibfnamefont{C.}~\bibnamefont{Pouchan}},
  \bibinfo{journal}{Phys.~Rev.~A} \textbf{\bibinfo{volume}{57}},
  \bibinfo{pages}{2470} (\bibinfo{year}{1998}).

\bibitem[{\citenamefont{Ho}(1981)}]{ho81}
\bibinfo{author}{\bibfnamefont{Y.~K.} \bibnamefont{Ho}},
  \bibinfo{journal}{Phys.~Rev.~A} \textbf{\bibinfo{volume}{23}},
  \bibinfo{pages}{2137} (\bibinfo{year}{1981}).

\bibitem[{\citenamefont{Herrick and Sinanoglu}(1975)}]{herrick75}
\bibinfo{author}{\bibfnamefont{D.~R.} \bibnamefont{Herrick}} \bibnamefont{and}
  \bibinfo{author}{\bibfnamefont{O.}~\bibnamefont{Sinanoglu}},
  \bibinfo{journal}{Phys.~Rev.~A} \textbf{\bibinfo{volume}{11}},
  \bibinfo{pages}{97} (\bibinfo{year}{1975}).

\bibitem[{\citenamefont{Fukuda et~al.}(1987)\citenamefont{Fukuda, Koyama, and
  Matsuzawa}}]{fukuda87}
\bibinfo{author}{\bibfnamefont{H.}~\bibnamefont{Fukuda}},
  \bibinfo{author}{\bibfnamefont{N.}~\bibnamefont{Koyama}}, \bibnamefont{and}
  \bibinfo{author}{\bibfnamefont{M.}~\bibnamefont{Matsuzawa}},
  \bibinfo{journal}{J.~Phys.~B} \textbf{\bibinfo{volume}{20}},
  \bibinfo{pages}{2959} (\bibinfo{year}{1987}).

\bibitem[{\citenamefont{Becke}(1988)}]{becke88}
\bibinfo{author}{\bibfnamefont{A.~D.} \bibnamefont{Becke}},
  \bibinfo{journal}{Phys.~Rev.~A} \textbf{\bibinfo{volume}{38}},
  \bibinfo{pages}{3098} (\bibinfo{year}{1988}).

\bibitem[{\citenamefont{Fischer}(1977)}]{fischer77}
\bibinfo{author}{\bibfnamefont{C.~F.} \bibnamefont{Fischer}},
  \emph{\bibinfo{title}{The Hartree-Fock Method for Atoms}}
  (\bibinfo{publisher}{{John Wiley}}, \bibinfo{address}{{New York}},
  \bibinfo{year}{1977}).

\bibitem[{\citenamefont{Yan et~al.}(1998)\citenamefont{Yan, Tambasco, and
  Drake}}]{yan98}
\bibinfo{author}{\bibfnamefont{Z.-C.} \bibnamefont{Yan}},
  \bibinfo{author}{\bibfnamefont{M.}~\bibnamefont{Tambasco}}, \bibnamefont{and}
  \bibinfo{author}{\bibfnamefont{G.~W.~F.} \bibnamefont{Drake}},
  \bibinfo{journal}{Phys.~Rev.~A} \textbf{\bibinfo{volume}{57}},
  \bibinfo{pages}{1652} (\bibinfo{year}{1998}).

\bibitem[{\citenamefont{Safronova and Bruch}(1998)}]{safronova98}
\bibinfo{author}{\bibfnamefont{U.~I.} \bibnamefont{Safronova}}
  \bibnamefont{and} \bibinfo{author}{\bibfnamefont{R.}~\bibnamefont{Bruch}},
  \bibinfo{journal}{Phys.~Scr.} \textbf{\bibinfo{volume}{57}},
  \bibinfo{pages}{519} (\bibinfo{year}{1998}).

\bibitem[{\citenamefont{Conneeely and Lipsky}(2002)}]{conneely02}
\bibinfo{author}{\bibfnamefont{M.~J.} \bibnamefont{Conneeely}}
  \bibnamefont{and} \bibinfo{author}{\bibfnamefont{L.}~\bibnamefont{Lipsky}},
  \bibinfo{journal}{At.~Data Nucl.~Data Tables} \textbf{\bibinfo{volume}{82}},
  \bibinfo{pages}{115} (\bibinfo{year}{2002}).

\bibitem[{\citenamefont{Conneeely and Lipsky}(2004)}]{conneely04}
\bibinfo{author}{\bibfnamefont{M.~J.} \bibnamefont{Conneeely}}
  \bibnamefont{and} \bibinfo{author}{\bibfnamefont{L.}~\bibnamefont{Lipsky}},
  \bibinfo{journal}{At.~Data Nucl.~Data Tables} \textbf{\bibinfo{volume}{86}},
  \bibinfo{pages}{35} (\bibinfo{year}{2004}).

\bibitem[{\citenamefont{Davis and Chung}(1990)}]{davis90}
\bibinfo{author}{\bibfnamefont{B.~F.} \bibnamefont{Davis}} \bibnamefont{and}
  \bibinfo{author}{\bibfnamefont{K.~T.} \bibnamefont{Chung}},
  \bibinfo{journal}{Phys.~Rev.~A} \textbf{\bibinfo{volume}{42}},
  \bibinfo{pages}{5121} (\bibinfo{year}{1990}).

\bibitem[{\citenamefont{Chung}(1991)}]{chung91}
\bibinfo{author}{\bibfnamefont{K.~T.} \bibnamefont{Chung}},
  \bibinfo{journal}{Phys.~Rev.~A} \textbf{\bibinfo{volume}{44}},
  \bibinfo{pages}{5421} (\bibinfo{year}{1991}).

\bibitem[{\citenamefont{Chung and Gou}(1995)}]{chung95}
\bibinfo{author}{\bibfnamefont{K.~T.} \bibnamefont{Chung}} \bibnamefont{and}
  \bibinfo{author}{\bibfnamefont{B.~C.} \bibnamefont{Gou}},
  \bibinfo{journal}{Phys.~Rev.~A} \textbf{\bibinfo{volume}{52}},
  \bibinfo{pages}{3669} (\bibinfo{year}{1995}).

\bibitem[{\citenamefont{Madsen et~al.}(2000)\citenamefont{Madsen, Schlagheck,
  and Lambropoulos}}]{madsen00}
\bibinfo{author}{\bibfnamefont{L.~B.} \bibnamefont{Madsen}},
  \bibinfo{author}{\bibfnamefont{P.}~\bibnamefont{Schlagheck}},
  \bibnamefont{and}
  \bibinfo{author}{\bibfnamefont{P.}~\bibnamefont{Lambropoulos}},
  \bibinfo{journal}{Phys.~Rev.~A} \textbf{\bibinfo{volume}{62}},
  \bibinfo{pages}{062719} (\bibinfo{year}{2000}).

\bibitem[{\citenamefont{Diehl et~al.}(1996)\citenamefont{Diehl, Cubaynes,
  Bizau, Journel, Rouvellou, Moussalami, Wuilleumier, Kennedy, Berrah, Blancard
  et~al.}}]{diehl96}
\bibinfo{author}{\bibfnamefont{S.}~\bibnamefont{Diehl}},
  \bibinfo{author}{\bibfnamefont{D.}~\bibnamefont{Cubaynes}},
  \bibinfo{author}{\bibfnamefont{J.-M.} \bibnamefont{Bizau}},
  \bibinfo{author}{\bibfnamefont{L.}~\bibnamefont{Journel}},
  \bibinfo{author}{\bibfnamefont{B.}~\bibnamefont{Rouvellou}},
  \bibinfo{author}{\bibfnamefont{S.~A.} \bibnamefont{Moussalami}},
  \bibinfo{author}{\bibfnamefont{F.~J.} \bibnamefont{Wuilleumier}},
  \bibinfo{author}{\bibfnamefont{E.~T.} \bibnamefont{Kennedy}},
  \bibinfo{author}{\bibfnamefont{N.}~\bibnamefont{Berrah}},
  \bibinfo{author}{\bibfnamefont{C.}~\bibnamefont{Blancard}},
  \bibnamefont{et~al.}, \bibinfo{journal}{Phys.~Rev.~Lett.}
  \textbf{\bibinfo{volume}{76}}, \bibinfo{pages}{3915} (\bibinfo{year}{1996}).

\bibitem[{\citenamefont{Berrington and Nakazaki}(1998)}]{berrington98}
\bibinfo{author}{\bibfnamefont{K.}~\bibnamefont{Berrington}} \bibnamefont{and}
  \bibinfo{author}{\bibfnamefont{S.}~\bibnamefont{Nakazaki}},
  \bibinfo{journal}{J.~Phys.~B} \textbf{\bibinfo{volume}{31}},
  \bibinfo{pages}{313} (\bibinfo{year}{1998}).

\bibitem[{\citenamefont{Azuma et~al.}(1995)\citenamefont{Azuma, Hasegawa,
  Koike, Kutluk, Nagata, Shigemasa, Yagishta, and Sellin}}]{azuma95}
\bibinfo{author}{\bibfnamefont{Y.}~\bibnamefont{Azuma}},
  \bibinfo{author}{\bibfnamefont{H.}~\bibnamefont{Hasegawa}},
  \bibinfo{author}{\bibfnamefont{F.}~\bibnamefont{Koike}},
  \bibinfo{author}{\bibfnamefont{G.}~\bibnamefont{Kutluk}},
  \bibinfo{author}{\bibfnamefont{T.}~\bibnamefont{Nagata}},
  \bibinfo{author}{\bibfnamefont{E.}~\bibnamefont{Shigemasa}},
  \bibinfo{author}{\bibfnamefont{A.}~\bibnamefont{Yagishta}}, \bibnamefont{and}
  \bibinfo{author}{\bibfnamefont{I.~A.} \bibnamefont{Sellin}},
  \bibinfo{journal}{Phys.~Rev.~Lett.} \textbf{\bibinfo{volume}{74}},
  \bibinfo{pages}{3768} (\bibinfo{year}{1995}).

\bibitem[{\citenamefont{Kiernan et~al.}(1995)\citenamefont{Kiernan, Lee,
  Sonntag, Sladeczek, Zimmermann, Kennedy, Mosnier, and Costello}}]{kiernan95}
\bibinfo{author}{\bibfnamefont{L.~M.} \bibnamefont{Kiernan}},
  \bibinfo{author}{\bibfnamefont{M.-K.} \bibnamefont{Lee}},
  \bibinfo{author}{\bibfnamefont{B.~F.} \bibnamefont{Sonntag}},
  \bibinfo{author}{\bibfnamefont{P.}~\bibnamefont{Sladeczek}},
  \bibinfo{author}{\bibfnamefont{P.}~\bibnamefont{Zimmermann}},
  \bibinfo{author}{\bibfnamefont{E.~T.} \bibnamefont{Kennedy}},
  \bibinfo{author}{\bibfnamefont{J.-P.} \bibnamefont{Mosnier}},
  \bibnamefont{and} \bibinfo{author}{\bibfnamefont{J.~T.}
  \bibnamefont{Costello}}, \bibinfo{journal}{J.~Phys.~B}
  \textbf{\bibinfo{volume}{28}}, \bibinfo{pages}{L161} (\bibinfo{year}{1995}).

\bibitem[{\citenamefont{Morishita and Lin}(2003)}]{morishita03}
\bibinfo{author}{\bibfnamefont{T.}~\bibnamefont{Morishita}} \bibnamefont{and}
  \bibinfo{author}{\bibfnamefont{C.~D.} \bibnamefont{Lin}},
  \bibinfo{journal}{Phys.~Rev.~A} \textbf{\bibinfo{volume}{67}},
  \bibinfo{pages}{022511} (\bibinfo{year}{2003}).

\bibitem[{\citenamefont{Zhou et~al.}(2000)\citenamefont{Zhou, Manson, Faucher,
  and {L.~Vo Ky}}}]{zhou00}
\bibinfo{author}{\bibfnamefont{H.~L.} \bibnamefont{Zhou}},
  \bibinfo{author}{\bibfnamefont{S.~T.} \bibnamefont{Manson}},
  \bibinfo{author}{\bibfnamefont{P.}~\bibnamefont{Faucher}}, \bibnamefont{and}
  \bibinfo{author}{\bibnamefont{{L.~Vo Ky}}}, \bibinfo{journal}{Phys.~Rev.~A}
  \textbf{\bibinfo{volume}{62}}, \bibinfo{pages}{012707}
  (\bibinfo{year}{2000}).

\bibitem[{\citenamefont{Chung and Gou}(1996)}]{chung96}
\bibinfo{author}{\bibfnamefont{K.~T.} \bibnamefont{Chung}} \bibnamefont{and}
  \bibinfo{author}{\bibfnamefont{B.~C.} \bibnamefont{Gou}},
  \bibinfo{journal}{Phys.~Rev.~A} \textbf{\bibinfo{volume}{53}},
  \bibinfo{pages}{2189} (\bibinfo{year}{1996}).

\bibitem[{\citenamefont{{L.~Vo Ky} et~al.}(1998)\citenamefont{{L.~Vo Ky},
  Faucher, Zhou, Hibbert, Qu, Li, and Bely-Dubau}}]{voky98}
\bibinfo{author}{\bibnamefont{{L.~Vo Ky}}},
  \bibinfo{author}{\bibfnamefont{P.}~\bibnamefont{Faucher}},
  \bibinfo{author}{\bibfnamefont{H.~L.} \bibnamefont{Zhou}},
  \bibinfo{author}{\bibfnamefont{A.}~\bibnamefont{Hibbert}},
  \bibinfo{author}{\bibfnamefont{Y.-Z.} \bibnamefont{Qu}},
  \bibinfo{author}{\bibfnamefont{J.-M.} \bibnamefont{Li}}, \bibnamefont{and}
  \bibinfo{author}{\bibfnamefont{F.}~\bibnamefont{Bely-Dubau}},
  \bibinfo{journal}{Phys.~Rev.~A} \textbf{\bibinfo{volume}{58}},
  \bibinfo{pages}{3688} (\bibinfo{year}{1998}).

\bibitem[{\citenamefont{Zhang and Chung}(1998)}]{zhang98}
\bibinfo{author}{\bibfnamefont{Y.}~\bibnamefont{Zhang}} \bibnamefont{and}
  \bibinfo{author}{\bibfnamefont{K.~T.} \bibnamefont{Chung}},
  \bibinfo{journal}{Phys.~Rev.~A} \textbf{\bibinfo{volume}{58}},
  \bibinfo{pages}{1098} (\bibinfo{year}{1998}).

\bibitem[{\citenamefont{Diehl et~al.}(1997)\citenamefont{Diehl, Cubaynes,
  Chung, Wuilleumier, Kennedy, Bizau, Journel, Blancard, {L.~Vo Ky}, Faucher
  et~al.}}]{diehl97}
\bibinfo{author}{\bibfnamefont{S.}~\bibnamefont{Diehl}},
  \bibinfo{author}{\bibfnamefont{D.}~\bibnamefont{Cubaynes}},
  \bibinfo{author}{\bibfnamefont{K.~T.} \bibnamefont{Chung}},
  \bibinfo{author}{\bibfnamefont{F.~J.} \bibnamefont{Wuilleumier}},
  \bibinfo{author}{\bibfnamefont{E.~T.} \bibnamefont{Kennedy}},
  \bibinfo{author}{\bibfnamefont{J.-M.} \bibnamefont{Bizau}},
  \bibinfo{author}{\bibfnamefont{L.}~\bibnamefont{Journel}},
  \bibinfo{author}{\bibfnamefont{C.}~\bibnamefont{Blancard}},
  \bibinfo{author}{\bibnamefont{{L.~Vo Ky}}},
  \bibinfo{author}{\bibfnamefont{P.}~\bibnamefont{Faucher}},
  \bibnamefont{et~al.}, \bibinfo{journal}{Phys.~Rev.~A}
  \textbf{\bibinfo{volume}{56}}, \bibinfo{pages}{R1071} (\bibinfo{year}{1997}).

\bibitem[{\citenamefont{Piangos and Nicolaides}(2003)}]{piangos03}
\bibinfo{author}{\bibfnamefont{N.~A.} \bibnamefont{Piangos}} \bibnamefont{and}
  \bibinfo{author}{\bibfnamefont{C.~A.} \bibnamefont{Nicolaides}},
  \bibinfo{journal}{Phys.~Rev.~A} \textbf{\bibinfo{volume}{67}},
  \bibinfo{pages}{052501} (\bibinfo{year}{2003}).

\bibitem[{\citenamefont{Azuma et~al.}(1997)\citenamefont{Azuma, Koike, Cooper,
  Nagata, Kutluk, Shigemasa, Wehlitz, and Sellin}}]{azuma97}
\bibinfo{author}{\bibfnamefont{Y.}~\bibnamefont{Azuma}},
  \bibinfo{author}{\bibfnamefont{F.}~\bibnamefont{Koike}},
  \bibinfo{author}{\bibfnamefont{J.~W.} \bibnamefont{Cooper}},
  \bibinfo{author}{\bibfnamefont{T.}~\bibnamefont{Nagata}},
  \bibinfo{author}{\bibfnamefont{G.}~\bibnamefont{Kutluk}},
  \bibinfo{author}{\bibfnamefont{E.}~\bibnamefont{Shigemasa}},
  \bibinfo{author}{\bibfnamefont{R.}~\bibnamefont{Wehlitz}}, \bibnamefont{and}
  \bibinfo{author}{\bibfnamefont{I.~A.} \bibnamefont{Sellin}},
  \bibinfo{journal}{Phys.~Rev.~Lett.} \textbf{\bibinfo{volume}{79}},
  \bibinfo{pages}{2419} (\bibinfo{year}{1997}).

\bibitem[{\citenamefont{Conneely and Lipsky}(2000)}]{conneely00}
\bibinfo{author}{\bibfnamefont{M.~J.} \bibnamefont{Conneely}} \bibnamefont{and}
  \bibinfo{author}{\bibfnamefont{L.}~\bibnamefont{Lipsky}},
  \bibinfo{journal}{Phys.~Rev.~A} \textbf{\bibinfo{volume}{61}},
  \bibinfo{pages}{032506} (\bibinfo{year}{2000}).

\bibitem[{\citenamefont{Vacek and Hansen}(1992)}]{vacek92}
\bibinfo{author}{\bibfnamefont{N.}~\bibnamefont{Vacek}} \bibnamefont{and}
  \bibinfo{author}{\bibfnamefont{J.~E.} \bibnamefont{Hansen}},
  \bibinfo{journal}{J.~Phys.~B} \textbf{\bibinfo{volume}{25}},
  \bibinfo{pages}{883} (\bibinfo{year}{1992}).

\bibitem[{\citenamefont{Bachau}(1996)}]{bachau96}
\bibinfo{author}{\bibfnamefont{H.}~\bibnamefont{Bachau}},
  \bibinfo{journal}{J.~Phys.~B} \textbf{\bibinfo{volume}{29}},
  \bibinfo{pages}{4365} (\bibinfo{year}{1996}).

\bibitem[{\citenamefont{Nicolaides and Piangos}(2001)}]{nicolaides01}
\bibinfo{author}{\bibfnamefont{C.~A.} \bibnamefont{Nicolaides}}
  \bibnamefont{and} \bibinfo{author}{\bibfnamefont{N.~A.}
  \bibnamefont{Piangos}}, \bibinfo{journal}{J.~Phys.~B}
  \textbf{\bibinfo{volume}{34}}, \bibinfo{pages}{99} (\bibinfo{year}{2001}).

\bibitem[{\citenamefont{Morishita and Lin}(2001)}]{morishita01}
\bibinfo{author}{\bibfnamefont{T.}~\bibnamefont{Morishita}} \bibnamefont{and}
  \bibinfo{author}{\bibfnamefont{C.~D.} \bibnamefont{Lin}},
  \bibinfo{journal}{Phys.~Rev.~A} \textbf{\bibinfo{volume}{64}},
  \bibinfo{pages}{052502} (\bibinfo{year}{2001}).

\bibitem[{\citenamefont{Chung}(2001)}]{chung01}
\bibinfo{author}{\bibfnamefont{K.~T.} \bibnamefont{Chung}},
  \bibinfo{journal}{Phys.~Rev.~A} \textbf{\bibinfo{volume}{64}},
  \bibinfo{pages}{052503} (\bibinfo{year}{2001}).

\bibitem[{\citenamefont{{F.~J.~G\'alvez and E.~Buend\'ia and
  A.~Sarsa}}(2006)}]{galvez06}
\bibinfo{author}{\bibnamefont{{F.~J.~G\'alvez and E.~Buend\'ia and A.~Sarsa}}},
  \bibinfo{journal}{Eur.~Phys.~J.~D} \textbf{\bibinfo{volume}{40}},
  \bibinfo{pages}{161} (\bibinfo{year}{2006}).

\bibitem[{\citenamefont{Wang et~al.}(2006)\citenamefont{Wang, Gou, and
  Chen}}]{wang06}
\bibinfo{author}{\bibfnamefont{Z.~B.} \bibnamefont{Wang}},
  \bibinfo{author}{\bibfnamefont{B.~C.} \bibnamefont{Gou}}, \bibnamefont{and}
  \bibinfo{author}{\bibfnamefont{F.}~\bibnamefont{Chen}},
  \bibinfo{journal}{Eur.~Phys.~J.~D} \textbf{\bibinfo{volume}{37}},
  \bibinfo{pages}{345} (\bibinfo{year}{2006}).

\bibitem[{\citenamefont{Fischer}(1993)}]{fischer93}
\bibinfo{author}{\bibfnamefont{C.~F.} \bibnamefont{Fischer}},
  \bibinfo{journal}{J.~Phys.~B} \textbf{\bibinfo{volume}{26}},
  \bibinfo{pages}{855} (\bibinfo{year}{1993}).

\bibitem[{\citenamefont{Weiss}(1968)}]{weiss68}
\bibinfo{author}{\bibfnamefont{A.~W.} \bibnamefont{Weiss}},
  \bibinfo{journal}{Phys.~Rev.} \textbf{\bibinfo{volume}{166}},
  \bibinfo{pages}{70} (\bibinfo{year}{1968}).

\bibitem[{\citenamefont{Bunge}(1980)}]{bunge80}
\bibinfo{author}{\bibfnamefont{C.~F.} \bibnamefont{Bunge}},
  \bibinfo{journal}{Phys.~Rev.~A} \textbf{\bibinfo{volume}{22}},
  \bibinfo{pages}{1} (\bibinfo{year}{1980}).

\bibitem[{\citenamefont{Yang and Chung}(1995)}]{yang95}
\bibinfo{author}{\bibfnamefont{H.~Y.} \bibnamefont{Yang}} \bibnamefont{and}
  \bibinfo{author}{\bibfnamefont{K.~T.} \bibnamefont{Chung}},
  \bibinfo{journal}{Phys.~Rev.~A} \textbf{\bibinfo{volume}{51}},
  \bibinfo{pages}{3621} (\bibinfo{year}{1995}).

\bibitem[{\citenamefont{Fischer}(1990)}]{fischer90}
\bibinfo{author}{\bibfnamefont{C.~F.} \bibnamefont{Fischer}},
  \bibinfo{journal}{Phys.~Rev.~A} \textbf{\bibinfo{volume}{41}},
  \bibinfo{pages}{3481} (\bibinfo{year}{1990}).

\bibitem[{\citenamefont{Beck et~al.}(1981)\citenamefont{Beck, Nicolaides, and
  Aspromallis}}]{beck81}
\bibinfo{author}{\bibfnamefont{D.~R.} \bibnamefont{Beck}},
  \bibinfo{author}{\bibfnamefont{C.~A.} \bibnamefont{Nicolaides}},
  \bibnamefont{and}
  \bibinfo{author}{\bibfnamefont{G.}~\bibnamefont{Aspromallis}},
  \bibinfo{journal}{Phys.~Rev.~A} \textbf{\bibinfo{volume}{24}},
  \bibinfo{pages}{3252} (\bibinfo{year}{1981}).

\bibitem[{\citenamefont{Hsu and Chung}(1995{\natexlab{a}})}]{hsu95}
\bibinfo{author}{\bibfnamefont{J.-J.} \bibnamefont{Hsu}} \bibnamefont{and}
  \bibinfo{author}{\bibfnamefont{K.~T.} \bibnamefont{Chung}},
  \bibinfo{journal}{Phys.~Rev.~A} \textbf{\bibinfo{volume}{52}},
  \bibinfo{pages}{R898} (\bibinfo{year}{1995}{\natexlab{a}}).

\bibitem[{\citenamefont{Hsu and Chung}(1995{\natexlab{b}})}]{hsu95a}
\bibinfo{author}{\bibfnamefont{J.-J.} \bibnamefont{Hsu}} \bibnamefont{and}
  \bibinfo{author}{\bibfnamefont{K.~T.} \bibnamefont{Chung}},
  \bibinfo{journal}{J.~Phys.~B} \textbf{\bibinfo{volume}{28}},
  \bibinfo{pages}{L649} (\bibinfo{year}{1995}{\natexlab{b}}).

\bibitem[{\citenamefont{Bunge}(1986)}]{bunge86}
\bibinfo{author}{\bibfnamefont{A.~V.} \bibnamefont{Bunge}},
  \bibinfo{journal}{Phys.~Rev.~A} \textbf{\bibinfo{volume}{33}},
  \bibinfo{pages}{82} (\bibinfo{year}{1986}).

\end{thebibliography}
\end{document}